\documentclass[prb,reprint,aps,
superscriptaddress,
10pt,
floatfix]{revtex4-1}
\usepackage{amsmath}
\usepackage{amssymb}
\usepackage{graphicx}
\usepackage{bm}
\usepackage[top=3cm, bottom=3.25cm, left=2.75cm, right=2.75cm]{geometry}
\usepackage{mathrsfs}
\usepackage{subfigure}

\def\be{\begin{equation}}
\def\ee{\end{equation}}
\def\bea{\begin{eqnarray}}

\def\bea{\end{eqnarray}}
\def\eea{\end{eqnarray}}
\def\bs{\begin{split}}
\def\es{\end{split}}
\def\ni{\noindent}

\def\bi{\begin{itemize}}
\def\ei{\end{itemize}}

\def\a{\alpha}
\def\b{\beta}

\def\p{\partial}

\def\e{\epsilon}
\def\vare{\varepsilon}

\def\d{\delta}
\def\o{\omega}

\def\f{\frac}

\begin{document}
\raggedbottom

\title{Non-perturbative Green’s function method to determine the electronic spectral function due to electron-phonon interactions: Application to a graphene model from weak to strong coupling}
\author{Jean Paul Nery}
\email{nery.jeanpaul@gmail.com}
\author{Francesco Mauri}
\email{francesco.mauri@uniroma1.it}
\affiliation{Dipartimento di Fisica, Universit\`a di Roma La Sapienza, I-00185 Roma, Italy}
\affiliation{Graphene Labs, Fondazione Istituto Italiano di Tecnologia, Via Morego, I-16163 Genova, Italy}
\date{\today}

\begin{abstract}
In solid state physics, the electron-phonon interaction (EPI) is central to many phenomena. The theory of the renormalization of electronic properties due to EPIs became well established with the theory of Allen-Heine-Cardona (AHC), which is usually applied to second order in perturbation theory. However, this is only valid in the weak coupling regime, while strong EPIs have been reported in many materials. As a result, and with AHC becoming more established through density-functional perturbation theory (DFPT), some non-perturbative (NP) methods have started to arise in the last years. However, they are usually not well justified and it is not clear to what degree they reproduce the exact theory. To address this issue, we present a stochastic approach for the evaluation of the non-perturbative interacting Green's function in the adiabatic limit, and show it is equivalent to the Feynman expansion to all orders in the perturbation. Also, by defining a self-energy, we can reduce the effect of broadening needed in numerical calculations, improving convergence in the supercell size. In addition, we clarify whether it is better to average the Green's function or self-energy. Then we apply the method to a graphene tight-binding model, and we obtain several interesting results: (i) The Debye-Waller term, which is normally neglected, does affect the change of the Fermi velocity $v_F$, and should be included to obtain accurate results. (ii) Although at room temperature second order perturbation theory (P2) agrees well with the NP change of $v_F$ and of the self-energy close to the Dirac point, at high temperatures there are significant differences. For other $\mathbf{k}$ points, the disagreement between the P2 and NP self-energies is visible even at low temperatures, raising the question of how well P2 works in other materials. (iii) Close enough to the Dirac point, positive and negative energy peaks merge, giving rise to a single peak. (iv) At strong coupling and high temperatures, a peak appears at $\o=0$ for several states, which is consistent with previous works on disorder and localization in graphene. (v) The spectral function becomes more asymmetric at stronger coupling and higher temperatures. Finally, in the Appendix we show that the method has better convergence properties when the coupling is strong relative to when it is weak, and discuss other technical aspects.
\end{abstract}

\maketitle

\section{Introduction}

The electron-phonon interaction (EPI) is a fundamental aspect of condensed matter physics. It determines the electrical conductivity in metals and the mobility in doped semiconductors. It drives the temperature dependence of the electronic bands, and thus of the bandgap in semiconductors, and it also distorts phonon dispersions through kinks and Kohn anomalies. It also leads to conventional superconductivity, among other important physical phenomena\cite{Giustino2017}.

EPIs are usually studied using lowest order perturbation theory, following the Allen-Heine-Cardona (AHC) approach\cite{Allen1976,Allen1981,Allen1983}, which is valid when the coupling is weak. However, strong EPIs have been measured in the last decades in perovskites \cite{Cohen1990,Saidi2016,Guzelturk2018}, numerous 2D materials \cite{Yildirim2013,Dey2016}, interfaces \cite{Zhang2017}, and quantum-dots \cite{Hameau1999}. Furthermore, it is ubiquitous in high temperature superconductors \cite{Lanzara2001}, and it has even been reported in graphite \cite{Sugawara2007} and twisted bilayer graphene \cite{Choi2018}. Thus, it plays a fundamental role in some of the most important systems currently studied in condensed matter physics.

In such systems, higher order terms are needed, but they tend to pose serious numerical challenges (namely, additional dense integrations and larger electron-phonon matrices). For example, using the standard perturbative approach, Ref.~\onlinecite{Lee2020} has calculated some fourth order Feynman diagrams in GaAs, and obtained that scattering rates are as large as half the usual second order Fan value (so even higher order terms might be needed for precise results). However, not all diagrams are included, and the cost of the fourth order diagrams is $10^4 - 10^5$ higher than the lowest-order ones\cite{Lee2020}. There are a few first-principles implementations of AHC \cite{Abinit2020, Yambo} which calculate electron-phonon matrix elements via DFPT \cite{Abinit2020, QE2017}, and some works have started to look further into non-perturbative methods.

A common non-perturbative approach consists of simulating an instantaneous snapshot of a system at a given temperature, including quantum fluctuations at $T=0$ (zero point motion), in which the atoms are displaced from their equilibrium positions; calculating the desired quantity in each of these configurations; and then averaging. In this approach, the phonons are introduced as static classical effective external parameters, rather than as a quantum dynamical particle of the system. In principle, however, both electrons and phonons should be internal quantum mechanical degrees of freedom in an exact description. It is not properly emphasized in the literature that, although this averaging approach is intuitive, it remains an uncontrolled approximation. The relation to the fully quantized theory should be adequately established, to understand its advantages and possible shortcomings.

Ref.~\onlinecite{Ponce2014} considered distorted configurations and recovered AHC, but limited the analysis to the electronic energies (as opposed to studying the spectral function) and to lowest order. In Ref.~\onlinecite{Zacharias2015} the authors use a stochastic approach to calculate the dielectric function as we just described (averaging over an ensemble). Then the same authors in Ref.~\onlinecite{Zacharias2016}, by looking at the form of the result, showed that the shift of the electronic energy (in the weak coupling regime) corresponds to AHC. For the electronic lifetime however, the expression is similar, but not the same as the standard result, which is attributed to the semiclassical rather than adiabatic expression of the dielectric function being used. It would thus be desirable to calculate an observable which is exact in the adiabatic limit of the fully quantized theory. Ref.~\onlinecite{Zacharias2016} also proposed a particular distorted configuration to reduce the number of configurations needed to achieve convergence. However, we will show that in order to use Wick's theorem and recover Feynman diagrams beyond second order in our approach, a Gaussian distribution is required. Other works have focused on the spectral function ~\cite{Ku2010,Popescu2010,Allen2013,Zacharias2020} . Allen \textit{et al.} \cite{Allen2013} rigorously show how to unfold states defined in the supercell (SC) to states defined in the primitive cell (PC), and write an expression for the spectral function in terms of distorted SC states (see Ref.~\onlinecite{Allen2013} for additional references and a more detailed discussion related to unfolding). Although the starting point is defined in terms of a Green's function, no further elements of Green's function theory are used, and no connection to AHC is established. Another non-perturbative approach is that of Ref.~\onlinecite{Gorelov2020}, which uses a path-integral quantum Monte Carlo approach to determine the bandgap, but there is no link to AHC either.

To address these issues, we develop a non-perturbative Green's function method and rigorously show how it relates to the standard expansion in terms of Feynman diagrams. Our method involves averaging a Green's function over configuration and defining a self-energy in the PC that is momentum dependent. It turns out to be the same type of method used to study impurity scattering as described in Mahan\cite{Mahan2000}, which also involves a diagrammatic expansion, and similar methods are used to study properties in disordered systems such as amorphous semiconductors and alloys. For example, the coherent potential approximation (CPA) averages the Green's function and defines an effective self-energy for the medium. There are many references on Green's functions methods to study disordered crystals, such as the review article of Elliott et al. \cite{Elliott1974} or the book by Economou\cite{Economou2006}. Our goal here however is to study the effect of phonons in otherwise periodic systems, rather than disorder.

In the first part of this paper, we describe how to define the Green's function and self-energy (Sec.~II.A). Then we show that to lowest order the self-energy coincides with the DW and Fan terms of AHC in the adiabatic limit (Sec.~II.B). Subsequently, we look at higher order terms, show that they can be represented diagrammatically, and see that such diagrams have the same shape as those of the exact theory (Sec.~II.C). In Sec.~II.D, we show that the expressions of the equivalent diagrams (of our theory and of the Feynman expansion) are exactly the same in the adiabatic limit, completing the proof. In Sec.~III, thanks to the introduction of the self-energy, we can modify the usual approach of Sec.~II.A, and obtain a spectral function with reduced error in the broadening parameter needed in numerical calculations.
 Then in Sec.~IV, we apply the method to a tight-binding graphene model. First, we present the spectral function for several couplings and temperatures, including the strong coupling regime, where bands merge and a non-perturbative method becomes indispensable. Then we compare AHC and our approach at the experimental coupling, and observe some differences even at room temperature. We also show how the bands and in particular the Fermi velocity changes close to the Dirac point.  Finally, we study how the spectral function becomes more asymmetric in the strong coupling regime. The Appendix, among other things, considers an alternative definition of the self-energy and includes some convergence studies.

\section{Theoretical aspects}

\subsection{Green's function and self-energy}
\label{sec:theory}

We are interested in determining the electronic spectral function due to EPIs. To do this, we consider an ensemble of distorted ionic configurations in a SC according to the phonons of the system. For each configuration, we determine a Green's function, which are then averaged to obtain the Green's function of the system. The averaged Green's function can be used to determine the spectral function and define a self-energy, as we describe in this subsection.

Let us consider a $N_1 \times N_2 \times N_3$ SC, and let $u_{li\a}$ be the displacements from the equilibrium position, where $l$ is the index of the PC in the SC, $i$ is the index of the ion in the PC, and $\a$ the Cartesian direction. In the harmonic approximation, the probability distribution of finding the system in a general ionic configuration $\{u_{li\a}\}$ is\cite{Rossano2005,Errea2013}

\small
\be
P(\{u_{li\a}\}) = A e^{-\sum_{\substack{li,mj,\\\a\b,\mathbf{q}\nu}} \f{\sqrt{M_i M_j}2\o_{\mathbf{q}\nu}}{2(2n_{\mathbf{q}\nu}+1)} \mathcal{E}^{li\a}_{\mathbf{q}\nu}\mathcal{E}^{mj\b}_{\mathbf{q}\nu} u_{li\a} u_{mj\b}}
\label{eq:distribution}
\ee
\normalsize

\ni where $\mathbf{q}$ is in $\mathcal{Q}_\mathrm{SC}$, the set of SC reciprocal lattice vectors in the primitive Brillouin zone (PBZ) (i.e. $\mathbf{q}$ is commensurate with the SC), $\nu$ is the phonon branch, $\o_{\mathbf{q}\nu}$ the phonon frequency of mode $\mathbf{q}\nu$, $n_{\mathbf{q}\nu}$ is the Bose-Einstein occupation factor (at a temperature $T$ and frequency $\o_{\mathbf{q}\nu}$), $M_i$ is the mass of atom $i$, $A$ is just the normalization constant, and $\mathcal{E}^{li\a}_{\mathbf{q}\nu}$ are the polarization vectors in the SC (which can be chosen real, since the corresponding dynamical matrix is symmetric and real) written in terms of the polarization vectors $\e^{i\a}_{\mathbf{q}\nu}$ in the PC,

\be
\mathcal{E}^{li\a}_{\mathbf{q\nu}} = \\
\begin{cases}
\hspace{1cm} e^{i \mathbf{q} \cdot \mathbf{R}_l} \e^{i\a}_{\mathbf{q}\nu} \hspace{0.83cm} \hspace{1mm} \mathrm{if} \hspace{1mm} \mathbf{q} \in D_1\\
\f{e^{i\mathbf{q} \cdot \mathbf{R}_l}\e^{i\a}_{\mathbf{q}\nu} + e^{-i\mathbf{q}\cdot\mathbf{R}_l}(\e^{i\a}_{\mathbf{q}\nu})^\ast}{\sqrt{2}}  \hspace{1mm} \mathrm{if} \hspace{1mm} \mathbf{q} \notin D_1, \mathbf{q} \in D_2 \\
\f{e^{i\mathbf{q} \cdot \mathbf{R}_l}\e^{i\a}_{\mathbf{q}\nu} - e^{-i\mathbf{q}\cdot\mathbf{R}_l}(\e^{i\a}_{\mathbf{q}\nu})^\ast}{\sqrt{2i}} \hspace{1mm} \mathrm{if} \hspace{1mm} \mathbf{q} \notin D_1, \mathbf{q} \notin D_2\\
\end{cases}
\label{eq:SCpol}
\ee

\ni where $D_1 \subset \mathcal{Q}_\mathrm{SC}$ is the set of points for which $\mathbf{q}$ and $-\mathbf{q}$ differ by a reciprocal lattice vector, and $D_2 \subset \mathcal{Q}_\mathrm{SC}$ is the irreducible BZ considering only time-reversal symmetry. In $D_1$ (first line), $\e^{i\a}_{\mathbf{q}\nu}$ is chosen real (the dynamical matrix is symmetric and real) and $e^{i \mathbf{q} \cdot \mathbf{R}_l}$ is just 1 or -1 . In the other cases we pick $(\e^{i\a}_{\mathbf{q}\nu})^\ast = \e^{i\a}_{-\mathbf{q}\nu}$. The second and third line correspond to the real and imaginary part, respectively, of $e^{i \mathbf{q} \cdot \mathbf{R}_l} e^{i\a}_{\mathbf{q}\nu}$.  So in all cases $\mathcal{E}^{li\a}_{\mathbf{q\nu}}$ is real. We write it in this way, explicitly in terms of exponentials with momentum $\mathbf{q}$ or $-\mathbf{q}$, to later use momentum conservation and establish more easily the connection with Feynman diagrams.
 
We consider an ensemble of stochastic distorted configurations which follows the distribution of Eq.~\eqref{eq:distribution},

\be
u^I_{li\a} = \f{1}{\sqrt{N}}\sum_{\mathbf{q}\nu} \f{\mathcal{E}^{li\a}_{\mathbf{q}\nu}}{\sqrt{2M_i \o_{\mathbf{q}\nu}}} \xi^I_{\mathbf{q}\nu},
\label{eq:u}
\ee

\ni where $I=1,...,N_c$ is the index of the configuration, $N=N_1N_2N_3$ is the number of cells in the SC, and

\be
\xi^I_{\mathbf{q}\nu}= \tilde{\xi}^I_{\mathbf{q}\nu} \sqrt{2n_{\mathbf{q}\nu}+1},
\label{eq:csi}
\ee

\ni where $\tilde{\xi}_{\mathbf{q}\nu}$ is a random number following a normal distribution  (centered at 0, of standard deviation 1). The temperature dependence enters through $n_{\mathbf{q}\nu}$. For each static distortion $u^I_{li\a}$, we will determine the electronic energies and wavefunctions. This means that the phonons are introduced as classical parameters, which do not dynamically interact with the electrons. But as we see later, this correctly reproduces the usual contribution of (quantum mechanical) phonons to the electronic Green's function (in the adiabatic limit), which we will now define.

Let us consider a momentum $\mathbf{k}$ in the PBZ where we want to determine the spectral function, and we first consider $\mathbf{k}$ in $\mathcal{Q}_\mathrm{SC}$. Let $\mathcal{H}_0$ be the undistorted Hamiltonian, and let $B_\mathrm{PC}^\mathbf{k}=\{|\mathbf{k}n\rangle,n=1,...,M\}$ be a basis of eigenstates in the PC with band index $n$, Bloch symmetry $\mathbf{k}$, and eigenvalues $\vare^0_{\mathbf{k}n}$, $\mathcal{H}_0 |\mathbf{k}n\rangle = \vare^0_{\mathbf{k}n} |\mathbf{k}n\rangle$ (in general there are infinite bands, but in practical calculations the number of bands $M$ is typically restricted to include a few above the Fermi level). In the SC, a Hamiltonian in general will not be diagonal in $\mathbf{k}$, and a basis is $\mathcal{B}_\mathrm{SC}=\{|\mathbf{q}n\rangle, \mathbf{q} \hspace{1mm} \mathrm{in} \hspace{1mm} \mathcal{Q}_\mathrm{SC},n=1,...,M\}$. 

Each configuration $I$ is described by a distorted static and single-particle Hamiltonian $\mathcal{H}^I$. Let $\{|J\rangle^I\}$ be a set of eigenstates of configuration $I$, with energies $\vare^I_J$. Ref.~\onlinecite{Allen2013} defined a distorted Green's function by starting from the usual second quantization expression for the retarded Green's function. Here instead, we define the retarded Green's function (the one normally used to compare to experiments) via

\be
\mathcal{G}^I_{\mathbf{k}n,\mathbf{k'}n'}(\o+i\d) =  \langle \mathbf{k}n| \f{1}{\o + i \delta - \mathcal{H}^I} |\mathbf{k'}n'\rangle,
\label{eq:Gkn}
\ee

\ni where $\omega$ corresponds to an electronic energy, and $\d$ is a positive parameter, that in principle (to get the exact result) is considered in the $\d \rightarrow 0$ limit after taking first the SC limit $N \rightarrow \infty$. Since we are interested in determining the spectral function in the original PC basis, we focus on $\mathbf{k}=\mathbf{k}'$ and define a Green's function $G^I$ in the PC. We use cursive for objects defined in the SC Hilbert space, and non-cursive for objects defined in the PC (see Table~\ref{tab:notation} for clarification on the notation). Inserting also the identity $\sum_J |J\rangle^I {^I}\langle J |$, we have 

\be
G^I_{\mathbf{k},nn'}(\o+i\d) = \sum_J \f{\langle \mathbf{k}n|J\rangle^I {^I}\langle J|\mathbf{k}n'\rangle}{\o + i \d - \vare^I_{J}}
\label{eq:Gnum}
\ee

\ni which coincides with the expression of Ref.~\onlinecite{Allen2013}. One can then define an averaged Green's function over the $N_c$ distorted configurations,

\be
\begin{split}
G_{\mathbf{k},nn'}(\o+i\d)= \langle \mathcal{G}^I_{\mathbf{k}n,\mathbf{k}n'}(\o+i\d)\rangle\\
\mathrm{with} \hspace{2mm} \langle \rangle = \lim_{N_c \rightarrow \infty} 1/N_c\sum_I^{N_c}
\end{split}
\label{eq:Gav}
\ee

\ni and determine the spectral function in the usual way,

\be
\begin{split}
A_\mathbf{k}(\o+i\d) & =-\f{1}{\pi} \Im m \textrm{Tr} G_\mathbf{k}(\o+i\d)\\
& = - \f{1}{\pi} \sum_n \Im m G_{\mathbf{k},nn}(\o+i\d)\\
& = \sum_n A_{\mathbf{k}n}(\o+i\d)
\end{split}
\label{eq:A}
\ee

\ni Experimentally the spectral function is typically accessed by angle-resolved photoemission experiments (ARPES). $A_{\mathbf{k}n}$ can be assigned to the spectral function of band $n$ if the bands are well separated. If there is a well defined peak, the maximum determines the quasiparticle energy, and the width determines the broadening (inverse lifetime). Using Eqs.~\eqref{eq:Gnum}-\eqref{eq:A}, the spectral function is\cite{Ku2010,Popescu2010}

\be
A_\mathbf{k}(\o) = \langle \sum_{Jn} | ^I\langle J|\mathbf{k}n\rangle|^2 \d(\o - \vare^I_{J})\rangle
\label{eq:Adelta}
\ee

\ni after taking the $\d \rightarrow 0, N \rightarrow \infty$ limit.

In the next subsections, we will rigorously show that $G_\mathbf{k}$ has the usual diagrammatic expansion in terms of Feynman diagrams, putting on firm grounds the approach we just described, which has been used somewhat heuristically in the literature. We now introduce a self-energy with two purposes: help establish the connection to the diagrammatic expansion, and to later modify the approach above to reduce the error in the broadening parameters that are needed in numerical calculations to describe the delta function in Eq.~\eqref{eq:Adelta}. The self-energy $\Sigma_\mathbf{k}$ is defined through the Dyson equation (with band indices $n,n'$)

\be
G_\mathbf{k}(\o+i\d) = \f{1}{(\o + i \d)\mathbb{I}_\mathrm{PC} - \vare^0_\mathbf{k} - \Sigma_\mathbf{k}(\o+i\d)},
\label{eq:Sigma}
\ee

\ni where $\mathbb{I}_\mathrm{PC}$ is just the $M \times M$ identity, and $\vare^0_{\mathbf{k},nn'}=\langle\mathbf{k}n| \mathcal{H}_0 |\mathbf{k}n'\rangle=\vare^0_{\mathbf{k}n} \d_{nn'}$. Since $G_\mathbf{k}$ has the usual diagrammatic expansion (in the adiabatic limit), $\Sigma_\mathbf{k}$ will correspond to the sum of irreducible diagrams. If $\mathbf{k}$ is not commensurate with the SC, there is a unique $\mathbf{K}=\mathbf{k+q_0}$ in the super Brillouin zone (SBZ) with $\mathbf{q_0}$ in $\mathcal{Q}_\mathrm{SC}$. Then the grids are just shifted by $\mathbf{K}$. That is, $\mathcal{B}_\mathrm{SC}=\{|\mathbf{K+q},n\rangle,\mathbf{q} \hspace{1mm} \mathrm{in} \hspace{1mm} \mathcal{Q}_\mathrm{SC},n=1,...,M\}$. Thus, the spectral function can be determined at any $\mathbf{k}$.

Before moving on to the proof, we want to mention an important conceptual point. Since $\Sigma_\mathbf{k}$ is a sum of irreducible diagrams, it is not Hermitian (it has complex eigenvalues) and is energy dependent. On the other hand, Eq.~\eqref{eq:Gkn} can be written in terms of $\mathcal{V}^I=\mathcal{H}^I-\mathcal{H}_0$, which is then also a Dyson equation, that relates the Green's function $\mathcal{G}^I$ with the self-energy $\mathcal{V}^I$, which is Hermitian and static. Why is this?

The key difference is that $\Sigma_\mathbf{k}$ only depends on $\mathbf{k}$, while Eq.~\eqref{eq:Gkn} and $\mathcal{V}^I$ depend on $\mathbf{k}$ and $\mathbf{k}'$, so the inverse of $\mathcal{V}^I$ in Eq.~\eqref{eq:Gkn} involves off-diagonal elements. Forcing the $\mathbf{k}$ diagonal part of $\mathcal{G}^I$ to be expressed in terms of a self-energy that is diagonal in $\mathbf{k}$ (defined in the smaller PC space), makes the self-energy acquire a more complicated structure. In other words, the reduced information contained in momentum is compensated by additional information in an imaginary part and $\o$ dependence.\\

\begin{table*}[]
\begin{tabular}{cccc}
\hline
Symbol                       & Description                                                                                                         & Hilbert space basis                                                                                        & First appearance         \\ \hline
 $\mathcal{H}_0$              & Undistorted Hamiltonian                                                                                             & $\mathcal{B}_\mathrm{SC}$                                                                                       & After Eq.~\eqref{eq:csi}        \\
$\mathcal{G}_0$              & Undistorted Green's function                                                                                        & $\mathcal{B}_\mathrm{SC}$                                                                                       & Eq.~\eqref{eq:G0}       \\
$\mathcal{H}^I$              & Distorted Hamiltonian                                                                                               & $\mathcal{B}_\mathrm{SC}$                                                                                       & Eq.~\eqref{eq:Gkn}         \\
$\mathcal{V}^I = \mathcal{H}^I-\mathcal{H}_0$              & Self-energy or ``External potential''                                                                                              & $\mathcal{B}_\mathrm{SC}$                                                                                       & After Eq.~\eqref{eq:Sigma}    \\
$\mathcal{G}^I$              & Distorted Green's function                                                                                          & $\mathcal{B}_\mathrm{SC}$                                                                                       & Eq.~\eqref{eq:DysonV0}         \\
$\mathcal{G}^I_{\mathbf{k}n,\mathbf{k}'n'}$              & \begin{tabular}[c]{@{}c@{}}Distorted Green's function matrix\\ elements in $\mathcal{B}_\mathrm{SC}$ basis \end{tabular} & $\mathcal{B}_\mathrm{SC}$                                                                                       & Eq.~\eqref{eq:Gkn}         \\
$\langle \rangle$            & Average over ensemble                                                                                               & $\mathcal{B}_\mathrm{SC}$                                                                                       & Eq.~\eqref{eq:Gav}       \\
$\langle \mathcal{O} \rangle$ & \begin{tabular}[c]{@{}c@{}}Average of some SC operator $\mathcal{O}$\\ The implied matrix element is $\mathcal{O}_{\mathbf{k}n,\mathbf{k}n'}$\end{tabular} & \begin{tabular}[c]{@{}c@{}}$\mathcal{B}_\mathrm{SC}$ for $\mathcal{O}$\\ $B_\mathrm{PC}^\mathbf{k}$ for $\langle \mathcal{O} \rangle$\end{tabular} & Eq.\eqref{eq:av}        \\
$G_{\mathbf{k},nn'}$                          & \begin{tabular}[c]{@{}c@{}}Averaged Green's function matrix \\elements in $B_\mathrm{PC}^\mathbf{k}$ basis \end{tabular} & $B_\mathrm{PC}^\mathbf{k}$                                                                                       & Eq.\eqref{eq:Gav}       \\
$G_\mathbf{k}$                        & Averaged Green's function & $B_\mathrm{PC}^\mathbf{k}$                                                                                       & Eq.\eqref{eq:Sigma}     \\
$\Sigma_\mathbf{k}$                   & Self-energy & $B_\mathrm{PC}^\mathbf{k}$                                                                                       & Eq.\eqref{eq:Sigma}     \\
$G$ & Averaged Green's function (implicit $\mathbf{k}$ index) & $B_\mathrm{PC}^\mathbf{k}$ & Eq.~\eqref{eq:Gimplicit} \\ 
$\Sigma$ & Self-energy (implicit $\mathbf{k}$ index) & $B_\mathrm{PC}^\mathbf{k}$ & Eq.~\eqref{eq:Sigmaimplicit} \\ 
\hline
\end{tabular}
\caption{Some of the symbols used in this work. We consider first $\mathbf{k}$ in $\mathcal{Q}_\mathrm{SC}$, the set of reciprocal SC lattice vectors, to compute the spectral function. Quantities in cursive are defined in the SC Hilbert space, with basis $\mathcal{B}_\mathrm{SC}=\{|\mathbf{q}n\rangle,\mathbf{q} \hspace{1mm} \mathrm{in} \hspace{1mm} \mathcal{Q}_\mathrm{SC},n=1,...,M\}$, with $M$ the number of bands in the PC. Averaged quantities are in the PC Hilbert space, with basis $B_\mathrm{PC}=\{|\mathbf{k}n\rangle,n=1,...,M\}$. In the SC we use other basis as well, namely the set of eigenstates $\{|J>^I\}$ in Eq.~\eqref{eq:Gnum}. When doing the average (over infinite configurations) 
 the same momenta $\mathbf{k}'=\mathbf{k}$ are implied unless otherwise specified. If $\mathbf{k}$ is not commensurate with the SC, then the SC basis is shifted by the $\mathbf{K}$ in the SBZ such that $\mathbf{K}=\mathbf{k+q_0}$ for some $\mathbf{q_0}$ in $\mathcal{Q}_\mathrm{SC}$.}
\label{tab:notation}
\end{table*}

\subsection{Comparison to AHC}

\subsubsection{AHC self-energy}

The standard Hamiltonian to second order in the electron-phonon interaction is given by\cite{Giustino2017}

\begin{widetext}

\be
\begin{split}
H & = \sum_{\mathbf{k}n} \vare_{\mathbf{k}n}c^\dagger_{\mathbf{k}n} c_{\mathbf{k}n} + \sum_{\mathbf{q}\nu} \o_{\mathbf{q}\nu}(a^\dagger_{\mathbf{q}\nu}a_{\mathbf{q}\nu} + \f{1}{2}) + \f{1}{\sqrt{N}} \sum_{\substack{\mathbf{k}nn' \\ \mathbf{q}\nu}} g^{\mathbf{q}\nu}_{\mathbf{k},nn'} c^\dagger_{\mathbf{k}+\mathbf{q}n'} c_{\mathbf{k}n}(a_{\mathbf{q}\nu} + a^\dagger_{-\mathbf{q}\nu}) \\
& + \f{1}{N}\sum_{\substack{\mathbf{k}nn' \\\mathbf{q}\nu\mathbf{q}'\nu'}} g^{\mathrm{DW},\mathbf{q}\nu\mathbf{q}'\nu'}_{\mathbf{k},nn'} c^\dagger_{\mathbf{k}+\mathbf{q}+\mathbf{q}'n'}c_{\mathbf{k}n}  \times (a_{\mathbf{q}\nu}+a^\dagger_{-\mathbf{q}\nu})(a_{\mathbf{q}'\nu'}+a^\dagger_{-\mathbf{q}'\nu'}),
\end{split}
\label{eq:Hfull}
\ee

\ni where, in the static approximation,

\be
\begin{split}
g^{\mathbf{q}\nu}_{\mathbf{k},nn'} & = \sum_{i\a} \langle \mathbf{k+q}n | \f{\p V}{\p u_{i\a}(\mathbf{q})}|\mathbf{k}n'\rangle \f{1}{\sqrt{2 M_i \o_{\mathbf{q}\nu}}} \e^{i\a}_{\mathbf{q}\nu}\\
g^{\mathrm{DW},\mathbf{q}\nu \mathbf{q}'\nu'}_{\mathbf{k},nn'} & = \f{1}{2}\sum _{i\a,j\b} \langle \mathbf{k}+\mathbf{q}+\mathbf{q}'n | \f{\p^2 V}{\p u_{i\a}(\mathbf{q})\p u_{j\b}(\mathbf{q}')} |\mathbf{k}n'\rangle \f{1}{\sqrt{2 M_i \o_{\mathbf{q}\nu}}} \e^{i\a}_{\mathbf{q}\nu}  \f{1}{\sqrt{2 M_j \o_{\mathbf{q}'\nu'}}} \e^{j\b}_{\mathbf{q}'\nu'}
\end{split}
\label{eq:g_gDW}
\ee
\end{widetext}

\ni and

\be
\f{\p}{\p u_{i\a}(\mathbf{q})} = \sum_l e^{i\mathbf{q} \cdot \mathbf{R}_l} \f{\p}{\p u_{li\a}},
\ee

\ni $g$ and $g^\mathrm{DW}$ can also be expressed in the primitive cell, using the periodic part of the wavefunctions (see Sec.~III.2 of Ref.~\onlinecite{Giustino2017}), but we write it in this way to make more explicit the connection with our method later.

Defining the retarded Green's function $G^R_{\mathbf{k}nn'}(t) = -i \theta(t) \langle \{	c_{\mathbf{k}n}(t),c^\dagger_{\mathbf{k}n'} \} \rangle$, one can obtain that to lowest order the self-energy is given by $\Sigma_{\mathbf{k},nn'}(\o) = \Sigma^\textrm{Fan}_{\mathbf{k},nn'}(\o) + \Sigma^\textrm{DW}_{\mathbf{k},nn'}$, where\\

\be
\Sigma^\textrm{DW}_{\mathbf{k},nn'}(\o) = \f{1}{N} \sum_{\mathbf{q}\nu} g^{\textrm{DW},\mathbf{q}\nu -\mathbf{q}\nu}_{\mathbf{k},nn'} (2 n_{\mathbf{q}\nu}+1),
\label{eq:DW}
\ee

\ni and in the adiabatic limit,

\be
\Sigma^\textrm{Fan}_{\mathbf{k},nn'}(\o) = \f{1}{N} \sum_{\mathbf{q}\nu,n''} \f{g^{-\mathbf{q}\nu}_{\mathbf{k}+\mathbf{q},nn''}g^{\mathbf{q}\nu}_{\mathbf{k},n''n'}}{\o +i\d - \vare^0_{\mathbf{k}+\mathbf{q}n''}}(2n_{\mathbf{q}\nu}+1).
\label{eq:Fan}
\ee

\ni The diagonal $n=n'$ part of this self-energy is what we referred to as the AHC self-energy in the adiabatic limit. In DFPT implementations of AHC, the second derivative $g^\mathrm{DW}$ so far has been avoided by using the rigid ion approximation (RIA) and the acoustic sum rule, which allows to write $\Sigma^\mathrm{DW}$ in terms of the first derivative $g$. Our method and others that use SC methods\cite{Ponce2014} do not need to use the RIA. AHC theory is usually applied by taking $n=n'$, and originates on the diagrammatic expansion of many-body perturbation theory, so it can actually be applied to higher order terms. So when we apply our method to the tight-binding model, we will usually refer to the (non-diagonal) Eqs.~\eqref{eq:DW} and \eqref{eq:Fan} and second order quantities in the ionic displacements as P2.

\subsubsection{NP second order self-energy}
\label{sec:NP_2}

In order to establish the connection of our method with perturbation theory, we start by writing Eq.~\ref{eq:Gkn} as another form of the Dyson equation (now writing the operator as opposed to the matrix element),

\be
\mathcal{G}^I = \mathcal{G}_0 + \mathcal{G}_0 \mathcal{V}^I \mathcal{G}^I,
\label{eq:DysonV0}
\ee

\ni where

\be
\mathcal{G}_0 = \f{1}{\o + i\d - \mathcal{H}_0}.
\label{eq:G0}
\ee

\ni Thus,

\be
\mathcal{G}^I = \mathcal{G}_0 + \mathcal{G}_0 \mathcal{V}^I \mathcal{G}_0 + \mathcal{G}_0 \mathcal{V}^I \mathcal{G}_0 \mathcal{V}^I \mathcal{G}_0 + ...
\label{eq:DysonV}
\ee

\ni and averaging,

\be
\langle \mathcal{G}^I \rangle = \mathcal{G}_0 + \mathcal{G}_0 \langle \mathcal{V}^I \rangle \mathcal{G}_0 + \mathcal{G}_0 \langle \mathcal{V}^I \mathcal{G}_0 \mathcal{V}^I \rangle \mathcal{G}_0 + ...
\label{eq:av}
\ee

\ni Defining $\Sigma^\mathrm{red} = \langle \mathcal{V}^I \rangle + \langle \mathcal{V}^I \mathcal{G}_0 \mathcal{V}^I \rangle +...$,

\be
G = G_0 + G_0\Sigma^\mathrm{red} G_0
\label{eq:Gimplicit}
\ee

\ni where $G^0_{\mathbf{k},nn'}=\mathcal{G}^0_{\mathbf{k}n,\mathbf{k}n'}$. The notation $\Sigma^\mathrm{red}$ will become more clear in the next subsection (it corresponds to a reducible self-energy).

We want to see that the self-energy $\Sigma^\mathrm{NP}$ of our approach, to lowest order, is equal to $\Sigma^\mathrm{Fan} + \Sigma^\mathrm{DW}$ of Eqs.~\eqref{eq:DW} and \eqref{eq:Fan}. So we now focus on an expansion of $\mathcal{G}^I$ in terms of the displacements of the ions $u_{li\a}$. We have assumed that the geometric sum $\mathcal{G}^I = \mathcal{G}_0(1 + \mathcal{V}^I \mathcal{G}_0 + (\mathcal{V}^I \mathcal{G}_0)^2+...)$ in Eq.~\eqref{eq:DysonV} converges, but so far the result is exact under the assumption of harmonic phonons. Perturbation theory in the displacements has not been used yet. If perturbation theory holds, we can write

\be
\begin{split}
\mathcal{V}^I & = \mathcal{V}^{I,(1)} + \mathcal{V}^{I,(2)} + ...\\ 
& = \sum_{li\a} \f{\p V}{\p u_{li\a}} u^I_{li\a}  \\
& + \f{1}{2} \sum_{\substack{li\a \\ mj\b}}\f{\p^2 V}{\p u_{li\a} u_{mj \b}}u^I_{li\a} u^I_{mj\b} 
 + ...
\end{split}
\label{eq:V}
\ee

\ni where we use the usual notation $V=V(\{u_{li\a}\})$ (as in Eq.~\eqref{eq:g_gDW}) for the potential seen as a function of the displacements (as opposed to $\mathcal{V}^I$, which is for a fixed distortion $u^I_{li\a}$). We will now see that the lower order terms of the self-energy correspond to the standard Fan and Debye-Waller (DW) terms. Writing indices explicitly in the $\mathcal{B}_\mathrm{SC}$ basis, these terms are:

\be
\begin{split}
(a) & \langle \sum_{\tilde{\mathbf{q}}n''} \mathcal{V}^{I,(1)}_{\mathbf{k}n,\mathbf{k}+\tilde{\mathbf{q}}n''} G^0_{\mathbf{k}+\tilde{\mathbf{q}}n''} \mathcal{V}^{I,(1)}_{\mathbf{k}+\tilde{\mathbf{q}}n''\mathbf{k}n'} \rangle\\
(b) & \langle \mathcal{V}^{I,(2)}_{\mathbf{k}n,\mathbf{k}n'} \rangle \\
\end{split}
\label{eq:DW_FAN}
\ee

\ni where the sum in $(a)$ just comes from the usual product of operators (in this case in the SC Hilbert space).

Using Eqs.~\eqref{eq:V} and \eqref{eq:u} in (a), we get (before averaging)

\begin{widetext}
\be
\begin{split}
\sum_{\tilde{\mathbf{q}}n''} \mathcal{V}^{I,(1)}_{\mathbf{k}n,\mathbf{k}+\tilde{q}n''} G^0_{\mathbf{k}+\tilde{\mathbf{q}}n''} \mathcal{V}^{I,(1)}_{\mathbf{k}+\tilde{\mathbf{q}}n''\mathbf{k}n'} =\f{1}{N}  \sum_{ \substack{li\a, mj\b \\ \tilde{\mathbf{q}}n'',\mathbf{q}\nu,\mathbf{q}'\nu'}}  & \langle \mathbf{k}n| \f{\p V}{\p u_{li\a}} \mathcal{E}^{li\a}_{\mathbf{q}\nu} |\mathbf{k} + \tilde{\mathbf{q}} n''\rangle \langle \mathbf{k} + \tilde{\mathbf{q}} n''| \f{\p V}{\p u_{mj\b}} \mathcal{E}^{mj\b}_{\mathbf{q'}\nu'}|\mathbf{k} n'\rangle \\
& \times \f{1}{\o + i\d - \vare^0_{\mathbf{k+q}n''}} \f{ \sqrt{2n_{\mathbf{q}\nu}+1} \sqrt{2n_{\mathbf{q}'\nu'}+1}}{2 \sqrt{\o_{\mathbf{q}\nu}\o_{\mathbf{q}'\nu'}} \sqrt{M_i M_j}}\tilde{\xi^I}_{\mathbf{q}\nu} \tilde{\xi^I}_{\mathbf{q}'\nu'}
\end{split}
\ee
\end{widetext}

\ni

The configurational average only affects the last two factors,

\be
\begin{split}
\lim_{N_c \rightarrow \infty} \f{1}{N_c}&\sum_{I=1}^{N_c} \tilde{\xi}^{I}_{\mathbf{q}\nu} \tilde{\xi}^{I}_{\mathbf{q}'\nu'} = \\ & = \begin{cases}
  0 \hspace{2mm} \mathrm{if} \hspace{2mm}\mathbf{q}\nu \neq \mathbf{q}'\nu'\\
\int dx_{\mathbf{q}\nu} x_{\mathbf{q}\nu}^2 \f{e^{-x_{\mathbf{q}\nu}^2/2}}{\sqrt{2\pi}}=1 \hspace{2mm} \mathrm{if} \hspace{2mm} \mathbf{q}\nu = \mathbf{q}'\nu'
\end{cases}
\end{split}
\label{eq:average}
\ee

\ni That is, each mode $\mathbf{q}\nu$ has associated its own random number, and the average of the product of different random numbers is just 0. Instead, when the modes are the same, the product of the random number with itself is positive, and the average (with the chosen normalization) is 1.

$\mathcal{E}^{li\a}_{\mathbf{q}\nu}$ is defined in Eq.~\eqref{eq:SCpol}, and is made up of terms with momentum $\mathbf{q}$ or $-\mathbf{q}$. Due to the usual momentum conservation, $\mathbf{k} \pm \mathbf{q} = \mathbf{k}+\tilde{\mathbf{q}}$, so $\tilde{\mathbf{q}} = \pm \mathbf{q}$ (so $\tilde{\mathbf{q}}$ corresponds as usual to the momentum wave vector). Momentum conservation holds for each of the matrix elements, and also for second order or higher derivatives of the potential (in which case the difference of the bra and ket momenta of the matrix element corresponds to the sum of the momenta of the multiple modes contained in the higher order terms). Using Eq.~\eqref{eq:average} and momentum conservation, we can now proceed to establish the connection with $\Sigma^\mathrm{Fan}$. The derivation is straightforward but a little bit cumbersome because of how the cases have to be divided between different $\mathbf{q}$'s to get real displacements.

Since $\mathbf{q}\nu = \mathbf{q}'\nu'$ from Eq.~\eqref{eq:average}, we can look at each of the cases of Eq.~\eqref{eq:SCpol} separately. For the second line (case), we have that the product of the SC polarization vectors is

\small
\be
\f{e^{i\mathbf{q} \cdot \mathbf{R}_l}\e^{i\a}_{\mathbf{q}\nu} + e^{-i\mathbf{q}\cdot\mathbf{R}_l}(\e^{i\a}_{\mathbf{q}\nu})^\ast}{\sqrt{2}} \f{e^{i\mathbf{q} \cdot \mathbf{R}_m}\e^{j\b}_{\mathbf{q}\nu} + e^{-i\mathbf{q}\cdot\mathbf{R}_m}(\e^{j\b}_{\mathbf{q}\nu})^\ast}{\sqrt{2}} 
\label{eq:inter}
\ee
\normalsize

\ni The usual contribution (corresponding to Eqs.~\eqref{eq:Fan} and \eqref{eq:g_gDW}), involves the factors\cite{Allen1981}

\be
e^{-i \mathbf{q} \cdot \mathbf{R}_l} (\e^{i\a}_{\mathbf{q}\nu})^\ast e^{i \mathbf{q} \cdot \mathbf{R}_m} \e^{j\b}_{\mathbf{q}\nu}.
\ee

\ni We want to see that the terms of Eq.~\eqref{eq:inter} can be written in this way.

The product of the first and last terms in Eq.~\eqref{eq:inter} can be written as

\be
\f{e^{-i(-\mathbf{q})\cdot\mathbf{R}_l}(\e^{i\a}_{-\mathbf{q}\nu})^\ast e^{i(-\mathbf{q})\cdot\mathbf{R}_m}\e^{j\b}_{-\mathbf{q}\nu}}{2}
\ee

\ni which gives half of the correct result for $-\mathbf{q}$. The product of the second and third term gives one half of the result for $\mathbf{q}$. In the same way, we can write Eq.~\eqref{eq:inter} but for the third case of Eq.~\eqref{eq:SCpol}, and we also get one half of the contributions for $\mathbf{q}$ and $-\mathbf{q}$. In addition, the non-cross terms (the ones with the same momenta) cancel out instead of adding up. Therefore, only terms with opposite momenta contribute

For the first case, the polarization vectors are real and $\mathbf{q} = -\mathbf{q} + \mathbf{G}$ by definition, so $e^{i \mathbf{q} \cdot \mathbf{R}_l} = e^{-i \mathbf{q} \cdot \mathbf{R}_l}$, and the contribution is
\vspace{1cm}

\begin{figure}[!h]
\includegraphics[width=0.4\textwidth]{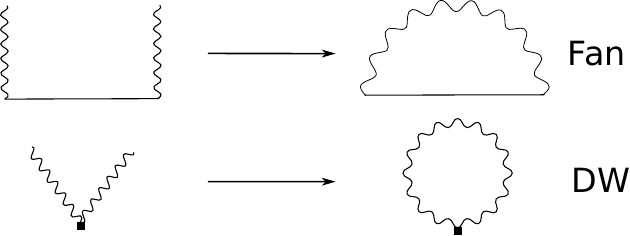}
\caption{Second order diagrams of $\Sigma^\mathrm{red}$ in Eq.~\eqref{eq:Gimplicit}. To the left we have $\mathcal{V}^{I,(1)} G^0 \mathcal{V}^{I,(1)}$ and $\mathcal{V}^{I,(2)}$. When doing the average over the ensemble, the phonon legs have to be paired, leading to the diagrams on the right, which coincide with AHC in the adiabatic limit.}
\label{fig:diagrams2}
\end{figure}

\be
e^{i \mathbf{q} \cdot \mathbf{R}_l} \e^{i\a}_{\mathbf{q}\nu} e^{i \mathbf{q} \cdot \mathbf{R}_m} \e^{j\b}_{\mathbf{q}\nu} = e^{-i \mathbf{q} \cdot \mathbf{R}_l} (\e^{i\a}_{\mathbf{q}\nu})^\ast e^{i \mathbf{q} \cdot \mathbf{R}_m} \e^{j\b}_{\mathbf{q}\nu}
\ee

\ni as we wanted to see. 

Therefore,

\begin{widetext}
\be
\begin{split}
\langle \sum_{\tilde{\mathbf{q}} n''} \mathcal{V}^{I,(1)}_{\mathbf{k}n,\mathbf{k}+\tilde{q}n''} G^0_{\mathbf{k}+\tilde{q}n''} \mathcal{V}^{I,(1)}_{\mathbf{k}+\tilde{\mathbf{q}}n''\mathbf{k}n'} \rangle  =\f{1}{N}  \sum_{ \substack{i\a, j\b \\ n'',\mathbf{q}\nu}} & \langle \mathbf{k}n| \f{\p V}{\p u_{i\a}(-\mathbf{q})} |\mathbf{k} + \mathbf{q} n''\rangle  \langle \mathbf{k} + \mathbf{q} n''| \f{\p V}{\p u_{j\b}(\mathbf{q})} |\mathbf{k} n'\rangle \\
& \times \f{1}{\o + i \d - \vare^0_{\mathbf{k}+\mathbf{q}n''}}\f{(\e^{i\a}_{\mathbf{q}\nu})^\ast\e^{j\b}_{\mathbf{q}\nu}}{2 \o_{\mathbf{q}\nu} \sqrt{M_i M_j}} (2n_{\mathbf{q}\nu}+1)\\
& = \f{1}{N} \sum_{n'',\mathbf{q}\nu} g^{-\mathbf{q}\nu}_{\mathbf{k+q},nn''} G^0_{\mathbf{k+q}n''} g^{\mathbf{q}\nu}_{\mathbf{k},n''n'} (2n_{\mathbf{q}\nu}+1)
\end{split}
\label{eq:NP_Fan}
\ee

\end{widetext}

\ni which corresponds to the usual Fan term, Eqs.~\eqref{eq:Fan} (and \eqref{eq:g_gDW}).
(Here we have used that the eigenstates $|\mathbf{k}n\rangle$

 The proof for Eq.~\eqref{eq:DW_FAN}(b) is similar. So aside from the $2n_{\mathbf{q}\nu}+1$ factor, the result corresponds to identifying $\mathcal{V}$ with $g$ and, without worrying about the sign, $\tilde{\mathbf{q}}$ with the phonon wavevector. Also, we see that averaging is necessary to obtain the standard diagrams, and not just a way to obtain smoother spectral functions.

The pairing of factors of $u$ can be done diagrammatically by assigning a ``leg'' to each factor and joining them. This is illustrated in Fig.~\ref{fig:diagrams2}. In the case of $\langle \mathcal{V}^{I,(2)} \rangle$, the two legs come out from one point. In the case of $\langle \mathcal{V}^{I,(1)} G^0 \mathcal{V}^{I,(1)} \rangle$, they come from two different points. In the next section we look at terms of higher order and their corresponding diagrams.\\

\subsection{Higher order terms}

\subsubsection{Wick's theorem}

For terms of order 2, we had two modes, which we labeled $\mathbf{q}\nu$ and $\mathbf{q}'\nu'$. For terms of order $n$, we use the label $q_l=\mathbf{q}_l \nu_l$, with $l=1,..,n$. Now, instead of the ``2-point correlator'' $\langle \tilde{\xi}_{q_1} \tilde{\xi}_{q_2} \rangle$ (omitting the $I$ index), we have to evaluate the $n$-point correlator $\langle \tilde{\xi}_{q_1}...\tilde{\xi}_{q_n} \rangle$.  If the distribution that generates the phonons is Gaussian, because of Wick's theorem\cite{Zee2010}, we can exactly write

\be
\langle \tilde{\xi}_{q_1}...\tilde{\xi}_{q_n} \rangle = \sum_P \langle \tilde{\xi}_{q_{P(1)}} \tilde{\xi}_{q_{P(2)}} \rangle...\langle \tilde{\xi}_{q_{P(n-1)}} \tilde{\xi}_{q_{P(n)}} \rangle
\ee

\ni where the sum is over all possible pairings $P$ of the indices. If all $q_l$ are different, the correlator is 0. If they can be grouped in pairs, but not more than two $q_l$ are the same, the correlator is 1. If more than two $q_l$ are the same, more than one pairing will contribute. (For example, for $n=4$ and a given set of vertices of Fig.~\ref{fig:diagrams4}, and $q_1=q_2=q_3=q_4$, there are 3 permutations, which equally contribute to the 3 diagrams corresponding to such vertices). For an odd number of terms, the result is 0. So $n$-point correlators reduce to products of 2-point correlators. Then, for higher order terms, legs can also be assigned to each $u$, and all possible diagrams are created by joining legs in all possible ways, just as for the standard diagrams.

The momenta conservation we mentioned earlier corresponds diagrammatically to the conservation of momenta in each vertex. When joining legs, the terms that contribute are those of opposite momenta, so the diagrams have the usual momenta structure: the leg carries \textit{out} a momenta $\mathbf{q}$ from one point (adds momenta $-\mathbf{q}$ to the vertex), and brings it back \textit{in} to another vertex (adds $\mathbf{q}$ to the vertex). As we see in more detail in Secs.~\ref{sec:higher_order} and \ref{sec:comparison_full}, this leads to the standard Feynman diagrams.

\subsubsection{Non-Gaussian distributions}

First, let us say a few more words about non-Gaussian distributions. In this case, higher-order moments have to be evaluated separately. That is, if we have more than two $q_l$ (an even number) that correspond to the same mode, then their correlator does not reduce to a product of 2-point correlators. For example, for $n=4$, we have

\be
\langle \tilde{\xi}_{q_1} \tilde{\xi}_{q_2} \tilde{\xi}_{q_3} \tilde{\xi}_{q_4} \rangle = \begin{cases} \langle \tilde{\xi}_{q_1} \tilde{\xi}_{q_1}\rangle \langle \tilde{\xi}_{q_3} \tilde{\xi}_{q_3} \rangle \hspace{1mm} \substack{\mathrm{if} \hspace{1mm} q_1=q_2,q_3=q_4\\ q_1 \neq q_3\\ \mathrm{or  \hspace{1mm} permutations}}  \\
\langle \tilde{\xi}_{q_1} \tilde{\xi}_{q_1} \tilde{\xi}_{q_1} \tilde{\xi}_{q_1}\rangle \hspace{5mm} \textrm{if all equal} \\
0 \hspace{22mm} \textrm{if all different}
\end{cases}
\ee

\ni If the distribution is such that the 2-point correlator is 1, because of Eq.~\eqref{eq:average}, then we recover Fan and DW terms. But $\langle \tilde{\xi}_{q_1} \tilde{\xi}_{q_1} \tilde{\xi}_{q_1} \tilde{\xi}_{q_1}\rangle$ will depend on the distribution, and higher order terms will not coincide with the Feynman diagrams.

Ref.~\onlinecite{Zacharias2020} proposed using a particular configuration to determine $\mathbf{k}$ averaged quantities, which would require the use of only one configuration instead averaging over $N_c$ configurations. Their configuration corresponds to setting $\tilde{\xi}_{\mathbf{q}\nu} = \pm 1$ (not following a Gaussian distribution), alternating the sign of nearby $\mathbf{q}$ points in the BZ to maximize the cancellation of terms that are not present in the exact thermally averaged expression to second order. In our proof, for the $\mathbf{k}$ resolved self-energy, we saw that for finite SCs, averaging is necessary to get Wick's theorem and match opposite momenta. Using one configuration, different branches and different momenta (which need not be opposite) remain correlated. Pictorically, the legs remain unpaired. Indeed, we tested the configuration of Ref.~\onlinecite{Zacharias2020} for different temperatures and couplings, and got the same type of unconverged spectral function as when using one random configuration with the Gaussian distribution. See Fig.~\ref{fig:ZG} after the introduction of the graphene model in Sec.~\ref{sec:mod_trick}. Thus, the displacement of Ref.~\onlinecite{Zacharias2020} does not seem useful to reduce the number of configurations for $\mathbf{k}$ resolved quantities.

\subsubsection{Higher order diagrams and 4th order as an example}
\label{sec:higher_order}

We mentioned that Wick's theorem leads to the usual diagrams. To understand higher-order terms in more detail, let us now look at the diagrams of order 4. The contributions to $\Sigma^\mathrm{red}$ are: 

\be
\begin{split}
& (a) \langle \mathcal{V}^{(2)} \mathcal{G}_0 \mathcal{V}^{(2)}\rangle \\
& (b)  \langle \mathcal{V}^{(2)} \mathcal{G}_0 \mathcal{V}^{(1)} \mathcal{G}_0 \mathcal{V}^{(1)}\rangle \\
& (c) \langle \mathcal{V}^{(1)} \mathcal{G}_0 \mathcal{V}^{(2)} \mathcal{G}_0 \mathcal{V}^{(1)}\rangle \\
& (b)' \langle \mathcal{V}^{(1)} \mathcal{G}_0 \mathcal{V}^{(1)} \mathcal{G}_0 \mathcal{V}^{(2)}\rangle \\
& (d) \langle \mathcal{V}^{(1)} \mathcal{G}_0 \mathcal{V}^{(1)} \mathcal{G}_0 \mathcal{V}^{(1)}\mathcal{G}_0 \mathcal{V}^{(1)}\rangle
\end{split}
\label{eq:fourth}
\ee

\begin{figure*}
\includegraphics[width=1.0\textwidth]{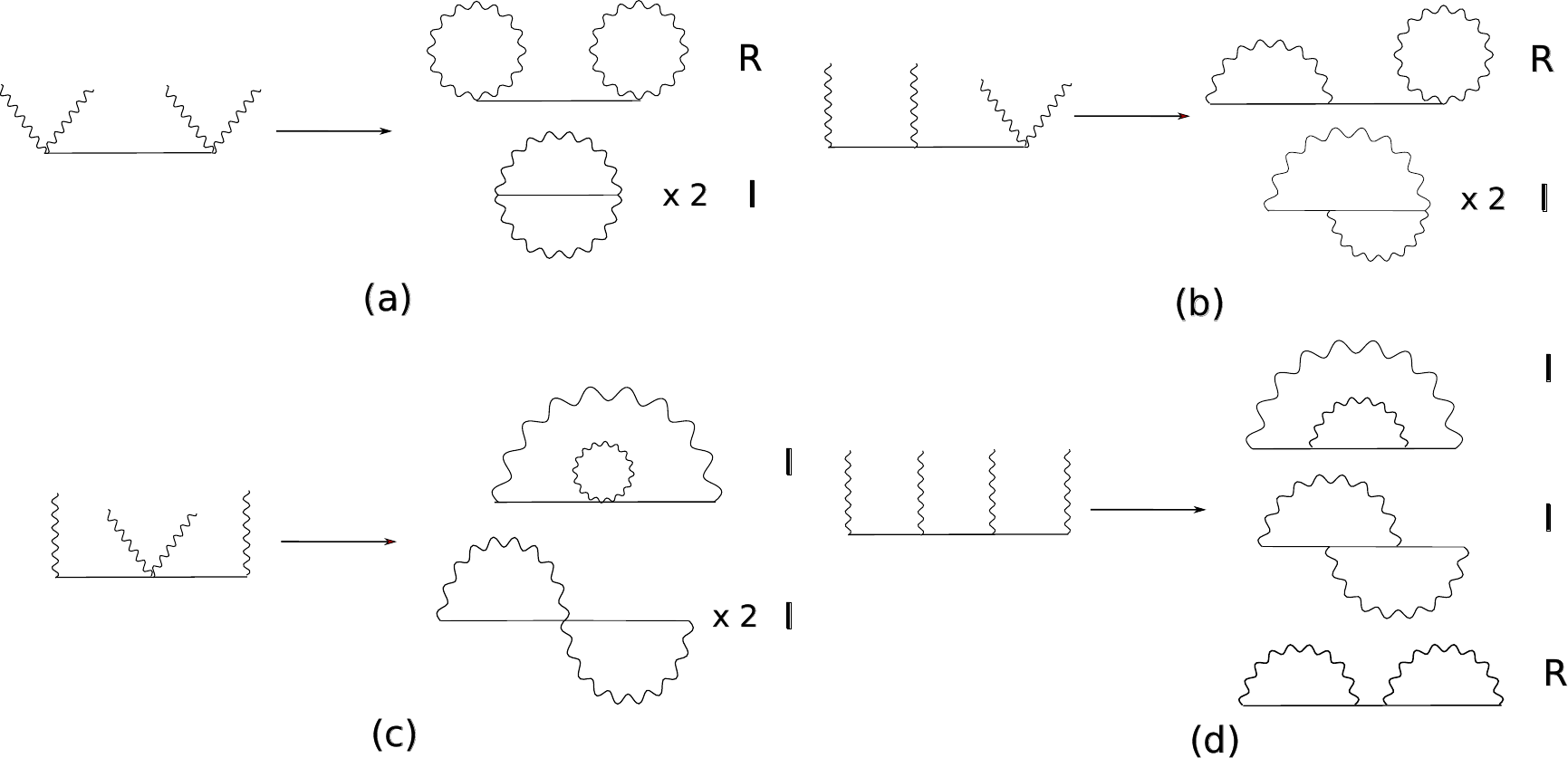}
\caption{Diagrams to order 4 in the displacements $u$ corresponding to Eq.~\eqref{eq:fourth}, without including indices for simplicity (to keep track of the momenta indices and momentum conservation, arrows should be added to each diagram). Vertices with one phonon line correspond to a $g$ factor, and the ones with two phonon lines to $g^\textrm{DW}$. The factors 2 correspond to different contractions that give rise to the same diagram, just as in the Wick contractions of the standard approach. The diagrams that determine $\Sigma$ are the irreducible ones I, as opposed to the reducible ones R.}
\label{fig:diagrams4}
\end{figure*}

\ni and the corresponding diagrams are illustrated in Fig.~\ref{fig:diagrams4}. Diagrams of (b)' are analogous to those of (b). `R' diagrams are reducible (can be cut in two to give other diagrams), while `I' diagrams are irreducible (they cannot be cut into two allowed diagrams)\cite{Mahan2000}. They correspond to a subset of the usual diagrams of the full theory (see the next subsection for more details), and the same holds for higher order terms.

Thus, if we define $\Sigma$ in the usual way, as the sum of the irreducible diagrams without the external legs $G_0$, we can write 

\be
\begin{split}
G & = G^0 + G^0 \Sigma G^0 + G^0 \Sigma G^0 \Sigma G^0 + ... \\
 & = G^0 + G^0 \Sigma G,
\end{split}
\label{eq:Sigmaimplicit}
\ee

\begin{figure}
\includegraphics[width=0.4\textwidth]{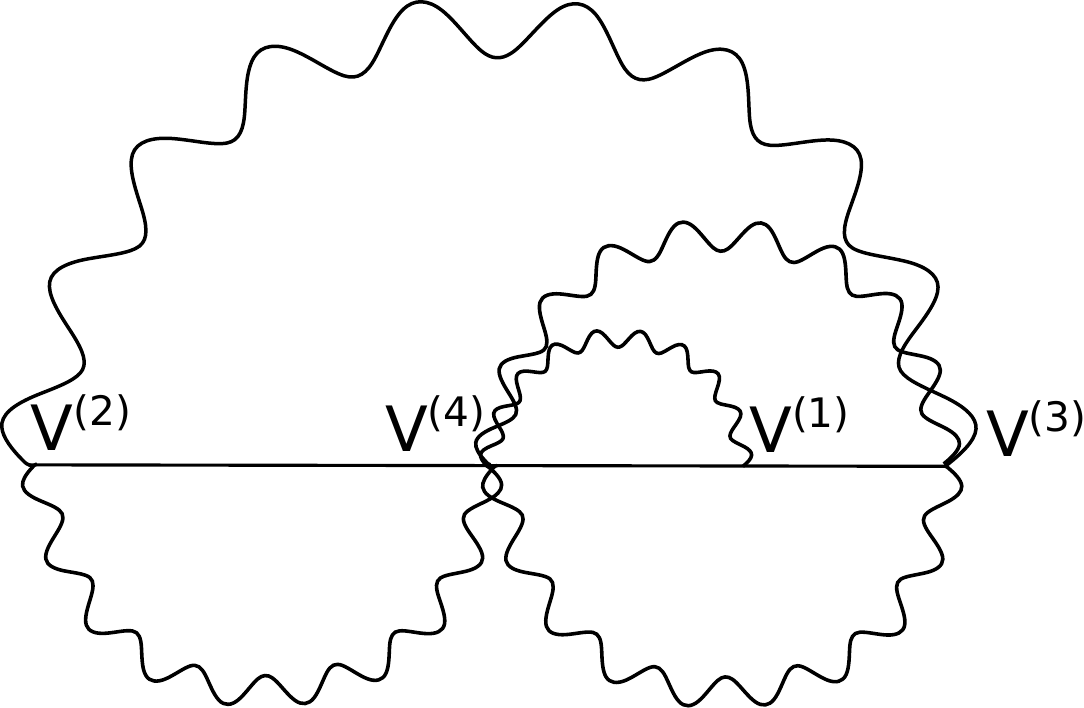}
\caption{Generic diagram corresponding to the self-energy in our theory (order 10), including also vertices with 3 and 4 legs, not present in the previous figures.}
\label{fig:diagram10}
\end{figure}

\ni which is just the Dyson's equation and coincides with definition Eq.~\eqref{eq:Sigma}. So $\Sigma$ is given by terms of the form $\langle \mathcal{V} \mathcal{G}_0 \mathcal{V} \mathcal{G}_0 ... \mathcal{V} \mathcal{G}_0 \rangle$ which are irreducible. Diagrammatically, this means drawing a straight (fermionic) line and vertices on top of it with one, two or more wavy (phononic) lines, and doing all possible contractions of the wavy lines that give rise to an irreducible diagram. The usual theoretical expression for the Hamiltonian, Eq.~\eqref{eq:Hfull}, includes one and two derivatives with respect to ionic displacements. In principle, all derivatives should be included in the Hamiltonian. Such terms are automatically included in our work and other numerical approaches that consider distorted ionic configurations. The term with $p$ derivatives corresponds in the diagrams to a vertex with $p$ legs. A generic diagram of our theory can be observed in Fig.~\ref{fig:diagram10}, where the order of the derivative is explicitly indicated in each vertex. We remind the reader that our method is non-perturbative, so that we can obtain the Green's function even when the variation of the potential with the displacements in not analytical (cannot be expressed using perturbation theory).
 
 The last thing we have to do is check that diagrams have exactly the same analytical expression as the corresponding standard Feynman diagram (in the fully quantized theory), in the adiabatic limit. Let us show this for a particular diagram, the one of the middle of Fig.~\ref{fig:diagram4}(d). From this analysis, it will then be easy to see that any diagram corresponding to our method coincides with the standard diagram. 
 
The diagram we just mentioned comes from $ \langle \mathcal{V}^{(1)} \mathcal{G}_0 \mathcal{V}^{(1)} \mathcal{G}_0 \mathcal{V}^{(1)} \mathcal{G}_0 \mathcal{V}^{(1)} \rangle$. Just writing indices explicitly for matrix element $\mathbf{k}n,\mathbf{k}n'$ before averaging, this is

\begin{widetext}

\be
\begin{split}
\sum_{\tilde{\mathbf{q}}_1 \tilde{\mathbf{q}}_2 \tilde{\mathbf{q}}_3} \sum_{n_1n_2n_3} \mathcal{V}^{(1)}_{\mathbf{k}n,\mathbf{k+\tilde{q}}_1 n_1} \mathcal{G}^0_{\mathbf{k+\tilde{q}}_1n_1}   \mathcal{V}^{(1)}_{\mathbf{k}+\mathbf{\tilde{q}}_1 n1,\mathbf{k}+\tilde{\mathbf{q}}_1 + \tilde{\mathbf{q}}_2 n_2} \mathcal{G}^0_{\mathbf{k}+\tilde{\mathbf{q}}_1 + \tilde{\mathbf{q}}_2 n_2} \mathcal{V}^{(1)}_{\mathbf{k} + \tilde{\mathbf{q}}_1 + \tilde{\mathbf{q}}_2 n_2,\mathbf{k}+ \tilde{\mathbf{q}}_1 +\tilde{\mathbf{q}}_2 +\tilde{\mathbf{q}}_3 n_3} \\ \times \mathcal{G}^0_{\mathbf{k} + \tilde{\mathbf{q}}_1 + \tilde{\mathbf{q}}_2 + \tilde{\mathbf{q}}_3 n_3} \mathcal{V}^{(1)}_{\mathbf{k} +\tilde{\mathbf{q}}_1 + \tilde{\mathbf{q}}_2 + \tilde{\mathbf{q}}_3 n_3,\mathbf{k}n'}
\end{split}
\label{eq:first_step}
\ee

\ni As before, each term $\mathcal{V}^{(1)}$ has a displacement written in momentum space, with a random factor $\tilde{\xi}_{\mathbf{q}_i\nu_i}$ for each mode. Taking the average and using Wick's theorem, we can look at the diagram we are interested in now (pairing terms 1 and 3, and 2 and 4). In Eq.~\eqref{eq:NP_Fan} we put all cases of Eq.~\eqref{eq:SCpol} together, paired momenta and used momenta conservation. Proceeding in the same way, we can write (with more generality in the indices),

\be
\langle \mathcal{V}^{(1)}_{\mathbf{k_1}n_1,\mathbf{k}_2 n_2} \mathcal{V}^{(1)}_{\mathbf{k}_3 n_3,\mathbf{k}_4 n_4} \rangle = \f{1}{N}\sum_\nu g^\nu_{\mathbf{k}_1 n_1,\mathbf{k}_2 n_2} g^\nu_{\mathbf{k}_3 n_3,\mathbf{k}_4n_2} (2 n_{\mathbf{k}_2-\mathbf{k}_1,\nu}+1) \d_{\mathbf{k}_2-\mathbf{k}_1,-(\mathbf{k}_4-\mathbf{k}_3)}
\label{eq:pairing}
\ee

\ni where we have switched to the notation $g^{\mathbf{q}\nu}_{\mathbf{k},nn'}$ to $g^{\nu}_{\mathbf{k+q}n,\mathbf{k}n'}$. Pairing terms 1 and 3, and 2 and 4 in Eq.~\eqref{eq:first_step}, and using Eq.~\eqref{eq:pairing} for each pairing, we get

\be
\begin{split}
\f{1}{N^2}\sum_{ \substack{\mathbf{q}_1 \mathbf{q}_2 \\ \nu_1 \nu_2}} \sum_{n_1 n_2 n_3} g^{\nu_1}_{\mathbf{k}n,\mathbf{k}+\mathbf{q}_1n_1} \mathcal{G}^0_{\mathbf{k+q}_1n_1} g^{\nu_2}_{\mathbf{k}+\mathbf{q}_1 n_1,\mathbf{k}+\mathbf{q}_1 + \mathbf{q}_2 n_2} \mathcal{G}^0_{\mathbf{k+q}_1+\mathbf{q}_2 n_2} & g^{\nu_1}_{\mathbf{k}+\mathbf{q}_1+\mathbf{q}_2 n_2,\mathbf{k} + \mathbf{q}_2 n_3}  \mathcal{G}^0_{\mathbf{k+q}_2 n_3} g^{\nu_2}_{\mathbf{k}+\mathbf{q}_2 n_3,\mathbf{k} n'} \\
& \times (2 n_{\mathbf{q}_1 \nu_1} + 1)(2 n_{\mathbf{q}_2 \nu_2} + 1).
\end{split}
\label{eq:diagram4_NP}
\ee

\end{widetext}

\ni Similarly, one can work out diagrams with higher order derivatives. The difference is that the matrix element $g^{(p)}$ of order $p$ has $p$ momenta indices. Therefore, if we want to extract the result more directly from a diagram, we see that momenta follows the usual rules, due to momenta conservation at each vertex and each phonon line carrying momentum in and out (i.e. the pairing of opposite momenta in each 2-point correlator after applying Wick's theorem). To label momenta indices correctly, arrows should be drawn in each straight or wavy line. Each straight line is associated with a $\mathcal{G}^0$, each vertex with the corresponding $g^{(p)}$ matrix element, and each wavy line with a mode $\mathbf{q}\nu$ and a $2n_{\mathbf{q}\nu}+1$ factor. In the next subsection, we show that the usual Feynman diagrams have the same expression in the adiabatic limit. First, we will see that Feynman rules applied to the middle diagram of Fig.~\ref{fig:diagrams4}(d) leads to the exact same expression of Eq.~\eqref{eq:diagram4_NP}.

\subsection{Comparison to the full theory}
\label{sec:comparison_full}

\begin{figure}
\includegraphics[width=0.4\textwidth]{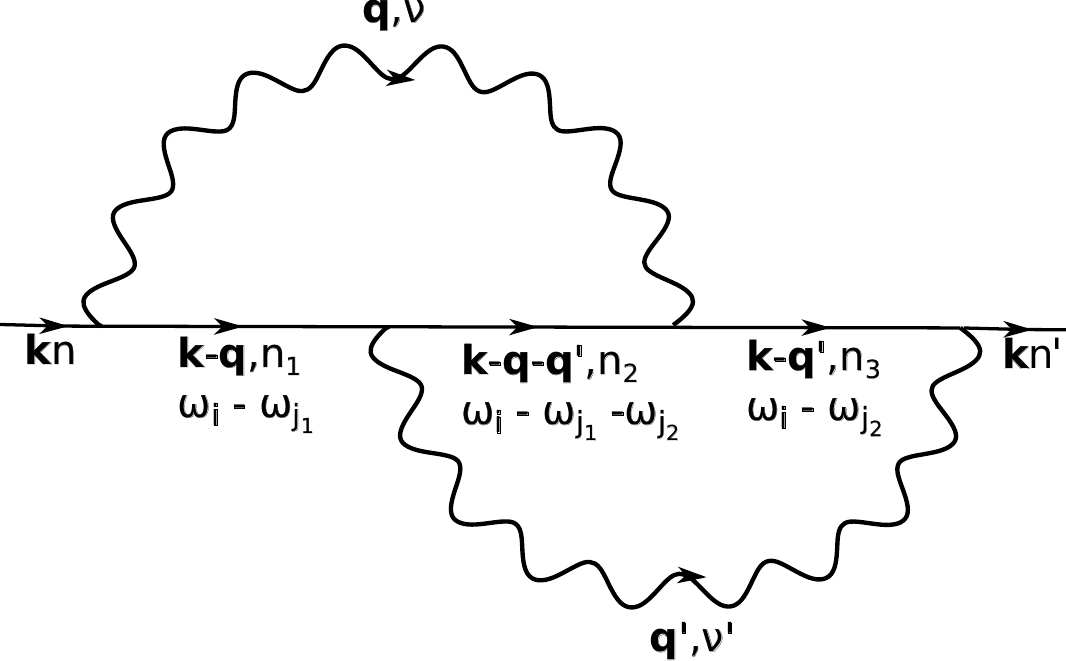}
\caption{Fourth order diagram of Fig.~\ref{fig:diagrams4}d including all indices. It is worked out in detail in the text.}
\label{fig:diagram4}
\end{figure}

The diagram we just analyzed in more detail in our method is analogous to a diagram in the fully quantized theory. Following Feynman rules in the Matsubara formalism, the diagram can be labeled with the indices of Fig.~\ref{fig:diagram4}, and written as

\begin{widetext}
\be
\begin{split}
-\f{-1}{(\b N)^2} \sum_{\substack{\mathbf{q}\nu \\ \mathbf{q}'\nu'}} \sum_{n_1 n_2 n_3} \sum_{j_1j_2}& g^{\mathbf{q}\nu}_{\mathbf{k-q},nn_1}g^{\mathbf{q}'\nu'}_{\mathbf{k}-\mathbf{q-q'},n_1n_2}g^{-\mathbf{q}\nu}_{\mathbf{k}-\mathbf{q}',n_2 n_3}g^{-\mathbf{q}'\nu'}_{\mathbf{k},n_3n'}\\
& \times D^0_{\mathbf{q}\nu}(\o_{j_1})D^0_{\mathbf{q}'\nu'}(\o_{j_2})\\
& \times G^0_{\mathbf{k}-\mathbf{q}n_1}(\o_i - \o_{j_1}) G^0_{\mathbf{k}-\mathbf{q}-\mathbf{q}'n_2}(\o_i - \o_{j_1}- \o_{j_2})G^0_{\mathbf{k}-\mathbf{q}'n_3}(\o_i - \o_{j_2})
\end{split}
\ee
\end{widetext}

\ni where

\be
\begin{split}
G^0_{\mathbf{k}n}(\o_j) & =\f{1}{i\o_j - \vare^0_{\mathbf{k}n} }\\
\end{split}
\ee

\ni and

\be
\begin{split}
D^0_{\mathbf{q}\nu}(\o_j) & = \f{1}{i\o_j - \o_{\mathbf{q}\nu}} - \f{1}{i\o_j + \o_{\mathbf{q}\nu}} \\
& = -\f{2 \o_{\mathbf{q}s}}{\o_j^2 + \o_{\mathbf{q}\nu}^2}
\end{split}
\ee

\ni are the unperturbed time-ordered electron and phonon Green's function in imaginary time, respectively.

After doing the sums over the Matsubara frequencies, the phonons frequencies $\o_{\mathbf{q}\nu}$ in $D^0$ and $G^0$ are set to 0. This is what we refer to as the adiabatic limit, which corresponds to the usual adiabatic limit in the Fan self-energy (dropping phonons frequencies  in the energy denominators).

The sum over Matsubara frequencies $\o_j$ is done in the standard way, by converting it into a complex integration with the Bose-Einstein factor $n$. Since $D^0_{\mathbf{q}\nu}(\o_j) \propto \o_{\mathbf{q}\nu} \rightarrow 0$ if $\o_{\mathbf{q}\nu} \rightarrow 0$, the only poles that contribute are those of $D^0$ (the poles of $G^0$ give terms with a $D^0 \rightarrow 0$ factor). This gives factors $-n(\o_{\mathbf{q}'\nu'})$ and $n(-\o_{\mathbf{q}'\nu'}) = -1 - n(\o_{\mathbf{q}'\nu'})$, and the frequencies $\o_j= \pm \o_{\mathbf{q}'\nu'}$ are dropped in the $G^0$'s (making the factors $-n(\o_{\mathbf{q}'\nu'})$ and $-n(\o_{\mathbf{q}'\nu'})-1$ proportional to the same term, simplifying the result). Performing these steps for $j_2$, we have

\begin{widetext}
\be
\begin{split}
\f{1}{N^2}\f{1}{\b } \sum_{\substack{\mathbf{q}\nu \\ \mathbf{q}'\nu'}} \sum_{n_1 n_2 n_3} \sum_{j_1}& g^{\mathbf{q}\nu}_{\mathbf{k-q},nn_1}g^{\mathbf{q}'\nu'}_{\mathbf{k}-\mathbf{q-q'},n_1n_2}g^{-\mathbf{q}\nu}_{\mathbf{k}-\mathbf{q}',n_2 n_3}g^{-\mathbf{q}'\nu'}_{\mathbf{k},n_3n'}\\
& \times D^0_{\mathbf{q}\nu}(\o_{j_1}) \\
& \times G^0_{\mathbf{k}-\mathbf{q}n_1}(\o_i - \o_{j_1}) G^0_{\mathbf{k}-\mathbf{q}-\mathbf{q}'n_2}(\o_i - \o_{j_1})G^0_{\mathbf{k}-\mathbf{q}'n_3}(\o_i) \\
& \times (-2 n_{\mathbf{q}'\nu'}-1)
\end{split}
\ee

\ni Notice how the procedure leads to a factor of -1 proportional to the usual factor $2n+1$.

Doing the sum over $j_1$, and switching again to the notation of Eq.~\eqref{eq:pairing} for $g$,

\be
\begin{split}
\f{1}{N^2} \sum_{\substack{\mathbf{q}\nu \\ \mathbf{q}'\nu'}} \sum_{n_1 n_2 n_3} & g^{\nu}_{\mathbf{k}n,\mathbf{k}-\mathbf{q}n_1}g^{\nu'}_{\mathbf{k}-\mathbf{q}n_1,\mathbf{k}-\mathbf{q}-\mathbf{q}'n_2}g^{\nu}_{\mathbf{k}-\mathbf{q}-\mathbf{q}'n_2,\mathbf{k}-\mathbf{q}'n_3}g^{\nu'}_{\mathbf{k}-\mathbf{q}'n_3,\mathbf{k}n'}\\
& \times G^0_{\mathbf{k}-\mathbf{q}n_1}(\o_i) G^0_{\mathbf{k}-\mathbf{q}-\mathbf{q}'n_2}(\o_i)G^0_{\mathbf{k}-\mathbf{q}'n_3}(\o_i) \\
& \times (2 n_{\mathbf{q}'\nu'}+1)(2 n_{\mathbf{q}\nu}+1)
\end{split}
\ee
\end{widetext}

\ni Doing the analytical continuation $i \o_i \rightarrow \o + i \delta$, we get the final expression. It coincides (changing the sign of the phonon momenta $\mathbf{q},\mathbf{q}'$ to $-\mathbf{q}_1,-\mathbf{q}_2$, and $\nu,\nu' \rightarrow \nu_1,\nu_2$) with Eq.~\eqref{eq:diagram4_NP}.

In general, we see that the $G^0$'s will be evaluated at $\o_i$ after doing all the Matsubara sums and each of the $D^0_{\mathbf{q}_l\nu_l}$ gives a factor $-(2n_{\mathbf{q}_l\nu_l}+1)$. Our method also has such factors for each phonon line. The momentum structure of the $g$ and $G^0$ with the Feynman rules is straightforward, and earlier we saw that our method follows the same rules (we also have momentum conservation, and pairings also correspond to the phonon momenta coming in and and out). Using the Feynman rules, the sign is less straightforward. It is $(-1).(-1)^v.(-1)^b.(-1)^f$, with $v$, $b$, $f$ the number of vertices, phonon propagators and fermionic propagators, respectively. The first -1 just comes from the definition of G. Each vertex has a $-$ sign, coming from the exponential $e^{-\beta H}$ that gives rise to the Feynman expansion. And from the definition of $G^0$ and $D^0$, with a $-$ in front, each contraction that gives a $G^0$ and $D^0$ give an additional sign. Furthermore, we saw that each Matsubara sum gives an additional $-$. Thus, $D^0$ together with the corresponding Matsubara sum gives a $+1$ factor. Finally, if there is one vertex, there is no internal $G^0$. For each additional vertex (-1 factor) there is an additional $G^0$ (additional $-1$). Since the sign of $\Sigma^\mathrm{DW}$ (one vertex) is +1, then the total sign is $+1$ in all diagrams, just as in our method. This completes the proof.

We see from the proof that the diagrams are exact for any SC (i.e. for a given $\mathbf{q}$-grid commensurate with the SC). There are no additional undesired terms that become negligible in the thermodynamic limit, or $\mathbf{q}$ points that are neglected. In the alternative self-energy of Sec.~\ref{sec:alternative} for example, there are terms missing in the sum over momenta. Averaging the most convenient quantity could also be relevant in the determination of other quantities through distorted configurations.

The results do not depend on the Fermi-Dirac distribution $f$, since there is no contribution from a fermionic Matsubara sum. This occurs because transforming the phonon into a parameter prevents electrons and phonons from exchanging energy; instead, the electronic cloud is continuously distorted according to the ionic displacement. Since there are no electronic excitations due to phonons, there is no need to know the probability that electrons are in a given state or that they will transition to an unoccupied state, so concentration plays no role. The phonons do keep a quantum character in the way they distort the crystal, via the factor $\sqrt{2n+1}$ in the ionic displacements.

\begin{figure}[!h]
\includegraphics[width=0.2\textwidth]{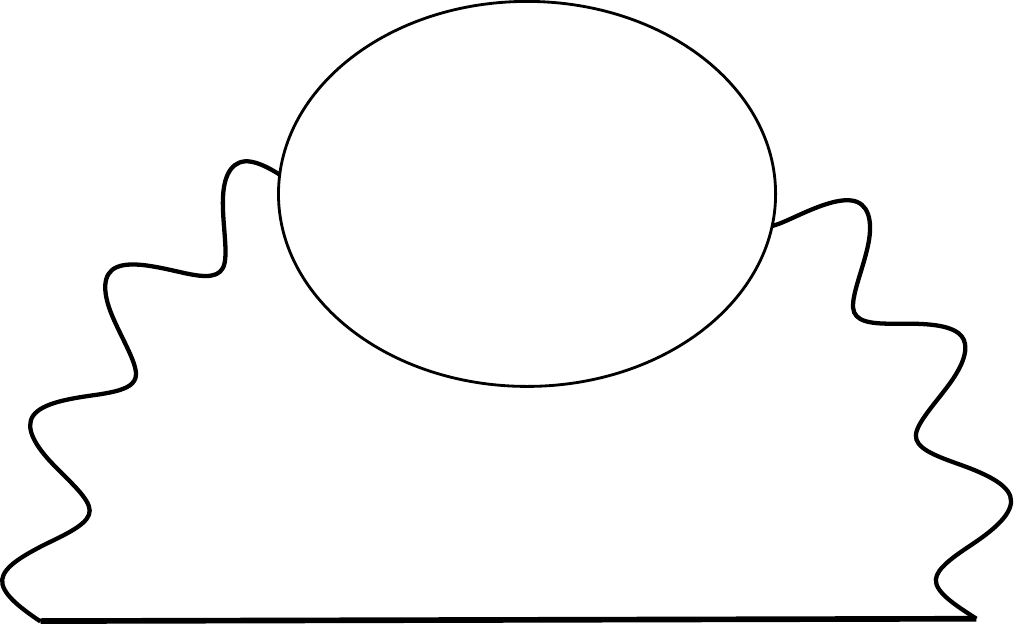}
\includegraphics[width=0.2\textwidth]{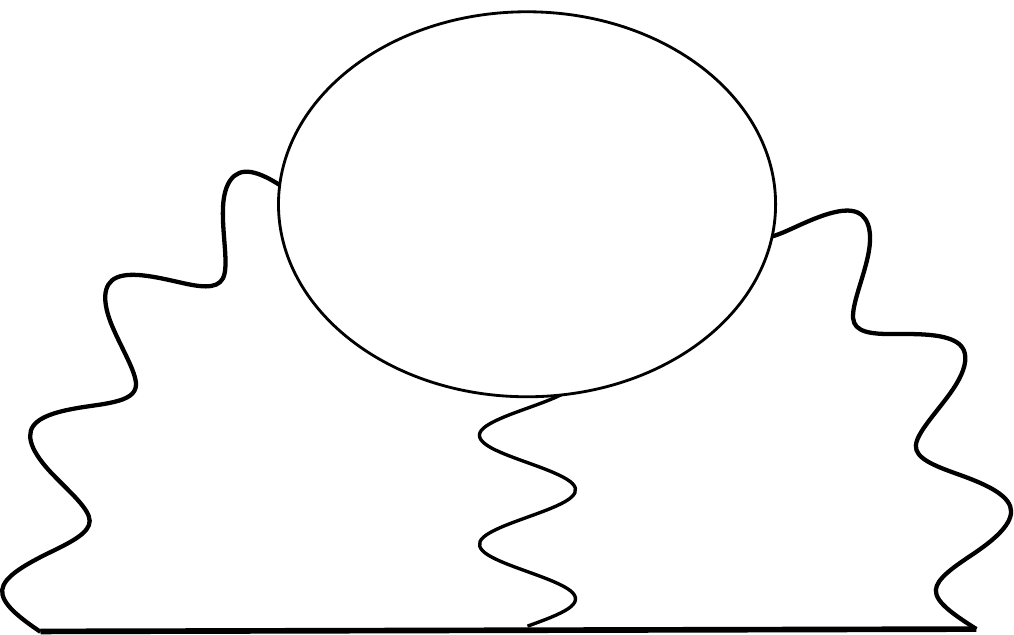}\\
(a) \hspace{2.8cm} (b)
\caption{(a) Diagrams with additional bubbles are actually included, since they correspond to a phonon renormalization, and the phonons corresponding to DFT calculations are screened quantities. Diagrams like (b) instead are not included, but correspond to anharmonicities, which we are not considering.}
\label{fig:bubbles}
\end{figure}

We proved that our method contains an infinite number of diagrams that coincide with the standard diagrams in the adiabatic limit. The exact theory has additional diagrams actually: those with a fermionic bubble, as in Fig.~\ref{fig:bubbles}. Diagrams of type (a), with a bubble or some structure in between two phonons lines, correspond to a phonon renormalization. Since we are using experimental phonons (phonons fitted to experimental data), all diagrams that correct the phonons are already included. That is, the phonon lines in this work correspond to renormalized phonon lines. Diagrams of type (b) are not included, but after integrating out the bubble, can be seen as an anharmonic term. Since we are not including anharmonicities, it is consistent in our approach to ignore these terms (of at least order 6). As a final remark, off-diagonal elements of the self-energy are automatically included in our method. That is, the usual diagonal approximation\cite{Antonius2015} is (omitting the $\mathbf{k}$ index) $G_{nn}=1/(\o - \vare^0_{n} - \Sigma_{nn})$, instead of $G_{nn}= (\o I - \vare^0 - \Sigma)^{-1}_{nn}$ as in our method. Off-diagonal terms should be included close to band-crossing points if terms beyond second order are relevant.

\section{Numerical implementation}
\label{sec:numerical_implementation}

In Sec.~\ref{sec:theory} we described the method, which involves determining the Green's function for each configuration of the ensemble, averaging, and determining the spectral function via Eqs.~\eqref{eq:Gkn}, ~\eqref{eq:Gav} and ~\eqref{eq:A}. In the limit of an infinite SC and $\d \rightarrow 0$, one gets Eq.~\eqref{eq:Adelta} for the spectral function, but in numerical calculations a broadening parameter needs to be introduced to describe the delta function, typically by considering a finite $\d$ in the previous equations.
The problem of introducing a finite $\d$ is that the peak is artificially broadened by $\sim \d$. To see this more explicitly, let us assume peaks are well separated and that we can use the diagonal approximation (to be clear, in this work we do not use these diagonal approximation and make no assumptions about the separation of the peaks). From Eq.~\eqref{eq:Sigma},

\be
\begin{split}
& A_{\mathbf{k}n}(\tilde{\o}) =\\ & -\f{1}{\pi} \f{\Sigma_{2,\mathbf{k}nn}(\tilde{\o})}{(\o-\vare^0_{\mathbf{k}n}-\Sigma_{1,\mathbf{k}nn}(\tilde{\o}))^2 + (\d+\Sigma_{2,\mathbf{k}nn}(\tilde{\o}))^2}
\label{eq:diag_approx}
\end{split}
\ee

\ni where $\Sigma_1$ is the real part and $\Sigma_2$ the imaginary part of $\Sigma$, and $\tilde{\o}=\o+i\d$ for a more compact notation. If the frequency dependence of the self-energy can be neglected in the region of the peak, then one gets the usual Lorentzian distribution, widened by exactly $\d$. But the larger $\d$ is, the more the shape of the spectral function gets distorted. For example, in a wider spectral function, the frequency dependence of the self-energy tends to become more relevant, and the peak becomes more asymmetric. Also, $\d$ can impact the shape significantly when peaks overlap (in this case of course Eq.~\eqref{eq:diag_approx} is not a good approximation).

To reduce the error introduced by $\d$, we define a new Green's function

\be
\tilde{G}_\mathbf{k} (\o)= \f{1}{\o\mathbb{I}_\mathrm{PC} - \vare^0_\mathbf{k} - \Sigma_\mathbf{k}(\o+i\d)},
\label{eq:Gnew}
\ee

\ni without a finite imaginary part outside of the self-energy, and a new spectral function

\be
\tilde{A}_\mathbf{k}(\o) = -\f{1}{\pi} \mathrm{Tr} \Im m \tilde{G}_\mathbf{k}(\o).
\label{eq:Anew}
\ee

\ni Both $\tilde{G}_\mathbf{k}$ and $\tilde{A}_\mathbf{k}$ depend on $\d$. In this way, we remove the artificial width from $G^0_\mathbf{k}$ (which is actually a delta function, and not a Lorentzian of width $\d$), but retain the width coming from $\Sigma_\mathbf{k}$. Since $\Sigma_\mathbf{k}$ corresponds to the usual sum of retarded diagrams, it has the usual analytical properties, so we should have $\tilde{A}_\mathbf{k} \geq 0$ (which is the case in all of our calculations). Later in Sec.~\ref{sec:mod_trick} and Fig.~\ref{fig:trick}, after we define the graphene model, we show with some numerical examples that this expression gives a better spectral function than the one calculated directly from Eq.~\eqref{eq:Sigma} (i.e., Eqs.~\eqref{eq:Gkn}, \eqref{eq:Gav}, and \eqref{eq:A}). In actual calculations, one also has to consider a finite ensemble. Here we consider enough configurations so as to get an error of about 1-2\%. See Fig.~\ref{fig:conv_delta}.

In the previous section, we defined a Green's function for each configuration, Eq.~\eqref{eq:Gnum}, and then averaged. Another possibility could be to define a self-energy for each configuration, and then average. Let us define a self-energy $\Pi^I_\mathbf{k}$ that only depends on $\mathbf{k}$ (as opposed to $\mathcal{V}^I_{\mathbf{k}\mathbf{k}'}$),

\be
G^I_\mathbf{k}(\o+i\d) =  \f{1}{(\o + i\d)\mathbb{I}_\mathrm{PC} - \vare^0_\mathbf{k} - \Pi^I_\mathbf{k}(\o+i\d)}.
\label{eq:Pi}
\ee

\ni In the Appendix we show that defining the self-energy as $\Pi_\mathbf{k} = \langle \Pi^I_\mathbf{k} \rangle$ coincides with $\Sigma_\mathbf{k}$ of Eq.~\eqref{eq:Sigma} in the thermodynamic limit, but for finite supercells $\Sigma_\mathbf{k}$ provides a better definition than $\Pi_\mathbf{k}$ to determine the Green's and spectral functions.

Therefore, our optimized method to determine the spectral function is the following. First, determine $G_\mathbf{k}$ according to Eqs.~\eqref{eq:Gnum} and \eqref{eq:Gav}, and $\Sigma_\mathbf{k}$ using Eq.~\eqref{eq:Sigma}. 
 Then, determine a new Green's function $\tilde{G}_\mathbf{k}$, Eq.~\eqref{eq:Gnew}, and finally the spectral function $\tilde{A}_\mathbf{k}$ via Eq.~\eqref{eq:Anew}.\\

\section{Graphene tight binding model}

\subsection{Model}
\label{sec:model}

We will now apply our method to a graphene tight-binding model\cite{Venezuela2011}. Let us consider (localized) $p_z$ orbitals $|ls\rangle$ in the SC, with $l$ again the index of the cell and $s=1,2$ of the atoms in each PC. We use as a basis the Bloch states $|\mathbf{k}s\rangle = \f{1}{\sqrt{N}} \sum_l e^{i\mathbf{k}\cdot(\mathbf{R}_l+\pmb{\tau}_s)}|ls\rangle$. If we only consider nearest neighbors (n.n.s), the matrix elements of the undistorted Hamiltonian in the PC $H^0_{\mathbf{k},ss'} = \langle \mathbf{k}s |H^0|\mathbf{k}s'\rangle$ are given by

\be
H_\mathbf{k}=
\begin{pmatrix}
0 & f(\mathbf{k}) \\
f^\ast(\mathbf{k}) & 0
\end{pmatrix}
\label{eq:tb}
\ee

\ni where $f(\mathbf{k})=-t_0\sum_i e^{i\mathbf{k}\cdot \pmb{\tau}_i}$, $t_0$ is the hopping parameter, and $\pmb{\tau}_i$ are the vectors that connect the $s=1$ atom with its n.n.s, which are all $s=2$ atoms. The eigenvalues are $\pm |f(\mathbf{k})|$, which give the characteristic conical bands around the Dirac point K.

 In the SC we use as a basis the states $|S\rangle = |ls\rangle$. The n.n.s correspond to different atoms and the matrix elements $\langle S|H| S'\rangle$ are just $-t_0$ (in the undistorted case). We consider that the hopping parameter $t$ changes linearly\cite{Ortenzi2016} with the distance $d_{SS'}$ between atoms $S,S'$ in the SC,

\be
t^I_{SS'} = t_0 + \eta (1 - d^I_{SS'}/d_0),
\label{eq:hopping}
\ee 
 
\ni with $d_0$ the equilibrium distance $1.42$ \AA. $\eta$ is an electron-phonon coupling parameter, that determines how much the hopping parameters changes for a given distortion.

\subsection{Comparison to the standard method and to the configuration of Ref.~\onlinecite{Zacharias2020}}
\label{sec:mod_trick}

We mentioned that the use of a finite SC implies the use of finite broadening parameters, and that a usual approach is to use a finite $\d$ in the expression of the Green's function, Eq.~\eqref{eq:Gnum}, and to calculate the spectral function using Eqs.~\eqref{eq:A} and Eq.~\eqref{eq:Sigma}. In Sec.~\ref{sec:numerical_implementation} we showed that this artificially increases the width, and to tackle this issue, we introduced a modified method, which uses Eq.~\eqref{eq:Gnew} and Eq.~\eqref{eq:Anew}.

\begin{figure}[!h]
\centering
\subfigure[]{\includegraphics[width=0.2\textwidth]{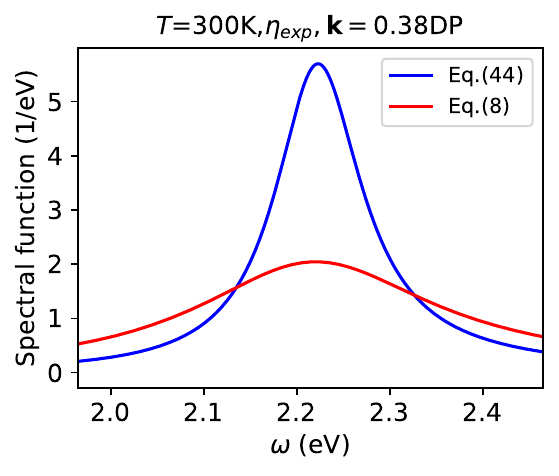}}
\subfigure[]{\includegraphics[width=0.2\textwidth]{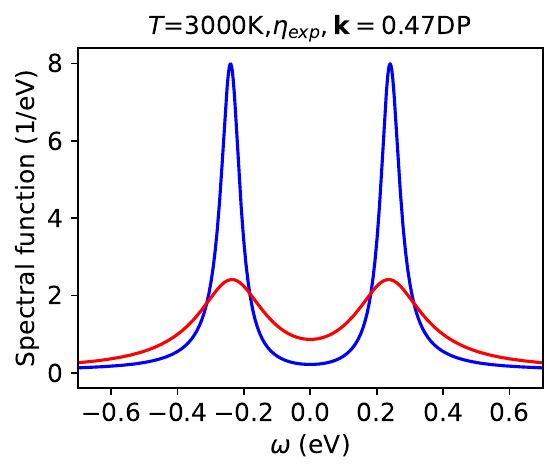}}
\subfigure[]{\includegraphics[width=0.2\textwidth]{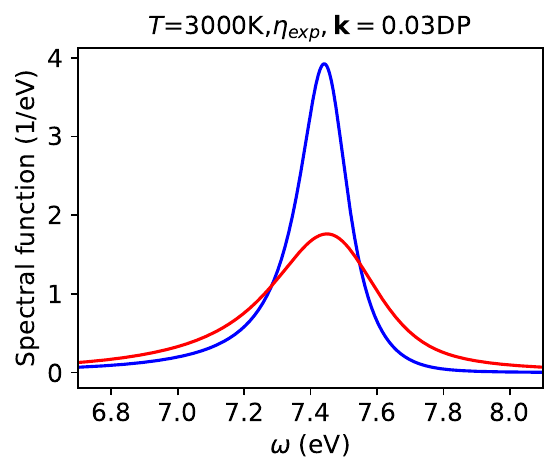}}
\subfigure[]{\includegraphics[width=0.2\textwidth]{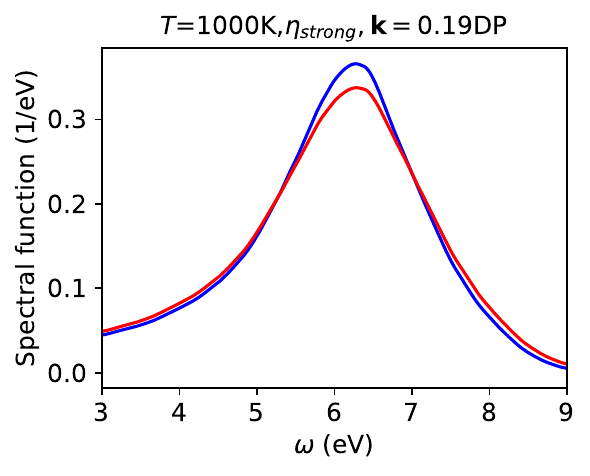}}
\vspace{-2mm}
\caption{Comparison between the spectral functions using our method Eq.~\eqref{eq:Anew} (blue) and the usual approach Eq.~\eqref{eq:A} (red), for different temperatures, $\mathbf{k}$ points and couplings, using $\d=0.1$ eV (lower values can lead to noise in some cases). DP stands for Dirac point (instead of K, to avoid confusion with Kelvin). Values of $\eta$ are defined at the beginning of Sec.~IV C. (a) For narrow linewidths, of the order of 0.1 eV, the red curve is considerably distorted.  (b) Close to K, peaks are actually separate, but in the red curve they are partially merged. (c) The asymmetry is also different between both curves. (d) Only when the linewidth is very big, here of the order of 2 eV ($\gg \d$), both curves are similar ($\d$ does not have much impact on the spectral function, as expected).}
\label{fig:trick}
\end{figure}

With the usual method, the width of the peak increases by about $\d$, so $\d$ should be a fraction of the width of the peak. For example, a common width is of the order of 0.1 eV, so $\d$ should be less than 0.01 eV, but then a large SC is required. Instead, with our method, we can use values that are a fraction of typical energy differences (of the order of the eV), such as $\d=0.1$ eV, and the SC can be smaller. A comparison of both methods can be seen in Fig.~\ref{fig:trick}. If the peak is very wide as in (d), then the usual method and our method give similar spectral functions. But if the width is comparable or lower than $\d$, then our method gives a much more accurate spectral function.

\begin{figure}[!h]
\centering
\includegraphics[width=0.23\textwidth]{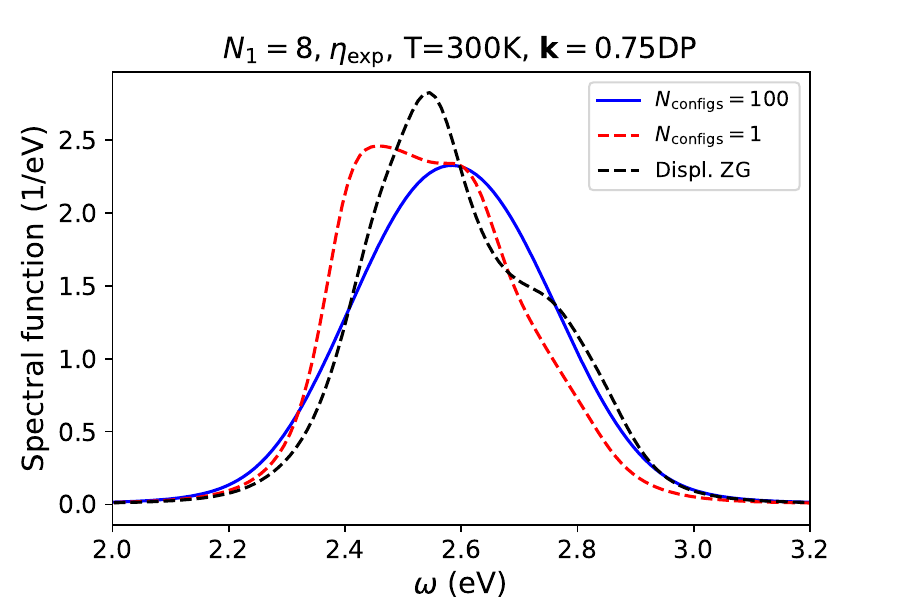}
\includegraphics[width=0.23\textwidth]{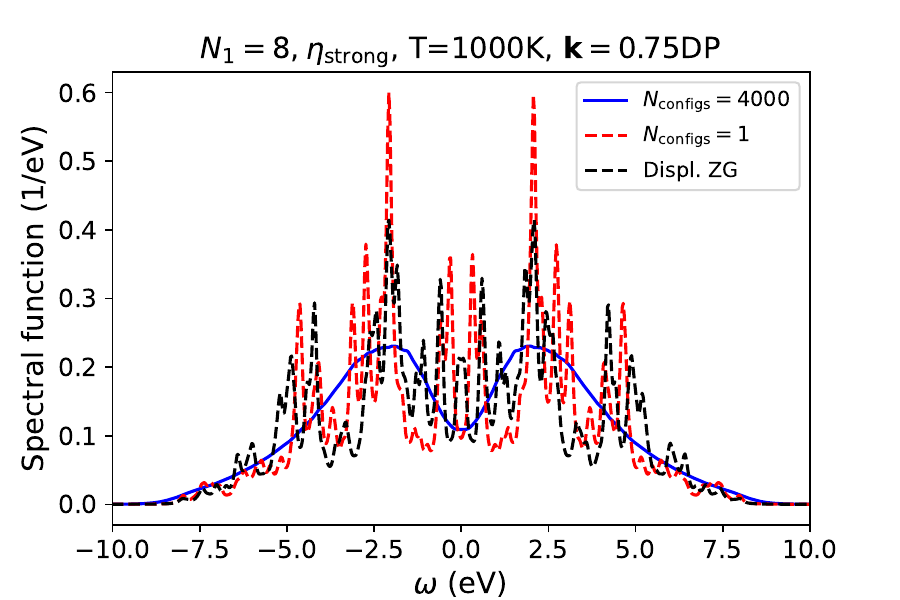}
\includegraphics[width=0.23\textwidth]{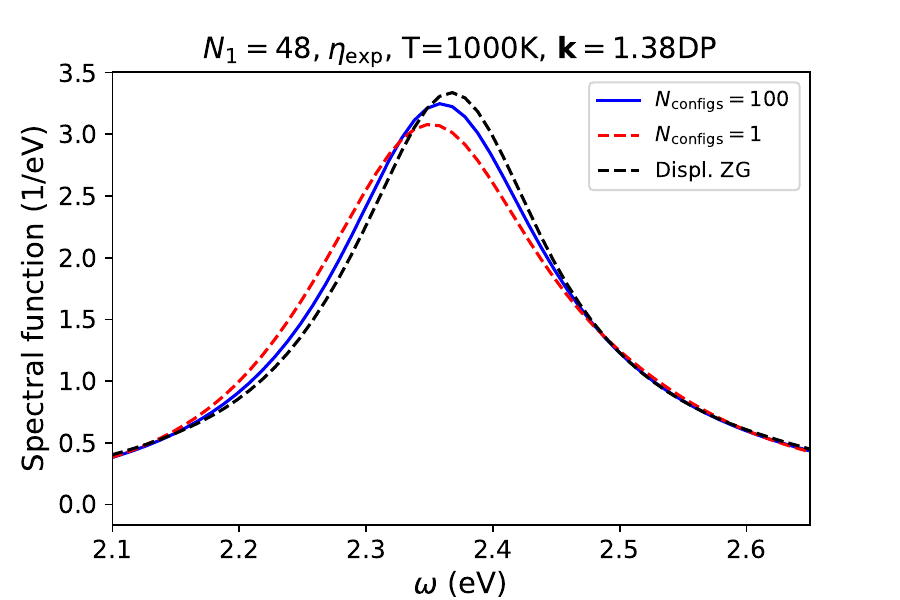}
\includegraphics[width=0.23\textwidth]{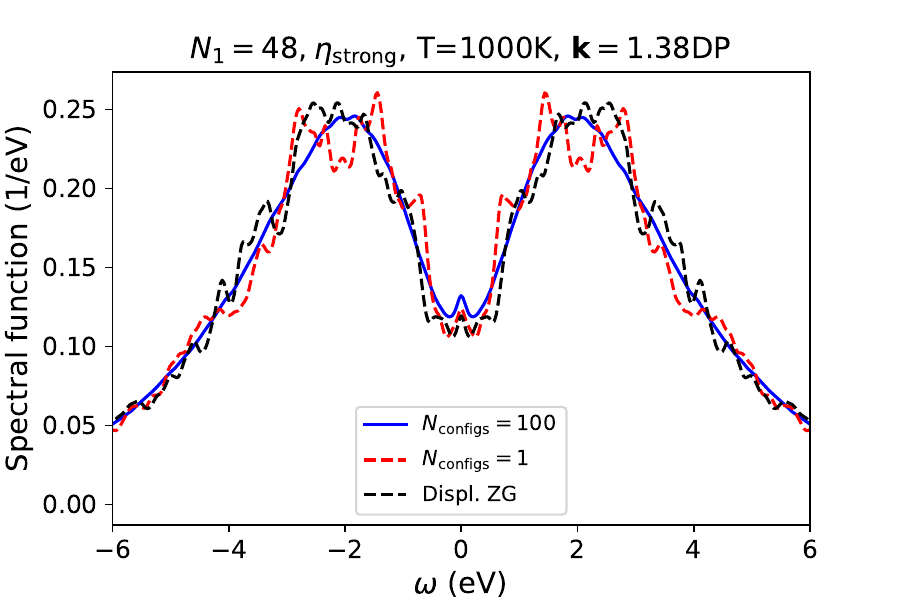}
\caption{Comparison between the displacement of Ref.~\onlinecite{Zacharias2020} and one configuration using the standard Gaussian distribution. The noise of both curves looks similar. The blue line shows the converged spectral function.}
\label{fig:ZG}
\end{figure}

After improving the spectral function for a given $\d$, we also wanted to see if the particular configuration of Ref.~\onlinecite{Zacharias2020} (in principle proposed for quantities that are $\mathbf{k}$ averaged) could be used to get converged results for $\tilde{A}_\mathbf{k}$. We generated the configuration for each SC and temperature using the implementation ZG.x in Quantum Espresso 7.0 \cite{QE2017}.
We considered both a small $N_1=8$ and a large $N_1=48$ SC, and experimental parameters and a strong coupling regime. In all cases, the spectral function is similar to the one obtained with just one standard Gaussian configuration. See Fig.~\ref{fig:ZG}. Thus, the configuration of Ref.~\onlinecite{Zacharias2020} does not seem to improve the convergence of the spectral function. 

\subsection{Results}
\label{sec:results}

Now that we have showed that our method improves the spectral function, we proceed to apply it more generally to the tight-binding model. In this work we use several values of the electron-phonon coupling constant $\eta$: a very weak coupling value $\eta_\textrm{weak}=10^{-4} \eta_\textrm{exp}$, an experimental value $\eta_\textrm{exp}$ (see Sec.~\ref{sec:params}), an intermediate value $\eta_\mathrm{inter}=2.5 \eta_\mathrm{exp}$, and a strong regime value $\eta_\textrm{strong}=5 \eta_\textrm{exp}$. 

\begin{figure*}
\centering
\includegraphics[width=0.45\textwidth]{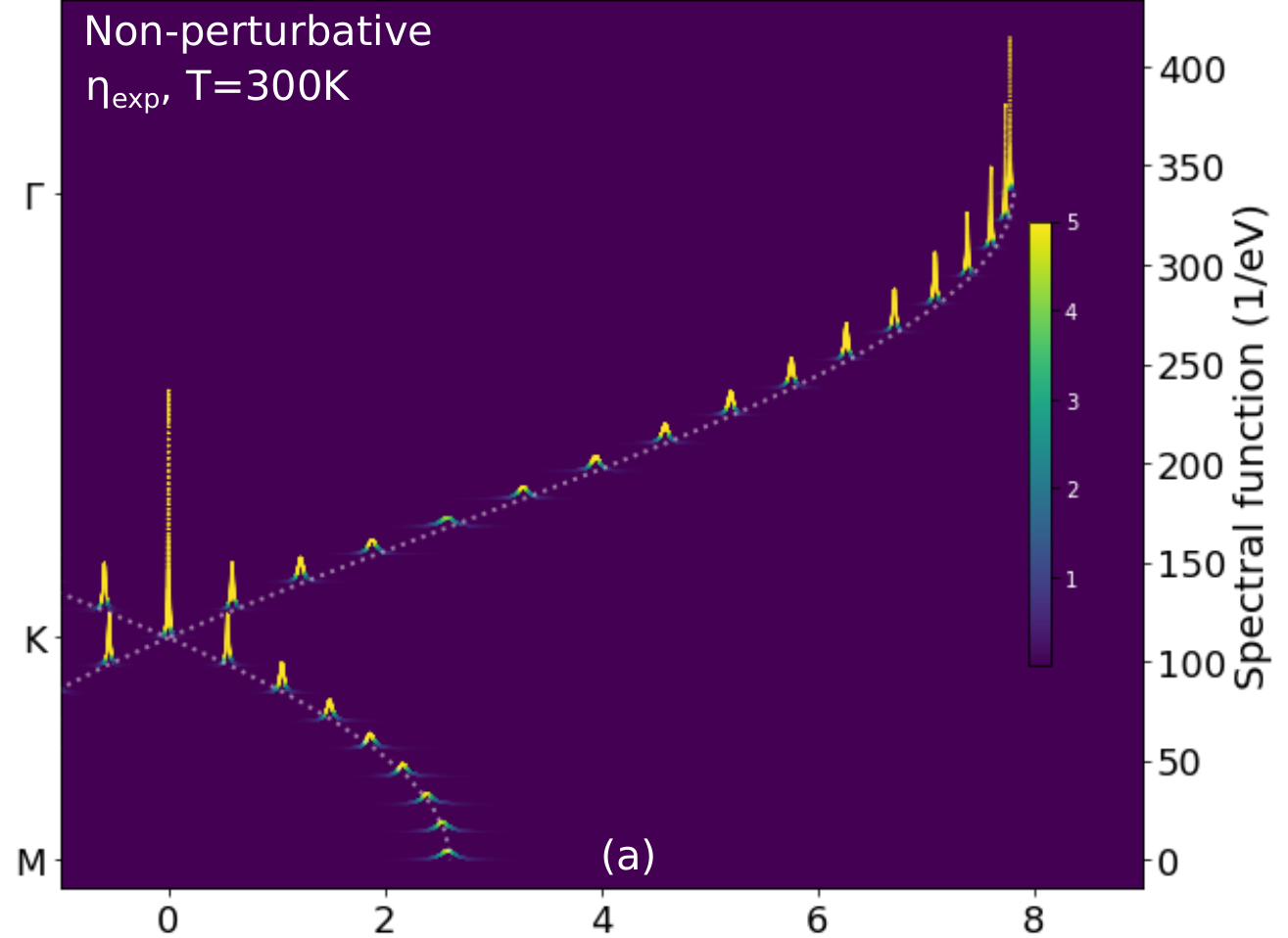}
\includegraphics[width=0.45\textwidth]{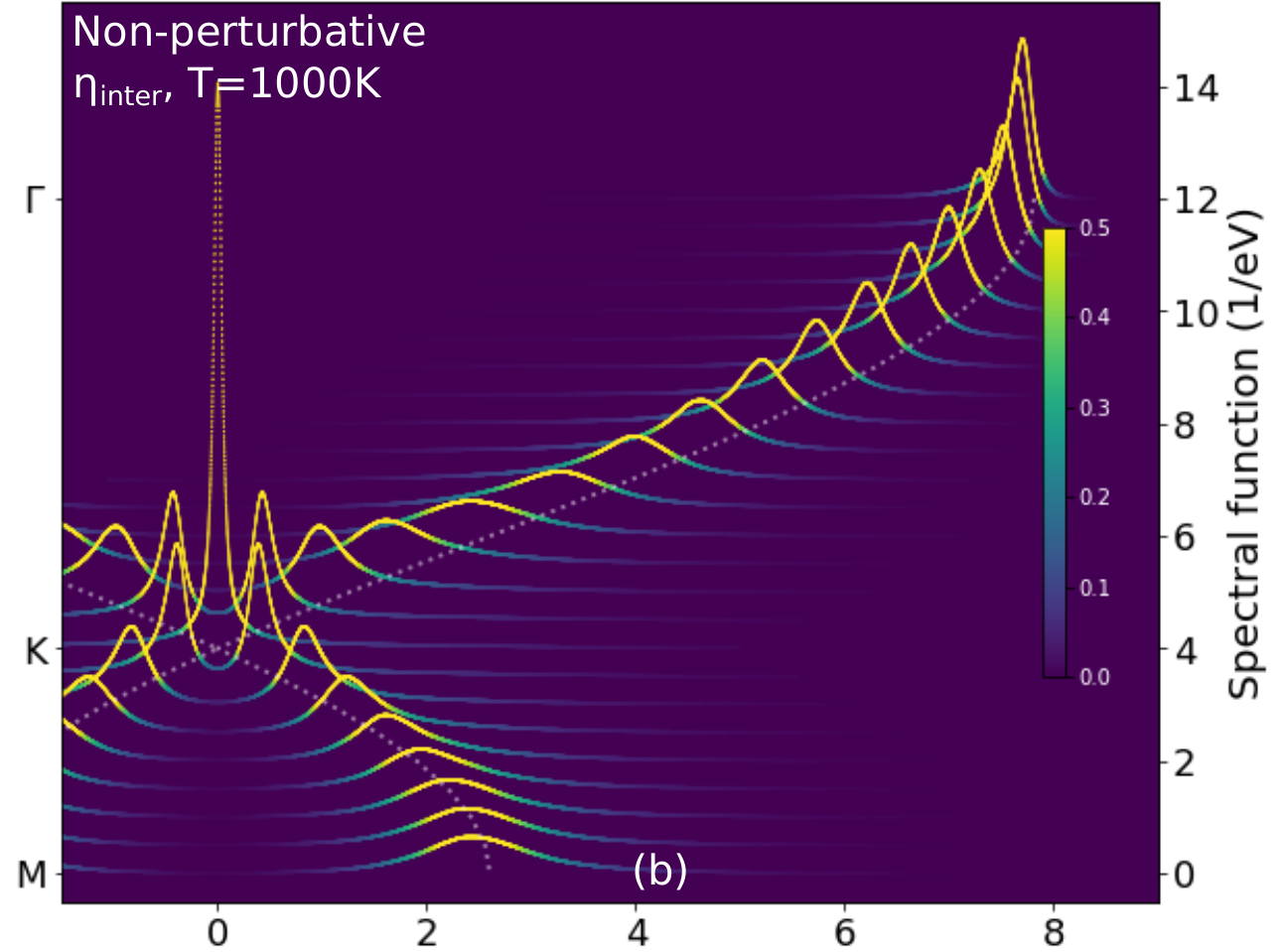}
\includegraphics[width=0.45\textwidth]{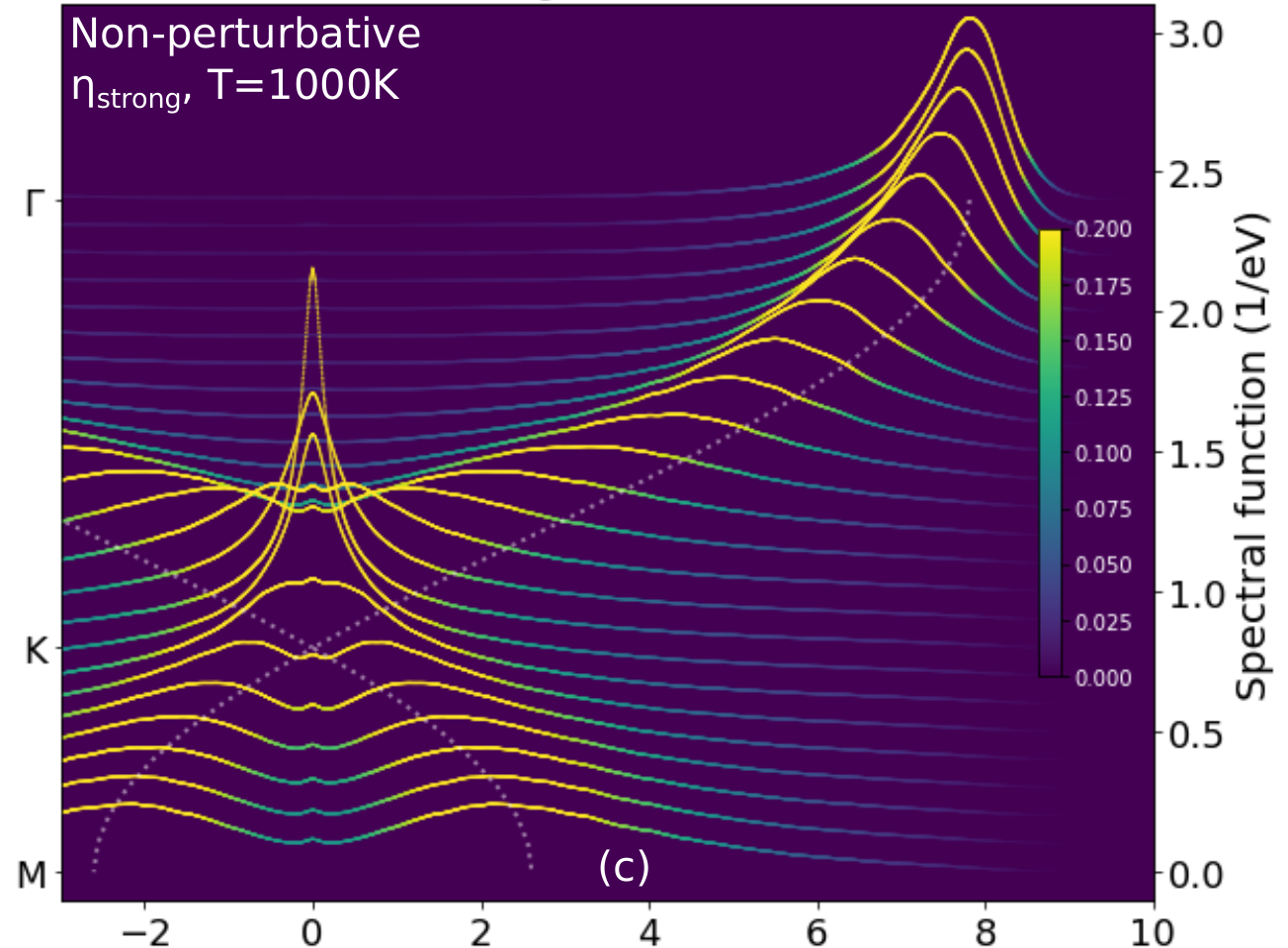}
\includegraphics[width=0.45\textwidth]{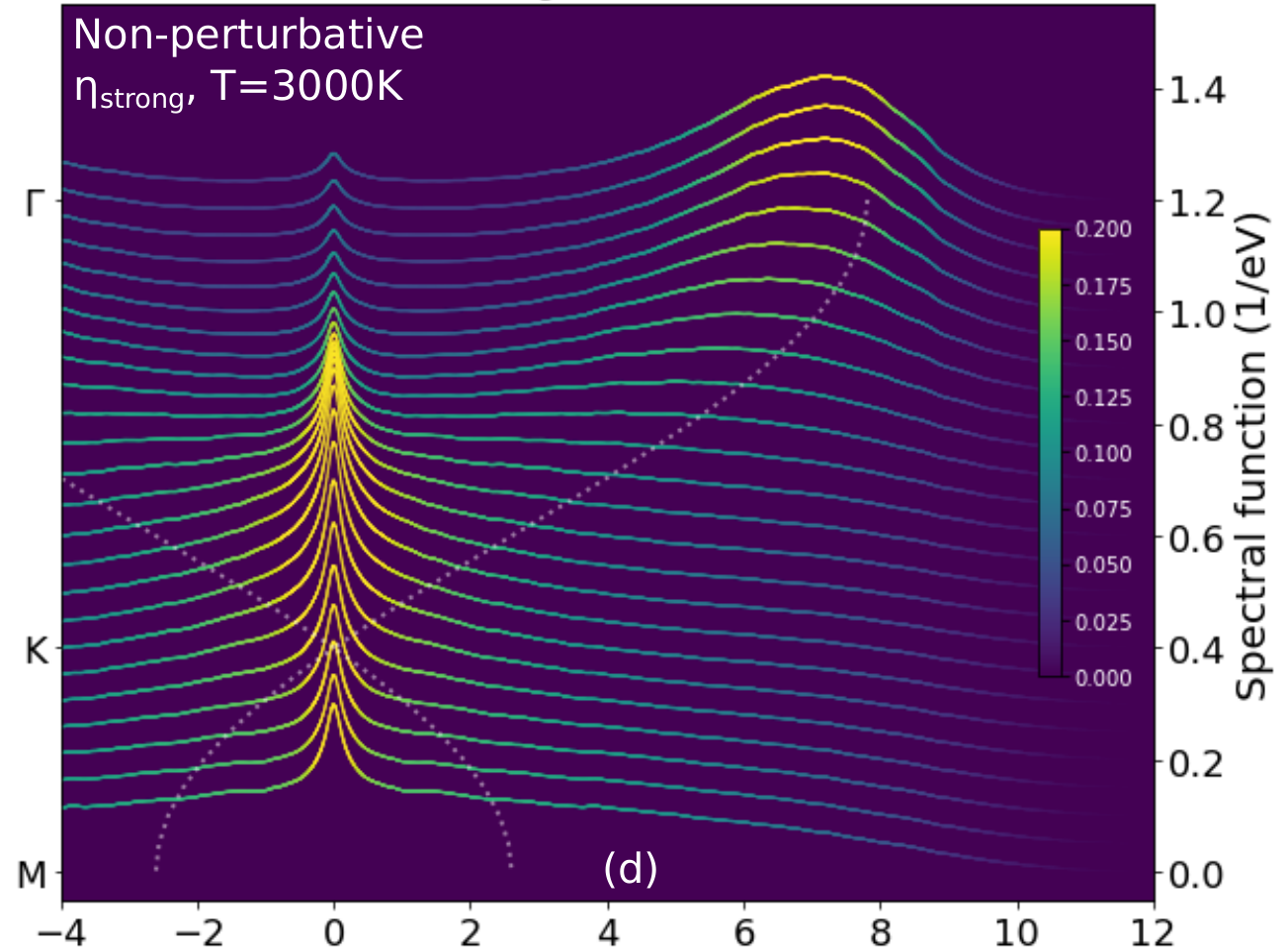}
\includegraphics[width=0.45\textwidth]{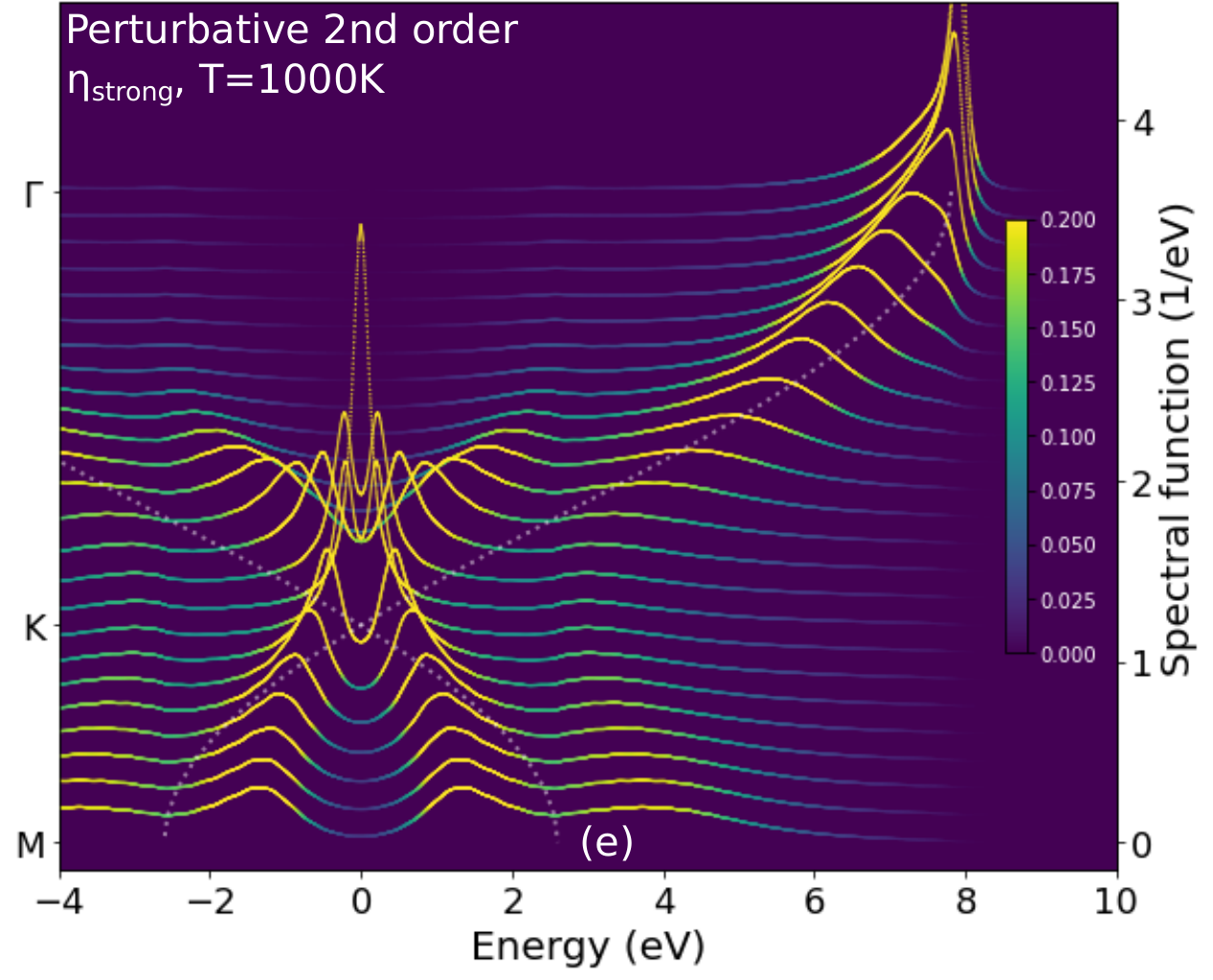}
\includegraphics[width=0.45\textwidth]{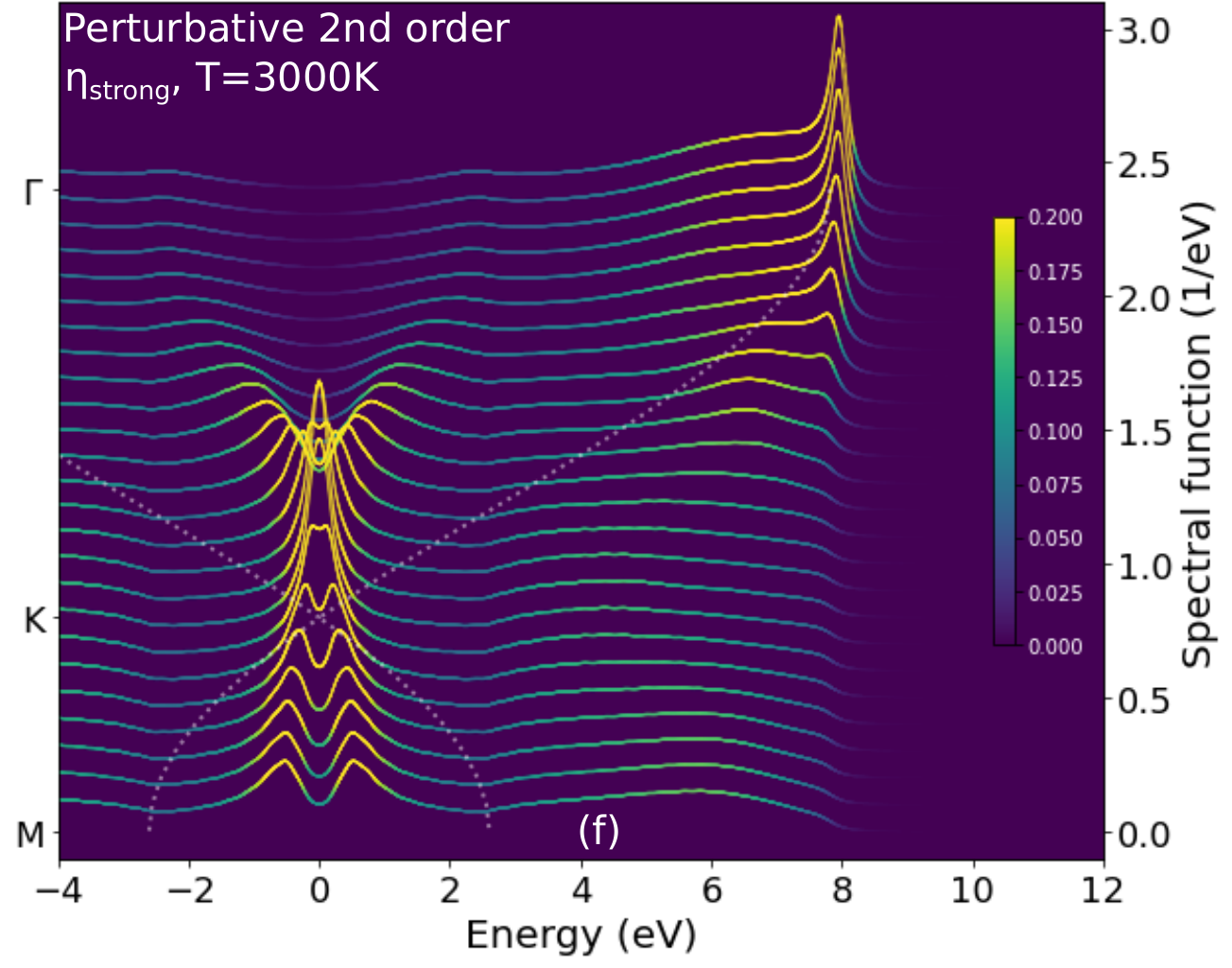}
\caption{Spectral functions $\tilde{A}_\mathbf{k}(\o)$ (scale on the right) as a function of the energy $\o$ for several values of $\mathbf{k}$ along the $\Gamma$-K-M path (left label), with our non-perturbative approach (NP), (a), (b), (c), and (d), and the perturbative approach to second order (P2), (e) and (f). The colorbar saturates below the maximum value to help visualize the peaks. Notice that the scale varies; peaks get broader at higher coupling and temperatures. The dashed line corresponds to $\vare^0_\mathbf{k}$. (a) At low temperatures and coupling, the width is of the order of $\sim$ 0.01-0.1 eV and peaks are very narrow. (b) At higher temperatures and coupling, the shift is larger and peaks become broader. Some overlap can be seen  between positive and negative peaks at the points closest to K. (c) At $\eta_\mathrm{strong}$ and T=1000 K, the bands of $\mathbf{k}$ points next to K merge, forming a single peak. A small peak also appears at $\o=0$ for several $\mathbf{k}$. (d) At $\eta_\mathrm{strong}$ and T=1000 K, even the bands at M merge, and a large peak forms at $\o=0$. (e) and (f) use the parameters of (c) and (d), respectively. There are several differences with the NP spectral functions: 1. There is no peak at $\o=0$. 2. For most states, there is a double peak structure, more easily seen in (e) close to M. So there are four peaks instead of two. 3. Energies do not go beyond the highest bare energy, of about 8 eV, so the weight is artificially restricted to the bare values. 4. There is no merging of peaks at 3000 K. 5. Peaks are not as wide. Thus, P2 completely fails to describe the strong coupling regime. (e) and (f) use $\d=0.2$ eV since noise is present at $\d=0.1$ eV.
}
\label{fig:ARPES_3D}
\end{figure*}

Let us first look at the spectral function in Fig.~\ref{fig:ARPES_3D} (scale on the right vertical axis) along the $\Gamma$-K-M path (left vertical axis), for several values of $\eta$ and temperature. At $\eta_\mathrm{exp}$ and $T=300$ K, the broadening is very small. At $\eta_\mathrm{inter}$ and $T=1000$ K, the peaks are significantly broadened, but qualitatively the spectrum still looks basically the same (except for a little of overlap at the $\mathbf{k}$ point closest to K). At $\eta_\mathrm{strong}$ and $T=1000$ K, two new qualitative differences become clearly visible. First, for points close to K, the negative and positive energy peaks merge together. This can be understood as a result of the energy renormalization lowering the energies close to K (see Fig.~\ref{fig:merge}) and the peaks getting broader at higher temperatures and stronger coupling. Second, a peak appears at $\o=0$, both for $\mathbf{k}$'s in which the main peaks are separated (closer to M), but also for $\mathbf{k}$'s in which the main peaks have merged (closer to the Dirac point, or DP). For $\eta_\mathrm{strong}$ and $T=3000$ K, the peaks of more $\mathbf{k}$'s become merged, and the peak at $\o=0$ becomes visible even at $\Gamma$ (where $\vare^0_\mathbf{k}$ and $-\vare^0_\mathbf{k}$ are the furthest apart).

P2 is not expected to be able to describe these qualitative changes of the features of the spectral function in the strong coupling regime (larger displacements, which occur at higher temperatures, and/or a strong coupling constant), nor to give accurate widths and shifts. Indeed, in Fig.~\ref{fig:ARPES_3D} (e) and (f) we can see the P2 spectral function varies significantly with respect to Fig.~\ref{fig:ARPES_3D} (c) and (d) (which use the same coupling and temperature). Compared to our exact NP approach, peaks do not merge properly, there is not a peak at $\o=0$, there is a double peak structure for each state instead of one, and the weight is artificially restricted to the bare values. Quantitatively, widths and shifts are also incorrect. In general, non-perturbative approaches or higher-order terms should be included in the strong regime.

Let us now look in more detail at the points mentioned in the abstract: (i) Contribution of DW to the change of the Fermi velocity. (ii) Failure of AHC (P2) even at experimental values of the coupling, specially at high temperatures. (iii) Merging of peaks at high temperature/coupling. (iv) Peak at $\o=0$. (v) Asymmetry of the spectral function.

\subsubsection{Contribution of DW to the Fermi velocity}

EPIs change the values of the electronic energies, and in particular, of states close to the Dirac point. Although all terms of the self-energy contribute to the electronic renormalization, the DW term has been frequently ignored\cite{Calandra2007,Park2007}. In the NP approach, all derivatives with respect to the ionic displacements are automatically taken into account, which means that DW (and higher order terms) are included as well. Although to lowest order it does not affect the linewidth or lifetime, since it is real, it does affect the Fermi velocity (it is $\mathbf{k}$ dependent). At the experimental values (blue curve in (a)), the Fermi velocity changes by about 4\%, and DW accounts for 20\% of this effect. (By Fan contribution, we mean the change given by the linearized version of Eq.\eqref{eq:hopping}, Eq.~\eqref{eq:t_linear}. By DW, we mean the rest of the contribution. In the weak coupling limit, they coincide with $\Sigma^\mathrm{Fan}$ and $\Sigma^\mathrm{DW}$, respectively.) See Table~\ref{tab:vF}. At $T = 3000$ K, the Fermi velocity changes by about 20\% and the contribution of the DW term is about 40\%. A first-principles calculation should be carried out to determine these values more accurately, but our results indicate that the DW cannot be neglected and plays a significant role at high temperatures.

\begin{table}[]
\begin{tabular}{cccc}
\hline
             & 300K & 1000K & 3000K \\ \hline
$\Delta v_F/v_F$ &  0.04     &   0.07     &    0.22    \\
DW   &  0.21     &  0.30      &   0.41     \\
$\Sigma_1^\mathrm{P2}/\Sigma_1^\mathrm{NP}$  &  0.98     &  0.94      &   0.83     \\ \hline
\end{tabular}
\caption{Change of the Fermi velocity $\Delta v_F$ relative to the bare value $v_F$, fraction of the DW contribution to $\Delta v_F$, and ratio of the P2 to the NP self-energy for different temperatures (which agrees very well with $\Delta v_F^\mathrm{P2}/\Delta v_F^\mathrm{NP}$). Results were obtained at $\mathbf{k}=0.97$ DP, with $\vare_0 = 0.3$ eV. At room temperature, of the 4\% change of $v_F$, 21\% corresponds to the DW term, and grows to 41\% at 3000 K. So the DW term, although normally neglected, should be included. The third line also shows that P2 agrees with NP at room temperature, but differs significantly at high temperatures. For $\mathbf{k}$ points further away from DP, P2 differs much more from NP (see Fig.~\ref{fig:P_vs_NP}).}
\label{tab:vF}
\end{table}

\subsubsection{Failure of AHC at $\eta_\mathrm{exp}$}

\begin{figure*}[ht!]
\centering
\subfigure[]{\includegraphics[width=0.4\textwidth]{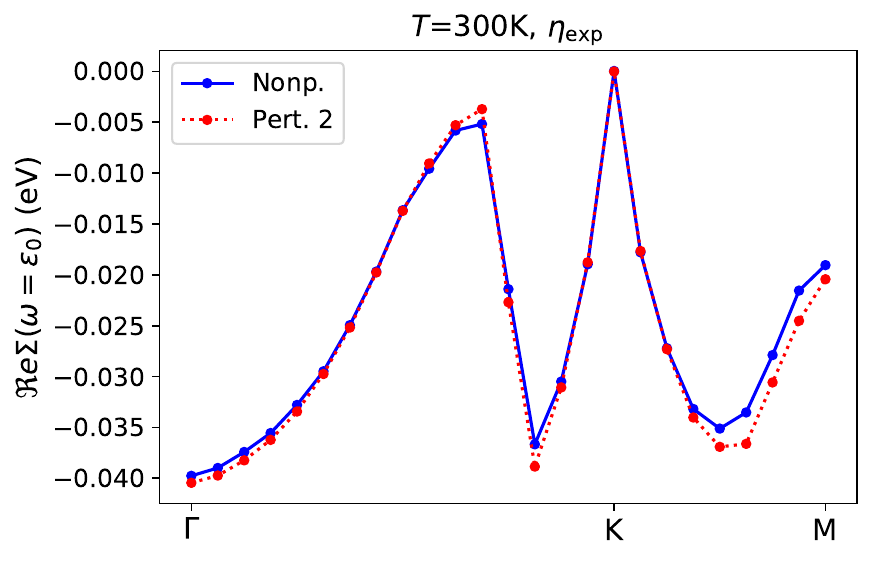}}
\subfigure[]{\includegraphics[width=0.4\textwidth]{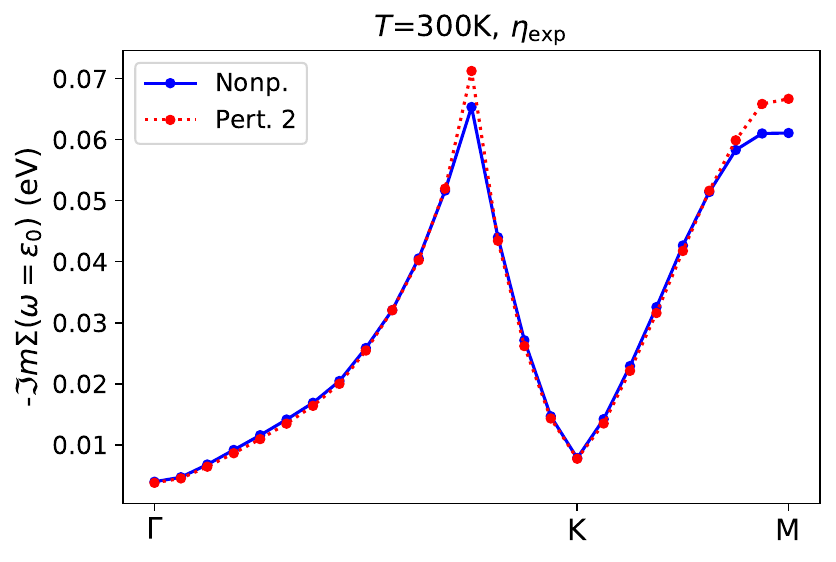}}
\subfigure[]{\includegraphics[width=0.4\textwidth]{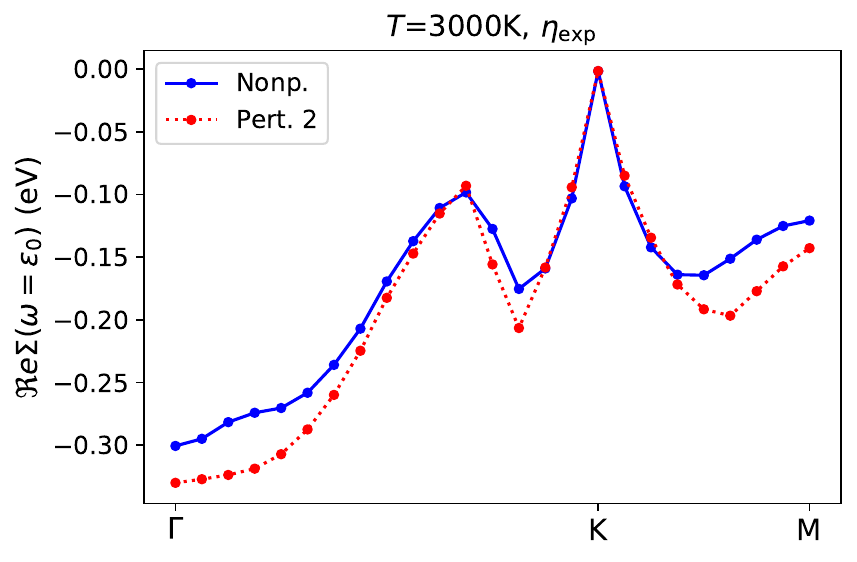}}
\subfigure[]{\includegraphics[width=0.4\textwidth]{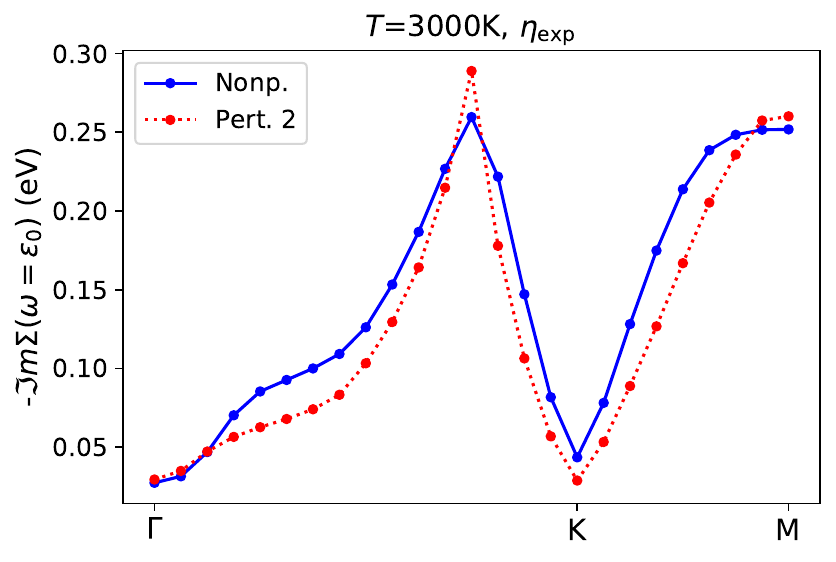}}
\caption{Comparison between the P2 and NP self-energy at $\eta_\mathrm{exp}$ (the lines are a guide to the eye) and $\d=0.2$ eV. Close to DP at $T=300$ K, they agree. But for some other $\mathbf{k}$ points there are visible differences. This is likely the case in other materials as well. In fact at high temperatures (commonly included in works that calculate $\Sigma$ to second order\cite{Ponce2015}), $\Sigma^\mathrm{P2}$ is far from $\Sigma^\mathrm{NP}$ for many $\mathbf{k}$ points. We show the negative part of the imaginary part since it corresponds to the linewidth.}
\label{fig:P_vs_NP}
\end{figure*}

We already saw that in the strong coupling regime (Fig.~\ref{fig:ARPES_3D} (c), (d) vs (e) and (f)), P2 can vary drastically from NP. Here we focus on $\eta_\mathrm{exp}$, where differences between P2 and NP are also visible. At higher temperatures, the agreement between P2 and NP to obtain $\Delta v_F$ gets worse. See third line of Table~\ref{tab:vF},  $\Delta v_F^\mathrm{P2}/\Delta v_F^\mathrm{NP} \simeq \Sigma_1^\mathrm{P2}/\Sigma_1^\mathrm{NP}$ (at $\eta_\mathrm{exp}$, the usual approximation $\Delta \vare = \Sigma_1$ works well). More in general, Fig.~\ref{fig:P_vs_NP} shows that at room temperature there are visible differences between P2 and NP. This means graphene is not in the weak coupling limit, but rather in what we can call an intermediate regime, in which higher order terms start to become relevant. This is significant, since it shows that the assumption of weak coupling might be ill-justified in many materials. It is hard to know a priori for which states P2 works well, and in which cases higher-orders terms are relevant. Commonly, P2 (AHC) is used because it is essentially the only available option, rather than because it has been shown that higher order terms are negligible. The difference becomes larger at higher temperatures.

\begin{figure}
\centering
\subfigure[]{\includegraphics[width=0.4\textwidth]{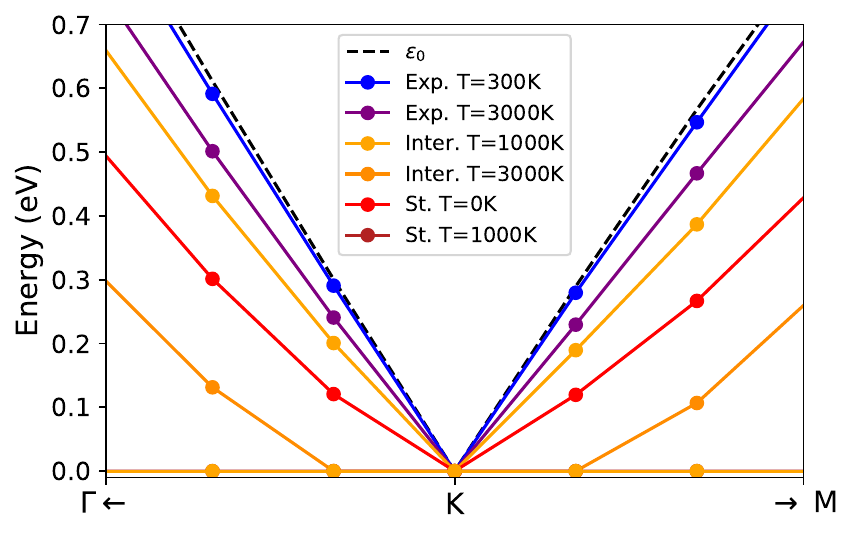}}
\subfigure[]{\includegraphics[width=0.4\textwidth]{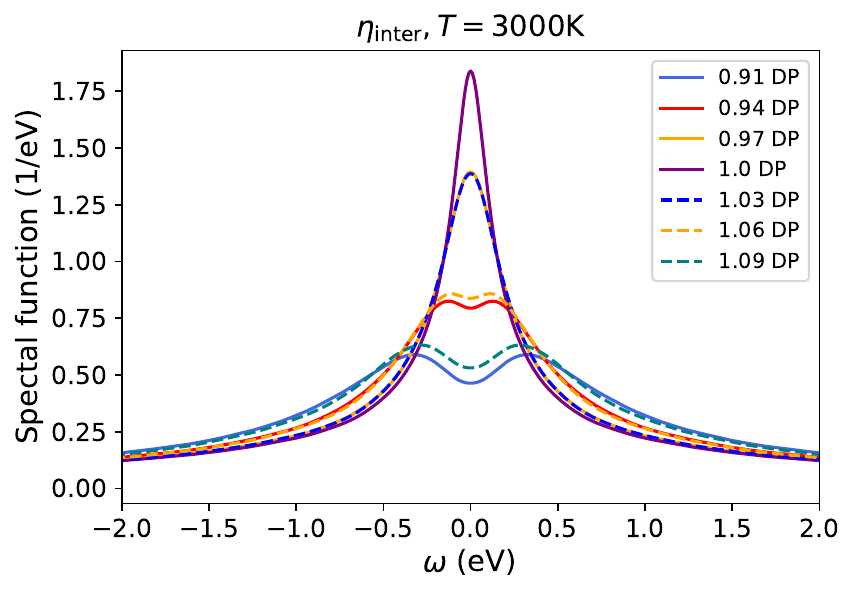}}
\vspace{-2mm}
\caption{Renormalization of the Fermi velocity and merging of bands close to the Dirac point.
(a) Energy of the QP peak maximum for several $\mathbf{k}$ points close to K. At $\eta_\mathrm{exp}$, the renormalization of $v_F$ increases with temperature (see Table~\ref{tab:vF} for more details). Lines connecting the calculated values are a guide to the eye.
At higher coupling and temperatures, the peak of the spectral function (the QP energy) of $\pi^\ast$ shifts to even lower energies ($\pi$ to higher energies). The peaks also get broader, and close enough to K, the peaks of of $\pi$ and $\pi^\ast$ merge, leaving only a maximum at $\o=0$. (b) This can be observed at $\mathbf{k}=0.97$ DP (between $\Gamma$ and DP) and $\mathbf{k}=1.03$ DP (between DP and M=1.5 DP). Slightly further away, at $\mathbf{k}=0.94$ DP and $\mathbf{k}=1.06$ DP, there are two distinct peaks, but they also merge substantially.}
\label{fig:merge}
\end{figure}

In non-polar systems like graphene, the matrix element in Eq.~\eqref{eq:Fan}, $g^{\mathbf{q}\nu}_{\mathbf{k},nn'}$ goes as $q^0$ for small $\mathbf{q}$, so the imaginary part of the self-energy at $T=0$ is proportional to $\int d^2q \d(\o - \vare_{\mathbf{k+q}n})$, which is proportional to the density of states. In 2D, the density of states (DOS) diverges logarithmically at a saddle point, which is the case of the M point in graphene. So when evaluating the imaginary part of the self-energy at the bare energy $\o=\vare_M$ to determine its width, it diverges. This is also reflected in the point between $\Gamma$ and K which has the same energy as M. In the non-adiabatic expression of the self-energy, there are additional phonon frequencies in the denominator, and the imaginary part will still be ill-defined for several values of $\o$. Thus, the spectral function at M is ill-defined using perturbation theory and is hard to converge close to M. On the other hand, NP is well defined, and the curves of Fig.~\ref{fig:P_vs_NP} are also smooth for lower values of $\d$, while P2 starts to show some noise.

\subsubsection{Merging of peaks}

Another effect that NP helps to describe is the merging of peaks at sufficiently high values of the coupling or temperature. See Fig.~\ref{fig:merge}. In (a), we can see how the peak of the spectral function gets closer to $\o=0$. When the peak is at 0 and $\mathbf{k}$ is not K, it means that the peaks have fully merged. For other $\mathbf{k}$ close to K, there are two maxima (see (b)), but the peaks still have significant overlap, and the full-width-half-maximum (FWHM) is not well defined (the value of the spectral function at $\o=0$ does not reach half of the maximum).
At stronger coupling and higher temperatures, peaks merge further away from K, as in Fig.~\ref{fig:ARPES_3D}(d). The fact that bands close to K get closer to the Fermi level is similar to what happens in semiconductors, in which the band renormalization due to phonons usually reduces the gap, and bands merging is analogous to the gap closure in solid hydrogen at high-pressures, attributed to strong electron-phonon coupling resulting from large quantum fluctuations of hydrogen\cite{Monacelli2020}. Hydrides more in general are also expected to have a large electron-phonon coupling $\lambda$, such as LaH$_{10}$ with $\lambda=3.6$ at 129 GPa \cite{Errea2020}, and NP effects might be relevant as well.

\subsubsection{Peak at $\o=0$}

\begin{figure}
\centering
\includegraphics[width=0.4\textwidth]{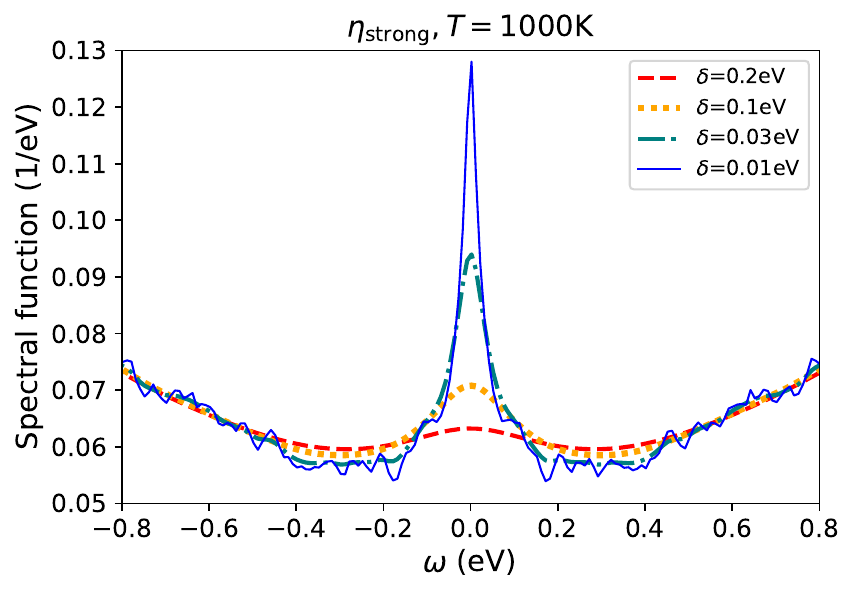}
\vspace{-2mm}
\caption{Peak in the spectral function at $\o=0$, for the state $\mathbf{k}=0.63$ DP. The peak appears in addition to the two other maxima at positive and negative energy (see Fig.~\ref{fig:ARPES_3D}(c)).
The width is similar to the value of $\d$ being used, so the actual width is less than 0.01 eV. This is consistent with the peak in the DOS of Ref.~\onlinecite{Zhu2016} and is related to a localization transition in the presence of off-diagonal disorder and chiral symmetry\cite{Ziegler2008,Zhu2016}.}
\label{fig:peak}
\end{figure}

An unexpected effect of the strong coupling regime is a peak at $\o=0$, which appears in addition to the positive and negative peaks. It can be observed in more detail for $\mathbf{k}=0.63$ DP (with DP the Dirac point, instead of K, to avoid confusion with Kelvin) in Fig.~\ref{fig:peak}. 
The peak gets narrower and higher for smaller values of $\d$ (the noise for the lower values of $\d$ can be eliminated by averaging over more configurations). This means that the width of the peak is smaller than $\d$. The same $\o=0$ peak can be observed for other $\mathbf{k}$ points, including points as in Fig.~\ref{fig:ARPES_3D}(d) where the positive and negative peaks merge together. 

Since the density of states (DOS) in the interacting case can be obtained by averaging the spectral function over all $\mathbf{k}$ points in the PBZ, a peak will be also present in the DOS. This is consistent with the results of Ref.~\onlinecite{Zhu2016}, in which disorder is considered via a random Gaussian hopping parameter. The authors also obtain that the peak becomes narrower with decreasing smearing parameter, and suggest that the DOS diverges in the thermodynamic limit. Interestingly, they observe that the peak disappears with a weak on-site disorder, so the chiral symmetry plays a crucial role in the divergence. Ref.~\onlinecite{Ziegler2008} points out that the DOS has a power law close to the Dirac point, and that a minimal (non-zero) conductivity is observed in graphene. If this still holds above the critical disorder, it would imply a transition from ballistic transport to localization. Ref.~\onlinecite{Zhu2016} follows the same argument, but observes a non-zero DOS below the critical disorder, and suggests a transition from diffusive transport to an insulating phase. It appears like the divergence at $\o=0$ is still not well understood\cite{Zhu2016}. In the context of Landau level broadening and the quantum Hall effect, a divergence is also observed when considering the so-called off-diagonal disorder (disorder in the hopping or interatomic coupling constant, involving different atoms) in the presence of a magnetic field. See Ref.~\onlinecite{Malla2019} and references therein.

\subsubsection{Asymmetry of the spectral function}

Another effect of the strong coupling regime is an increase of the asymmetry of the peak of the spectral function. In order to characterize it, we make use of the following line-shape\cite{Stancik2008},

\be
\begin{split}
A(\o) & = \f{A_0}{\pi} \f{\Gamma(\o)}{(\o-\o_0)^2 + \Gamma(\o)^2}\\
 \textrm{with} \hspace{2mm} & \Gamma(\o)  = \f{2\Gamma_0}{1+ e^{a(\o-\o_0)}}.
\end{split}
\label{eq:A_as}
\ee

\begin{figure}
\centering
\includegraphics[width=0.4	\textwidth]{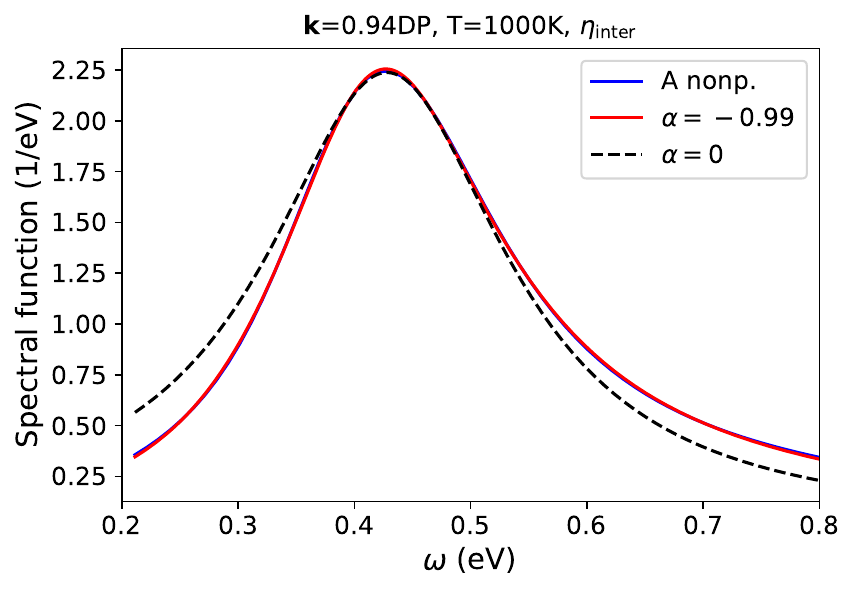}
\vspace{-2mm}
\caption{Illustration of how the spectral function looks asymmetric at higher temperatures or coupling, for a particular $\mathbf{k}$ point. The spectral function calculated with our method (blue) is very well fitted with Eq.~\ref{eq:A_as} (red) with $\a=-0.99$. The dashed line corresponds to a Lorentzian lineshape ($\a=0$), to better visualize the asymmetry.}
\label{fig:asymmetry_part}
\end{figure}

\ni That is, a Lorentzian with a width that varies sigmoidally. We can define $\alpha = a \Gamma_0$ as the dimensionless parameter the quantifies the asymmetry of a spectral function (see Sec.~\ref{sec:asymmetry}). In Fig.~\ref{fig:asymmetry_part}, we show the spectral function (blue) for a particular $\mathbf{k}$ point close to the Dirac point, together with the fit using Eq.~\ref{eq:A_as} (red). The agreement is very good, and the asymmetry of the peak can be visualized by plotting also a Lorentzian curve (dashed-black). In the Appendix, we can also see that the asymmetry increases with temperature (Fig.~\ref{fig:alpha_T_and_eta}(a)) and coupling (Fig.~\ref{fig:alpha_T_and_eta}(b)) for all the considered $\mathbf{k}$'s. This might be a general feature of crystals in the strong coupling regime, related to the asymmetry of bands above and below the energy of given state.

\section{Conclusions}

We have developed a non-perturbative (NP) method to determine the electronic Green's function in presence of electron-phonon interactions, and rigorously proved how it relates to the standard Feynman diagrammatic approach to all orders. Such a proof to all orders and involving both the real and imaginary part of the self-energy 
was missing in the literature. The diagrams coincide exactly with those of the standard theory, without any additional spurious terms. In particular, the method reproduces AHC to second order in the adiabatic limit. The diagrams include most of the diagrams of the standard theory, except for diagrams containing bubbles, which are unimportant when using fixed harmonic phonons fitted to experimental data. In addition, we showed that averaging over an ensemble is necessary to get the pairing of legs through Wick's theorem and recover the usual Feynman diagrams, and not just a procedure to get a smooth spectral function. Increasing the number of configurations improves the calculation until a certain point, when convergence to the Feynman diagram is achieved. The smoothness of the spectral function still depends on choosing $\d$ that is large enough for a given SC. The smallest value of $\d$ for a given SC that gives a smooth spectral function can be larger than the linewidth of the state under consideration, unless the SC is very large. To solve this issue, by using the self-energy, we presented an optimized method with a reduced error in the broadening parameter. Basically, the method removes the artificial broadening introduced in the unperturbed spectral function (which should be a delta function), and keeps the broadening contained in the self-energy. In this way, the broadening parameter does not need to be extremely small, and does not require very large SCs.

We then applied the method to a graphene tight-binding model, and observed that AHC (or P2 in this work) completely fails in the strong coupling regime. We obtained several interesting results, most of them related to the strong coupling regime. (i) First of all, in the tight-binding model, the role of the DW term in the change of the Fermi velocity (lineshift close to the Dirac point) is comparable to the Fan contribution, as is usually the case. This holds at room temperature, where P2 is accurate, and at higher temperatures, where P2 and NP start to differ. (ii) At $\eta_\mathrm{exp}$, for several $\mathbf{k}$ points further away from the Dirac point, differences in the self-energy are visible at room temperature, and significant at higher temperatures. (iii) As the coupling and temperature increase, positive and negative energy peaks of the spectral function merge for $\mathbf{k}$ close enough to the Dirac point. In the strong coupling regime, the lineshift and linewidth are so large, that peaks merge for a wide range of $\mathbf{k}$, including points far away from the Dirac point. (iv) In the strong regime, a sharp peak becomes visible at $\o=0$, resulting in spectral functions with 3 peaks, instead of the standard 2 peaks associated to the $\pi$ and $\pi^\ast$ bands. Previous works on disordered graphene identify this peak with a diffusion or ballistic transport to localization transition. (v) With increasing temperature and coupling, the peaks become more asymmetric. This could be a generic feature of strongly coupled systems. High resolution ARPES might be needed to identify this effect.

By putting our approach on solid grounds (and comparing to the exact conical model in Sec.~\ref{sec:conical}), it becomes more clear under which conditions non-perturbative methods can be used. The method should give good results in non-polar semiconductors and insulators, and in metals and semi-metals (as we explicitly showed here for graphene) not close to the Fermi level (where the Fermi-Dirac factors are relevant). Another interesting approach to study the electron-phonon interaction that goes beyond AHC and which has gained traction in the last years, is the cumulant method\cite{Story2014,Nery2018}. It is non-adiabatic and it includes higher order terms by basically putting the AHC self-energy in an exponential. The disadvantage is that higher-order terms are included in an approximate way. On the other hand, non-perturbative methods like the one described here are adiabatic, but include all higher-order terms exactly. Combining both methods could lead to very precise results, and in fact, adiabatic and non-adiabatic methods have already been used conjunctly in polar materials with Fr\"ohlich-type models\cite{Nery2016,Zacharias2020}.

An analysis like the one we performed here can probably be applied to the dielectric function or other correlation functions, in order to establish a clear connection with non-adiabatic methods. In future work, we plan to apply our method to systems in the strong coupling regime, like high-pressure hydrogen\cite{Monacelli2020,Gorelov2020}, using first principles calculations. 

As a final remark, the fact that for some states the self-energy differs appreciably between the perturbative and non-perturbative approaches, is a salient result of this work. It seems hard to know a priori for which states lowest order perturbation theory might work, and in which cases higher-order terms are relevant. So it is likely that the accurate determination of the electronic properties of many other materials, specially above room temperature, also requires going beyond second order perturbation theory.

\section*{Acknowledgements}
JPN thanks Matthieu Verstraete, Jos\'e Lorenzana and Sergio Ciuchi for useful discussions. This work has received funding from the European Union's Horizon 2020 research and innovation programme under grant agreement 881603, and from the MORE-TEM ERC-SYN project, grant agreement 951215.

\appendix
\section*{Appendix}

\renewcommand\thefigure{\thesection.\arabic{figure}}    
\setcounter{figure}{0}  
\renewcommand{\thefigure}{A\arabic{figure}}

\setcounter{equation}{0}  
\renewcommand{\theequation}{A\arabic{equation}}

\renewcommand{\thesubsection}{A\arabic{subsection}}

\renewcommand{\appendixname}{}

\subsection{Calculation details}
\label{sec:params}

\begin{figure}[!h]
\centering
\includegraphics[width=0.45\textwidth]{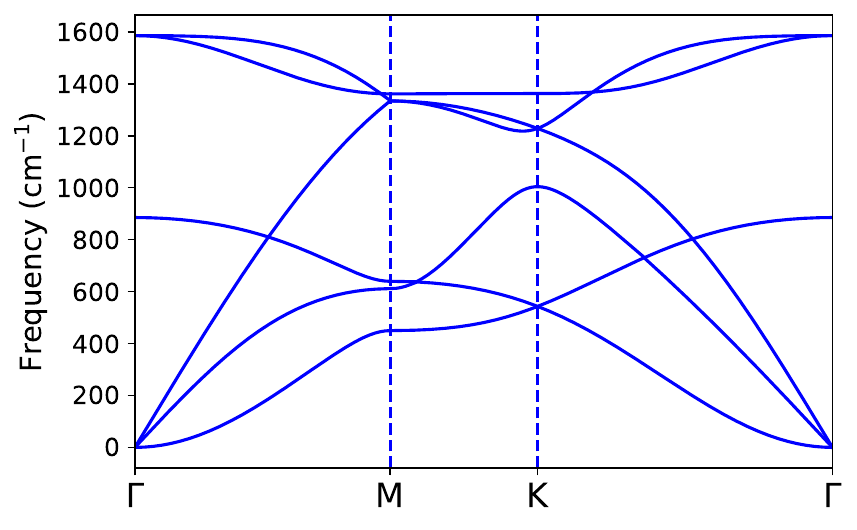}
\caption{Phonon dispersion}
\label{fig:phonons}
\end{figure}

The phonons used to generate the ensemble of random configurations were obtained using the interatomic potential of Ref.~\onlinecite{Libbi2020}, which is fitted to DFT-PBE calculations. For completeness, the phonon dispersion is shown in Fig.~\ref{fig:phonons}. The lattice parameter is $a=2.467$ \AA. Displacements are considered both in and out of plane.

In the tight-binding model, the hopping constant $t_0$ is picked so that the slope around the Dirac point is $\beta = 5.52$ eV \AA as in Ref.~\onlinecite{Calandra2007}, so $\sqrt{3}/2 a t_0 = \beta$. We also use $\eta_\mathrm{exp} = 4.42$ eV$/$\AA, obtained from $\langle g^2_\Gamma \rangle = \hbar/(2 M \o_\Gamma) 9/4 \eta^2$ (see Ref.~\onlinecite{Venezuela2011}, Sec.~II C) and $\langle g^2_\Gamma \rangle = 0.0405$ eV$^2$ (see Ref.~\onlinecite{Calandra2007} after Eq.~(6)).

 Recall from the main section that if $\mathbf{k}$ is not in $\mathcal{Q}_\mathrm{SC}$, then the grids have to be shifted by $\mathbf{K}$, with $\mathbf{K}$ the only wavevector in the SBZ such that $\mathbf{k}=\mathbf{K}+\mathbf{q_0}$ for some $\mathbf{q_0}$ in $\mathcal{Q}_\mathrm{SC}$. To determine the Green's function, for each configuration, $\mathcal{H}^I$ has to be diagonalized to get the eigenstates $|J\rangle^I$ needed in Eq.~\eqref{eq:Gnum}. Then the inner product can be done for any $\mathbf{k}=\mathbf{K}+\mathbf{q}$ (using Eq.~\eqref{eq:innerproduct} in our model). So by diagonalizing an ensemble of Hamiltonians $\mathcal{H}^I$, the Green's function can be determined in a set of points $\{\mathbf{k}=\mathbf{K}+\mathbf{q},\mathbf{q} \hspace{1mm} \mathrm{in} \hspace{1mm} \mathcal{Q}_\mathrm{SC} \}$ without diagonalizaing additional Hamiltonians.
 
Calculations are done in a $48 \times 48 \times 1$ supercell, with a smearing parameter $\d=0.1$ eV, and 100 configurations, unless otherwise stated. Also $N_1=N_2$ in all calculations, so the supercells are $N_1 \times N_1 \times 1$ and we only specify $N_1$.  Using $\mathbf{K}=\Gamma$ and commensurate points, this gives 25 $\mathbf{k}$ points along $\Gamma$-DP-M (i.e. $\Gamma$-K-M), as in Fig.~\ref{fig:ARPES_3D}. The closest point to DP has a bare energy of 0.6 eV. To determine the change of the Fermi velocity, the calculation is done  at
$\mathbf{K}$ = DP/32, to get the energy at $\mathbf{k} =$ 31/32 DP = 0.97 DP,
 with $\vare_0=0.3$ eV ($\mathbf{K} =$ 0.97 DP can actually be directly used as well).

\subsection{Tight binding model}
\label{sec:TB}

Calculations in the SC in the tight-binding case are a little bit subtle, because the basis depends on the positions of the atoms. In the SC, we use $|S\rangle^I = |ls\rangle^I$ as a basis, and in the PC we consider $|\mathbf{k}s\rangle^I = \f{1}{\sqrt{N}} \sum_{l} e^{i\mathbf{k}\cdot(\mathbf{R}_l + \pmb{\tau}_s)} |ls\rangle^I$. Thus, in order to work out the inner product of Eq.~\eqref{eq:Gnum}, we need to determine $^I\langle S | \mathbf{k}s'\rangle^I$,

\be
\begin{split}
^I\langle S | \mathbf{k}s'\rangle^I & = \f{1}{\sqrt{N}}\sum_{l'} e^{i \mathbf{k} \cdot (\mathbf{R}_{l'}+\pmb{\tau}_{s'})} {}^I\langle ls | l's'\rangle ^I \\
& = \f{1}{\sqrt{N}} e^{i \mathbf{k} \cdot (\mathbf{R}_{l}+\pmb{\tau}_{s})}\d_{ss'}
\end{split}
\label{eq:innerproduct}
\ee

\ni In the distorted SC, the localized basis of orbitals is naturally shifted. Consequently, the PC basis has to be shifted as well (considering it has to be defined in the SC Hilbert space for the inner product to be well defined), so there is actually no dependence on the basis in the matrix element of Eq.~\eqref{eq:Gnum}. So, the information on how the interaction changes is contained in the hopping parameter, and the proof of the main section holds. What really matters in the model is how the distance between atoms changes in each configuration and the hexagonal topology (which atoms interact with each other); where the basis is centered plays no role. For incommensurate $\mathbf{k=K+q_0}$, $e^{i \mathbf{K} \cdot \pmb{\tau}_S} |S\rangle ^I$ is the SC basis and the result of the inner product (second line of Eq.~\eqref{eq:innerproduct}) has $\mathbf{q_0}$ instead of $\mathbf{k}$.

\subsection{Self-energy}

\begin{figure}
\includegraphics[width=0.22\textwidth]{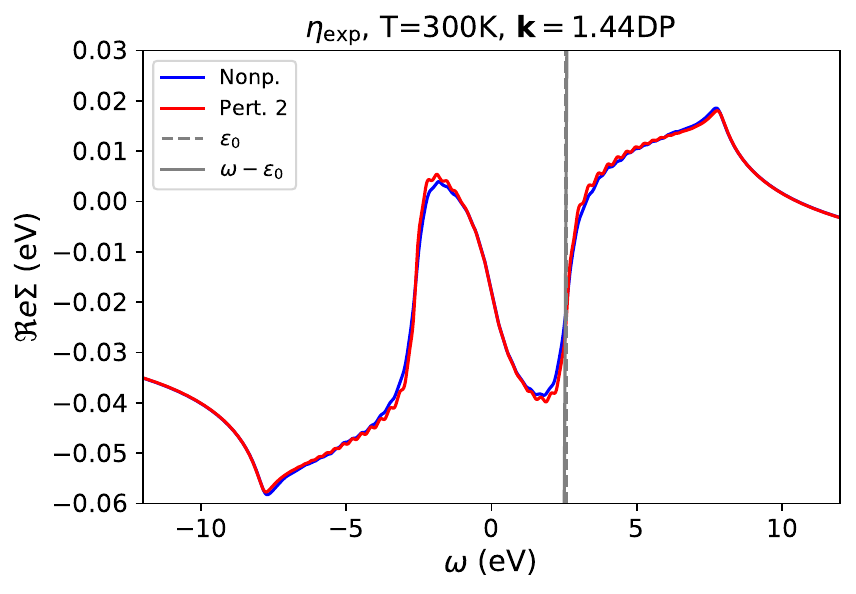}
\includegraphics[width=0.22\textwidth]{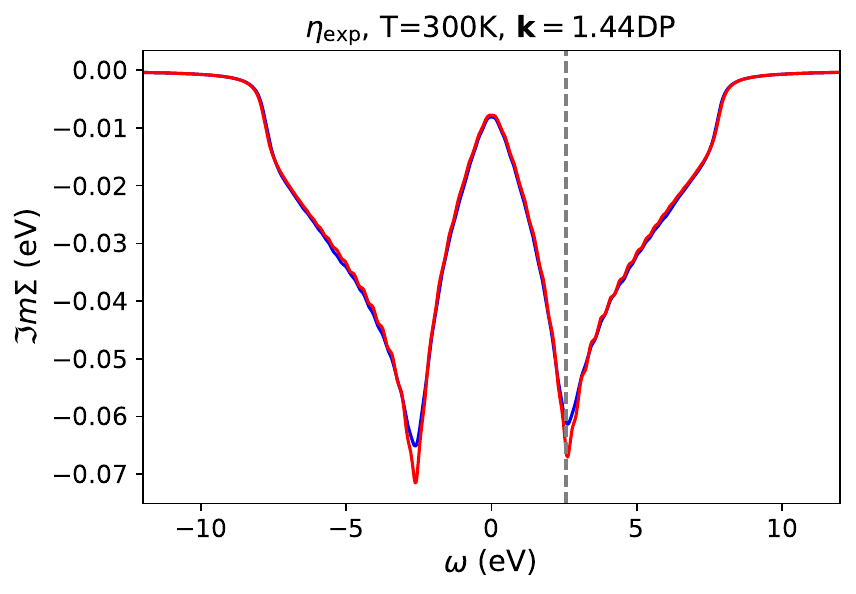}\\
(a)\\
\includegraphics[width=0.22\textwidth]{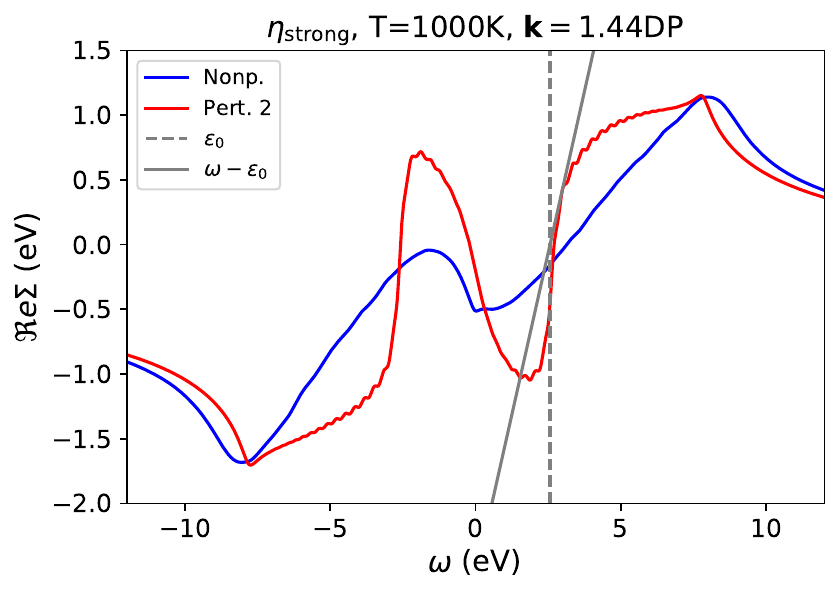}
\includegraphics[width=0.22\textwidth]{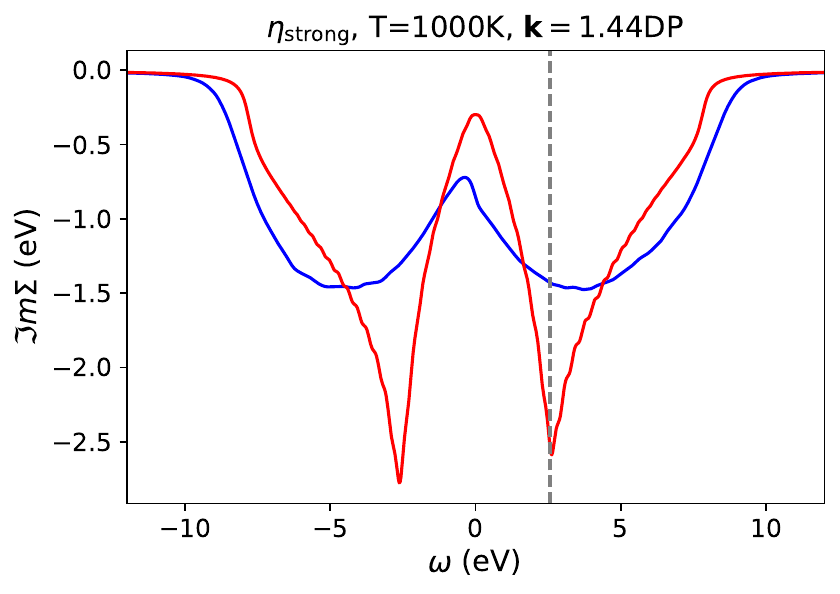}\\
(b)\\
\caption{Real and imaginary part of the self-energy. (a) Parameters of Fig.~\ref{fig:ARPES_3D}(a). In this case, P2 and NP give very similar results, and the $\o-\vare_0$ line (which looks almost vertical) intersects $\Re e \Sigma$ at only one point, resulting in just one peak (which is the usual behavior). We plot the full range of energy, rather than a window around the bare energy, to better visualize the comparison to the strong case. (b) Parameters of Fig.~\ref{fig:ARPES_3D}(c) and (e). The shape of P2 would be exactly the same as in (a), if the temperature were the same.  In any case, the shape changes little; the main change is in the amplitude (compare the $y$-axis scale). Here as well as for other $\mathbf{k}$, $\o-\vare_0$ intersects or is very close to $\Re e \Sigma$ in two points, resulting in the anomalous double peak structure of Fig.~\ref{fig:ARPES_3D}(e). On the other hand, the shape of NP changes significantly, and  $\o-\vare_0$ intersects $\Re e \Sigma$ at one point. Here $\d=0.2$ eV to obtain a smoother curve for P2.}
 \label{fig:self-energy}
\end{figure}

To get a better understanding of why the P2 and NP spectral functions in Fig.~\ref{fig:ARPES_3D} are so different, we plot here also the real and imaginary part of the self-energy. See Fig.~\ref{fig:self-energy}. In P2, the shape of the self-energy at a given temperature does not change with coupling, since the frequency dependence is only contained in the Fan term. The magnitude just changes as $\eta^2$. This explains for example why in P2 there is no peak at $\o=0$. Instead, a small peak is visible in $\Re e \Sigma$ at $\o=0$ at $\eta_\mathrm{strong}$. See also the caption of Fig.~\ref{fig:self-energy}.

\subsection{Alternative self-energy definition}
\label{sec:alternative}

In the main section, we defined the self-energy by averaging the Green's function $\mathcal{G}^I$. This is a good definition, since we saw it reproduces AHC and also higher-order terms of the standard diagrammatic expansion. Here, we consider an alternative definition, by averaging $\Pi^I_\mathbf{k}$ of Eq.~\eqref{eq:Pi}. Writing momentum indices explicitly (and band indices implicitly, and omitting the $I$ index to simplify notation) Eq.~\eqref{eq:DysonV0} can be written

\begin{figure}
\centering
\includegraphics[width=0.22\textwidth]{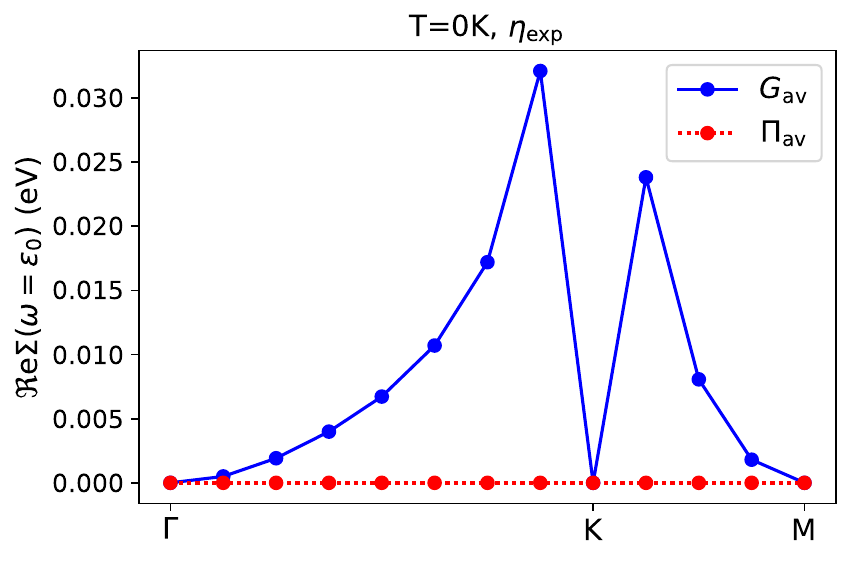}
\includegraphics[width=0.22\textwidth]{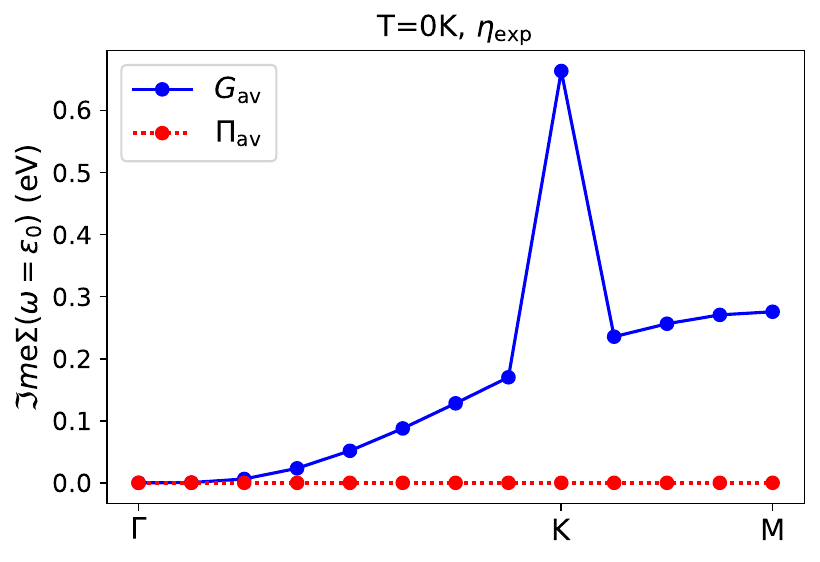}\\
(a) $1 \times 1 \times 1$\\
\includegraphics[width=0.22\textwidth]{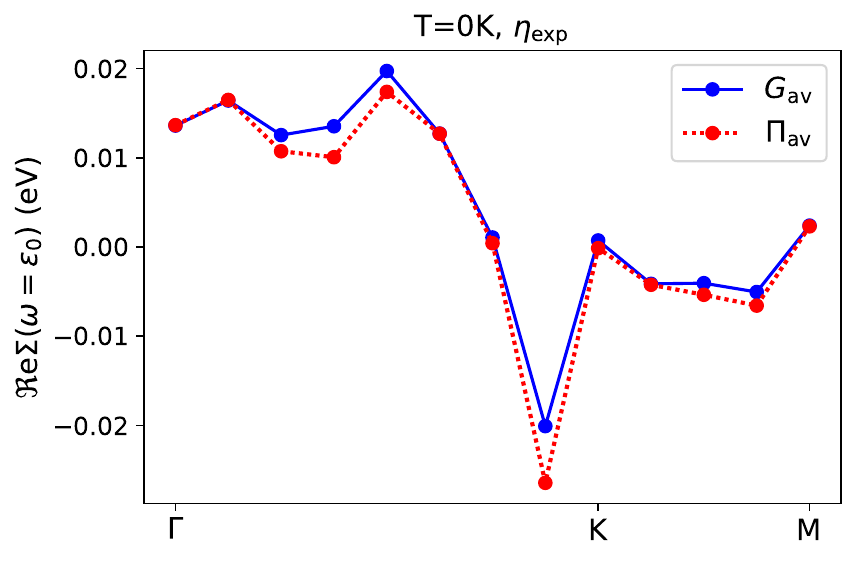}
\includegraphics[width=0.22\textwidth]{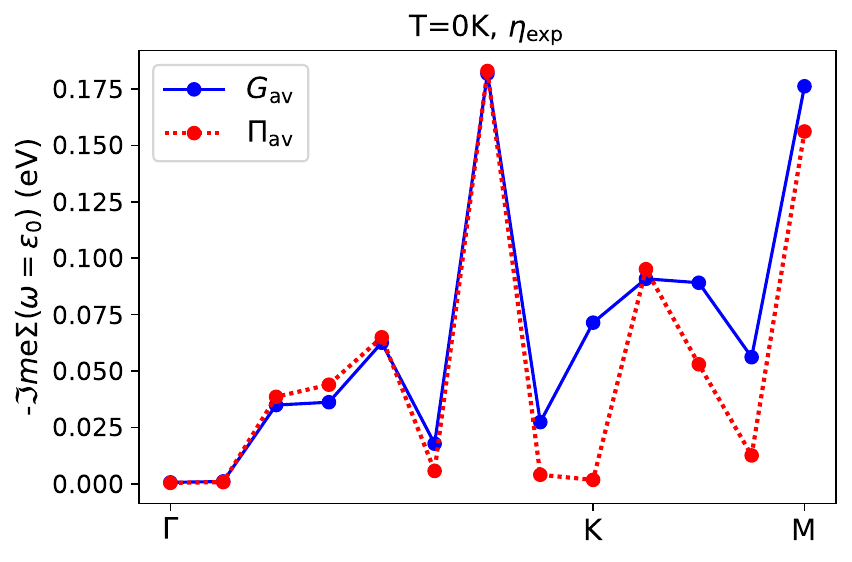}\\
(b) $4 \times 4\times 1$\\
\includegraphics[width=0.22\textwidth]{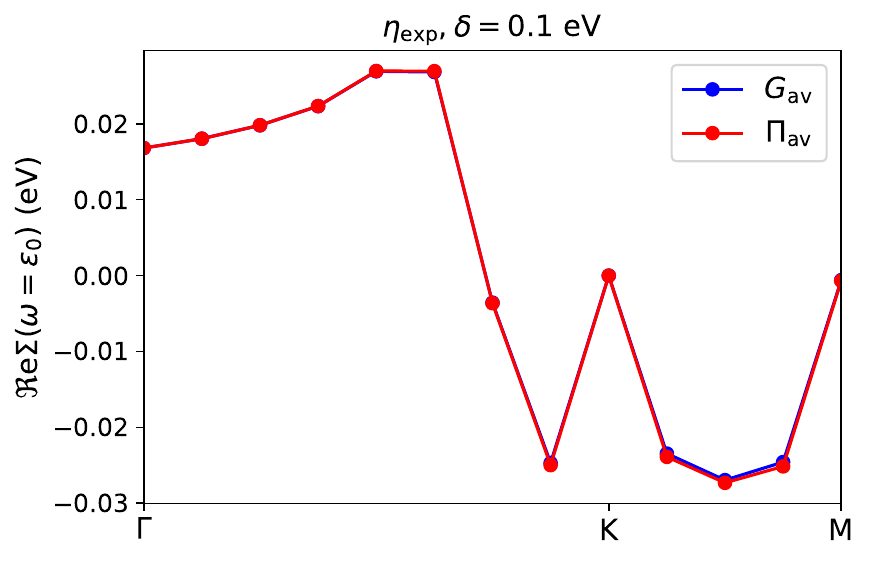}
\includegraphics[width=0.22\textwidth]{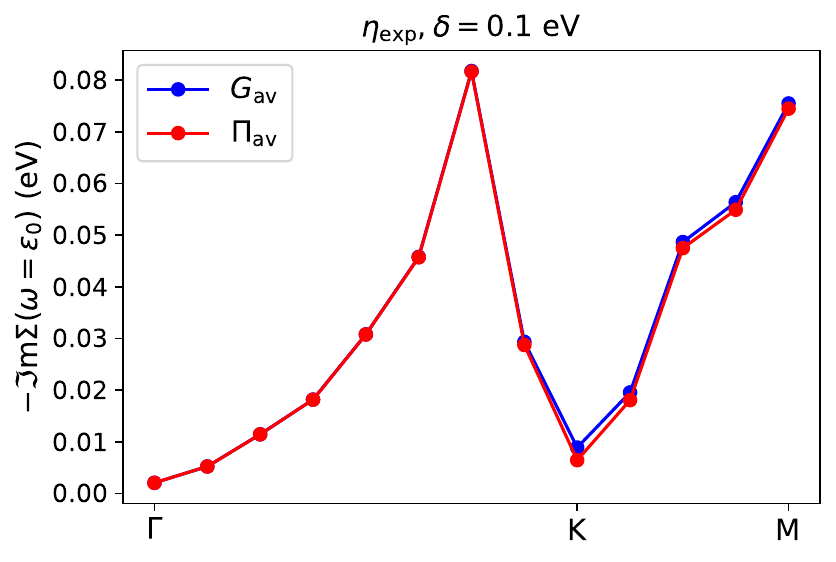}\\
(c) $24 \times 24\times 1$\\
\caption{Comparison between both definitions of the self-energy for different supercells (linearized Hamiltonian Eq.~\eqref{eq:t_linear}). Terms with internal momenta $\mathbf{k}_i$ equal to the external momenta $\mathbf{k}$ are not included in $\Pi_\mathbf{k}$ (that is, the phonon momenta $\mathbf{q}_i$ equal 0). So in the $1 \times 1\times 1$ case, $\Pi_\mathbf{k}=0$ for any $\mathbf{k}$ because $\mathbf{k}_i=\mathbf{k}$ are the only terms. For larger supercells, the difference between both definitions gets smaller.}
\end{figure}

\be
\mathcal{G}_{\mathbf{k}\mathbf{k}} = \mathcal{G}^0_{\mathbf{k}\mathbf{k}} + \sum_{\mathbf{k}'} \mathcal{G}^0_{\mathbf{k}\mathbf{k}} \mathcal{V}_{\mathbf{k}\mathbf{k}'} \mathcal{G}_{\mathbf{k}'\mathbf{k}},
\label{eq:Dyson3}
\ee

\ni From Eq.~\eqref{eq:Pi},

\be
G_{\mathbf{k}} = G^0_\mathbf{k} +  G^0_\mathbf{k} \Pi_\mathbf{k} G_{\mathbf{k}}
\label{eq:DysonPi2}
\ee

\begin{figure}[!ht]
\centering
\includegraphics[width=0.22\textwidth]{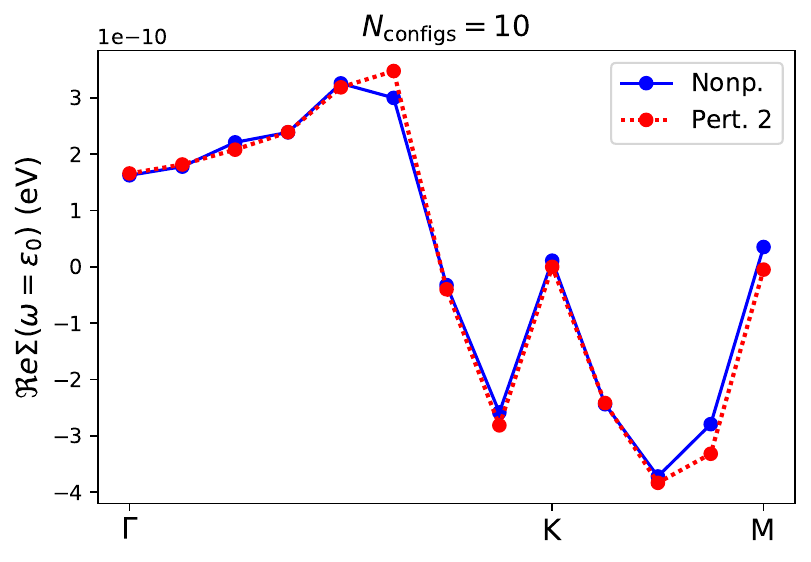}
\includegraphics[width=0.22\textwidth]{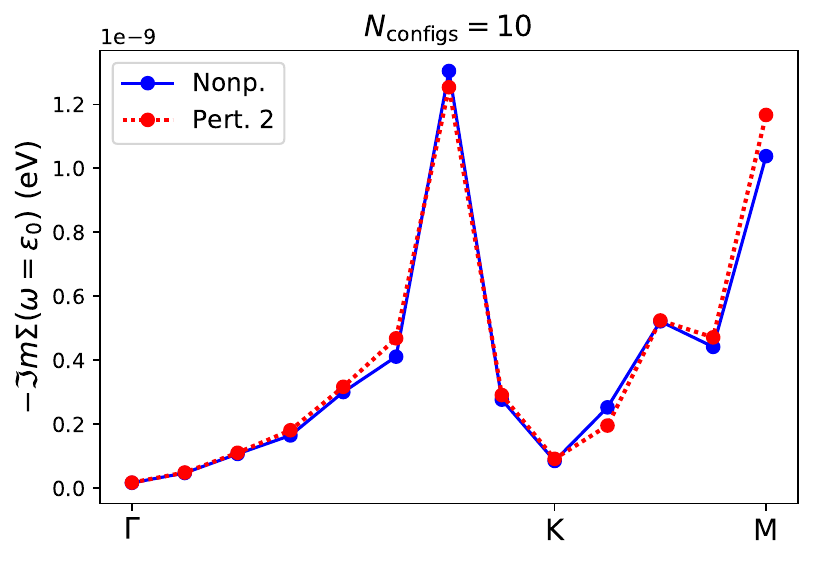}
 \includegraphics[width=0.22\textwidth]{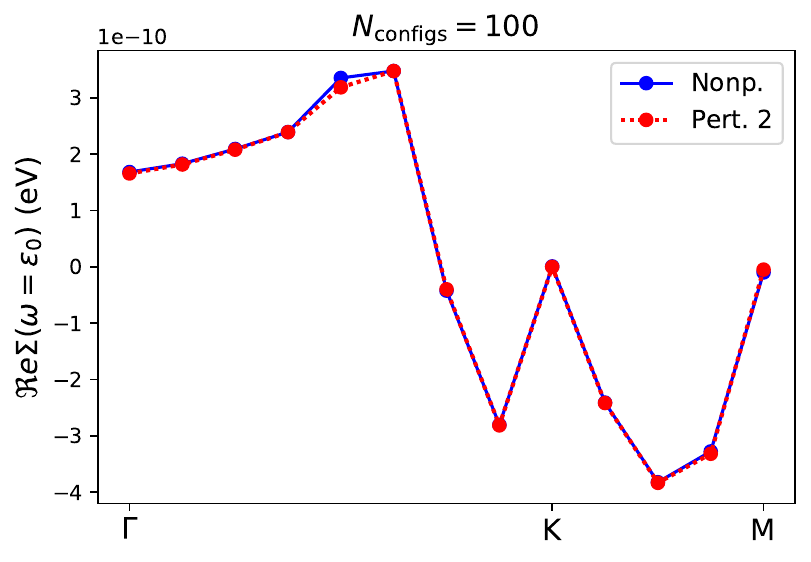}
\includegraphics[width=0.22\textwidth]{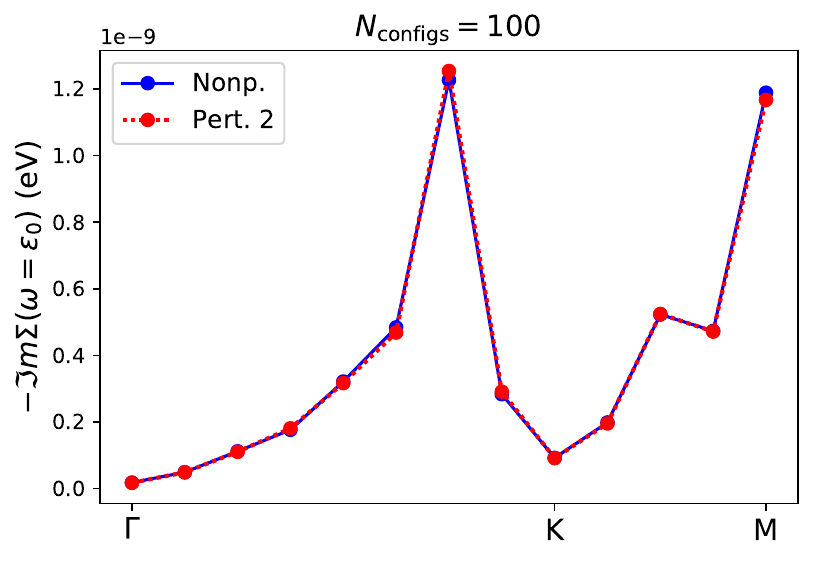}
\includegraphics[width=0.22\textwidth]{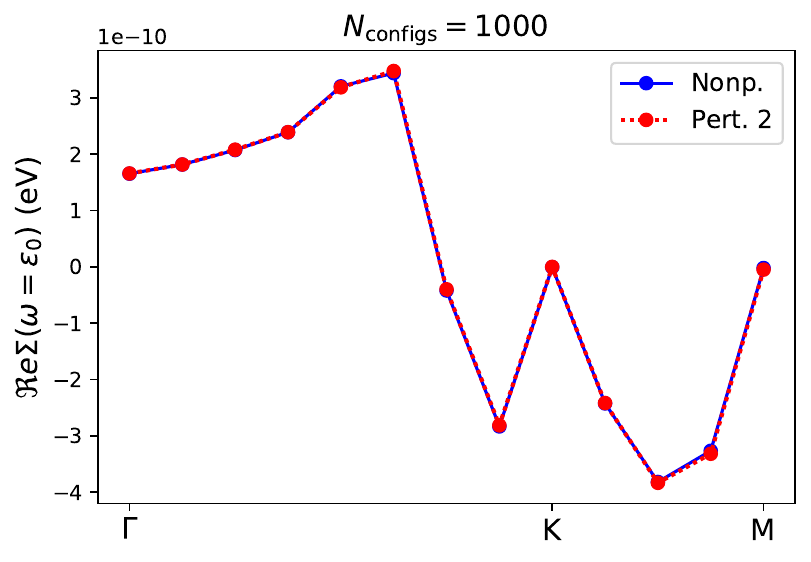}
\includegraphics[width=0.22\textwidth]{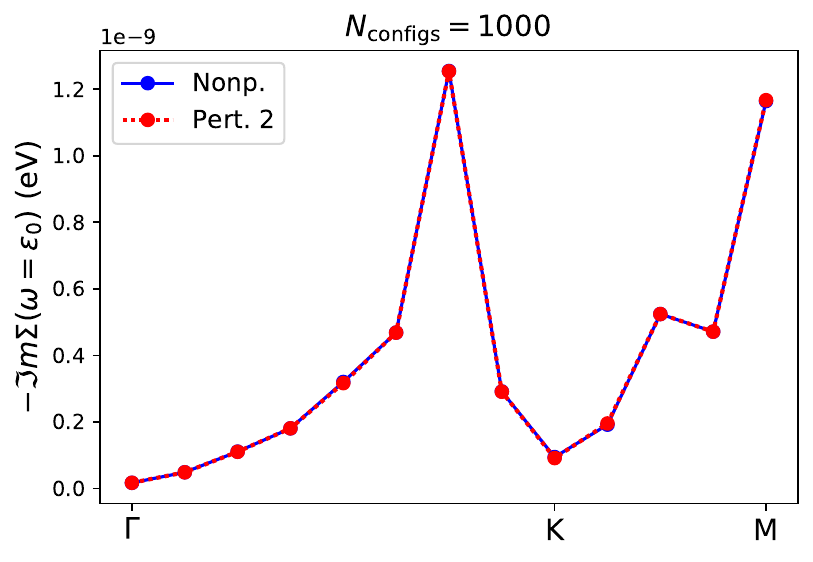}
\caption{Comparison at $\eta_\mathrm{weak}$ between P2 and NP (linearized Hamiltonian Eq.~\eqref{eq:t_linear}), for $N_1=24$ at $T=0$ K, for different number of configurations. At $N_\mathrm{cfgs}=1000$ there is perfect agreement, and even $N_\mathrm{cfgs}=10$ is already good for most $\mathbf{k}$ points.}
\label{fig:conv_N24}
\end{figure}

\ni These equations are matrix equations in the band indices $n,n'$ ($\mathcal{G}_{\mathbf{k'}\mathbf{k}}$, $\mathcal{G}^0_{\mathbf{k}\mathbf{k}}$, $\mathcal{V}_{\mathbf{kk'}}$, $G^0_\mathbf{k}$, $G_{\mathbf{k}\mathbf{k}}$, and $\Pi_\mathbf{k}$ are matrices in such indices). We remind the reader that calligraphic symbols $\mathcal{G}$,$\mathcal{G}^0$,$\mathcal{V}$ are defined in the SC (Hilbert space) and have two momentum indices, while $G,G^0$ are defined in the PC and have one momentum index. In the case of $\mathcal{G}^0$, which is diagonal in $\mathbf{k}$, we can write $\mathcal{G}^0_{\mathbf{k}\mathbf{k}}=\mathcal{G}^0_\mathbf{k}$.

In order to obtain an expression for $\Pi_\mathbf{k}$, we write these equations as

\begin{align}
\mathcal{G}_{\mathbf{k}\mathbf{k}} & = \mathcal{G}^0_\mathbf{k} + \mathcal{G}^0_\mathbf{k} \mathcal{V}_{\mathbf{k}\mathbf{k}} \mathcal{G}^0_\mathbf{k} + \sum_{\mathbf{k}_1} \mathcal{G}^0_\mathbf{k} \mathcal{V}_{\mathbf{k}\mathbf{k}_1} \mathcal{G}^0_{\mathbf{k}_1} \mathcal{V}_{\mathbf{k}_1\mathbf{k}} \mathcal{G}^0_\mathbf{k} + ... \label{Gexp1}\\
G_{\mathbf{k}} & = G^0_\mathbf{k} + G^0_\mathbf{k} \Pi_\mathbf{k} G^0_\mathbf{k} +  G^0_\mathbf{k} \Pi_\mathbf{k} G^0_{\mathbf{k}} \Pi_\mathbf{k} G^0_\mathbf{k} + ...\label{Gexp2}
\end{align}

\ni By definition, $\mathcal{G}_\mathbf{k}=G_\mathbf{k}$ and $\mathcal{G}^0_\mathbf{k}=G^0_\mathbf{k}$ (see after Eqs.~\eqref{eq:Gkn} and \eqref{eq:Gimplicit}). In Eq.~\eqref{Gexp2}, $G^0$ only appears with index $\mathbf{k}$. So, in Eq.~\eqref{Gexp1}, we separate the sums $\sum_{\mathbf{k}_i}$ as $\sum_{\mathbf{k}_i \neq \mathbf{k}} + \sum_{\mathbf{k}_i = \mathbf{k}}$. The terms with $\mathbf{k}_i \neq \mathbf{k}$ can be grouped together, and it can be seen that \cite{CT1996}

\be
\begin{split}
\Pi_\mathbf{k} = & \Pi_\mathbf{k}^{(1)} + \Pi_\mathbf{k}^{(2)} + \Pi_\mathbf{k}^{(3)} + ... \hspace{3cm}\\
\\\
\textrm{with:} & \hspace{2mm} \Pi_\mathbf{k}^{(1)} =  V_{\mathbf{k}\mathbf{k}}\\
& \hspace{2mm}  \Pi_\mathbf{k}^{(n+1)} =  \sum_{\mathbf{k}_1 \neq \mathbf{k},...,\mathbf{k}_n \neq \mathbf{k}} \mathcal{V}_{\mathbf{k}\mathbf{k}_1} \mathcal{G}^0_{\mathbf{k}_1}... \mathcal{G}^0_{\mathbf{k}_n} \mathcal{V}_{\mathbf{k}_n\mathbf{k}}
\end{split}
\label{TaylorPi}
\ee

\ni Averaging, terms have to be grouped in pairs as before, so we recover the same type of structure as $\Sigma = \langle \mathcal{V}^I \rangle_\textrm{irred} + \langle \mathcal{V}^I G^0 \mathcal{V}^I \rangle_\textrm{irred} +...$, because $\mathbf{k}_i \neq \mathbf{k}$ (reducible diagrams cannot be present). So by defining (putting back the $I$ index)

\be
\Pi_{\mathbf{k},nn'} = \langle \Pi^I_{\mathbf{k},nn'} \rangle
\label{eq:Piav}
\ee

\ni we see that $\Pi$ almost coincides with $\Sigma$ in Eq.~\eqref{eq:Sigma}, except for the fact that internal momenta cannot be $\mathbf{k}$. Although both definitions coincide in the thermodynamic limit, for finite supercells $\Sigma$ provides a better definition than $\Pi$ to determine the Green's and spectral functions.

\subsection{Comparison between AHC and NP in the very weak coupling limit}
\label{sec:comparison}

In order to check that the method is correctly implemented, we compared our method to AHC, by using the analytical expression of the electron-phonon matrix element Eqs.~B4 and B50 of Ref.~\onlinecite{Venezuela2011} together with Eq.~\eqref{eq:Fan} of our work. To do so, we linearized the distorted $H^I$,

\be
t^I_{SS'} = t_0 - \f{\eta}{d_0} \pmb{\tau}_i \cdot (\mathbf{u}^I_{S'}-\mathbf{u}^I_S)
\label{eq:t_linear}
\ee

\ni to omit the DW term, and used $\eta_\mathrm{weak}=10^{-4} \eta_\mathrm{exp}$, to make sure there are no contributions from higher order terms.

\begin{figure}
\centering
\subfigure[]{\includegraphics[width=0.22\textwidth]{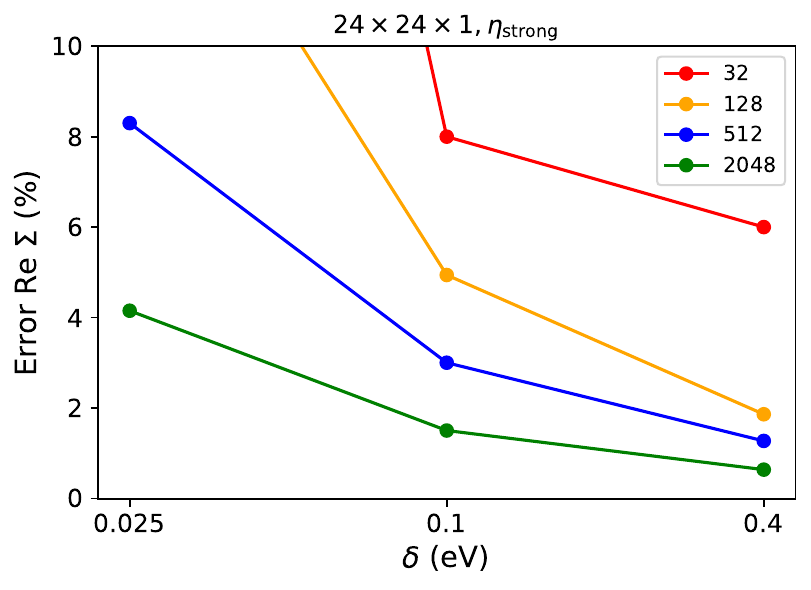}}
\subfigure[]{\includegraphics[width=0.22\textwidth]{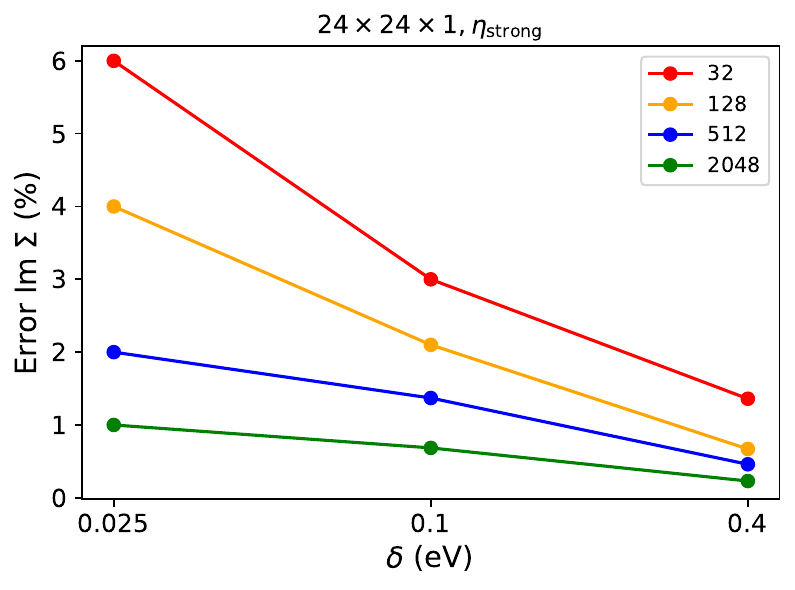}}
\subfigure[]{\includegraphics[width=0.25\textwidth]{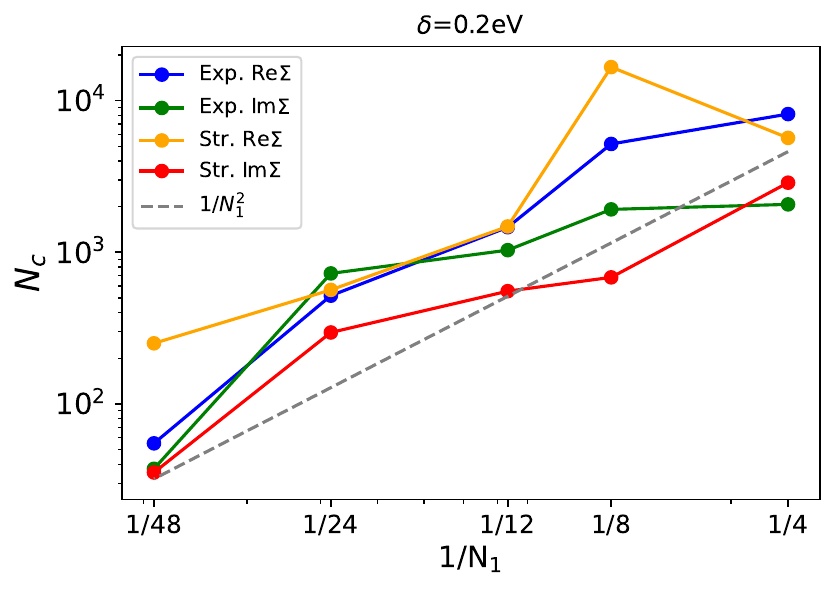}}
\vspace{-2mm}
\caption{(a) and (b): Convergence of the error of the self-energy as a function of $\d$, for different number of configurations (indicated in the label). The error is obtained by doing the ratio between the self-energy (at the bare energy) at successive number of configurations (for the commensurate points along the $\Gamma$-K-M line), and averaging. The error at 2048 configurations is obtained by dividing the error of 512 by $\sqrt{2048/512}=2$. (c) Number of configurations needed to achieve less than 1\% error, as a function of $1/N_1$. The largest number of configurations used for $N_1=4,8,12,24,48$ is 18432, 4608, 2056, 512, 128, respectively. The dashed line is $1/N_1^2$, so the number of configurations needed roughly decreases with the number of atoms. The imaginary part has less error than the real part. 
}
\label{fig:conv_delta}
\end{figure}

In Fig.~\ref{fig:conv_N24}, we can see the comparison between the perturbative AHC self-energy (black) and our non-perturbative approach (blue), for different number of configurations $N_\textrm{cfg}$, for several $\mathbf{k}$ points along the $\Gamma$-K-M path. For $N_\textrm{cfg}=10$ the curves are quite similar, but there are clear differences. At $N_\textrm{cfg}=100$ the values are the same for most points and at $N_\textrm{cfg}=1000$ they overlap perfectly (here the self-energy is evaluated at the bare value of the energy, but the same holds for any $\o$). More configurations are needed for smaller supercells (see Fig.~\ref{fig:conv_delta}).

\subsection{Comparison between P2 and exact analytical results}
\label{sec:conical}

\begin{figure}[h]
\includegraphics[width=0.4\textwidth]{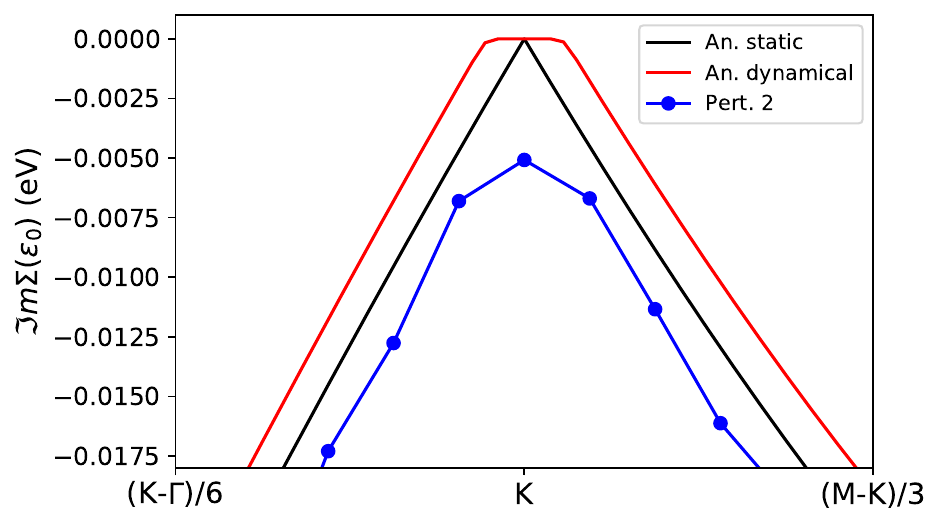}
\vspace{-2mm}
\caption{Comparison of $\Im m \Sigma (\vare_0)$ between the perturbative tight-binding model (with a $N_1=48$ SC and $\d=0.1$ eV), and the exact (analytical) conical model\cite{Calandra2007}, in the static and dynamical cases, where $\vare_0$ is the bare energy of the corresponding $\mathbf{k}$ state in the $\Gamma$-K-M line. The static result is similar to the dynamical one, except close to the Fermi level, where the dynamical part has additional features due to the Fermi-Dirac factors. In the tight-binding model, the dispersion is also conical close to K, and the contribution from states that are further away (where the dispersion is not conical) is smaller because energy denominators are larger. So, the perturbative result in the conical and tight-binding models should be very similar in the $N \rightarrow \infty$ and $\d \rightarrow 0$ limit. The difference gives an estimate of the error incurred by using a finite SC.
} 
\label{fig:analytical_vs_pert}
\end{figure}

To better understand how our results compare to an exact adiabatic and non-adiabatic calculation ($N_1,N_2 \rightarrow +\infty$ and $\d \rightarrow 0$), let us look at Fig.~\ref{fig:analytical_vs_pert}. It includes the analytical imaginary part of the self-energy in a conical model\cite{Calandra2007}, with the same slope around the Dirac point as in the tight-binding model, in the adiabatic and non-adiabatic cases. The self-energy in the tight-binding model should coincide very close to K, where the dispersion is linear. First, we notice that far away from K, the values are quite similar, so the adiabatic approximation works well. Second, the use of a finite SC and a finite $\d$ introduces some error. But we see that the adiabatic exact calculation (black curve) and the perturbative result (blue curve) are similar, again, not too close to K (where the Fermi level lies).

\subsection{Convergence of NP}

\begin{figure}
\centering
\includegraphics[width=0.22\textwidth]{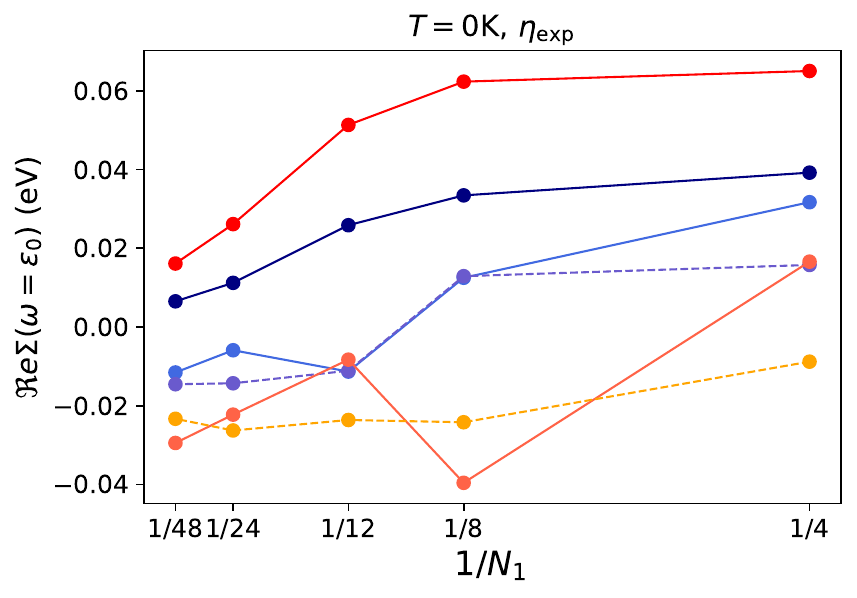}
\includegraphics[width=0.22\textwidth]{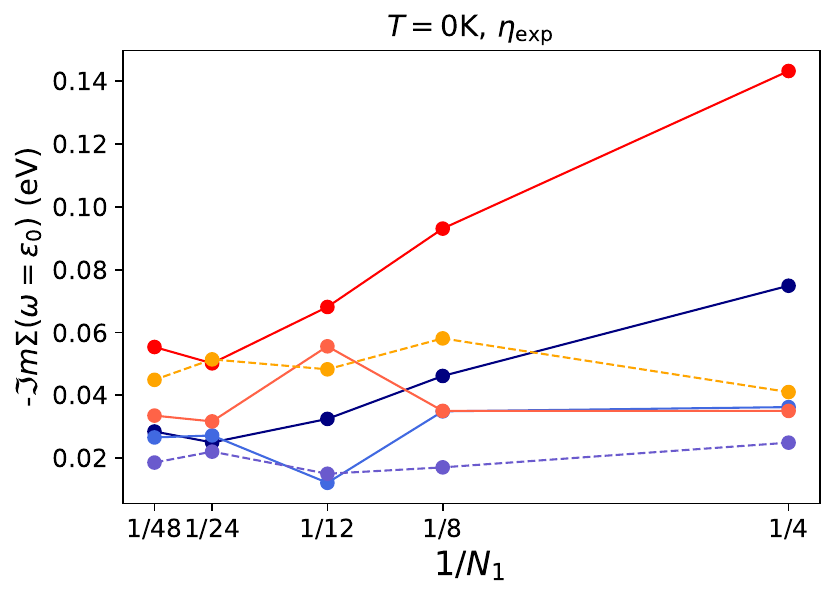}
\includegraphics[width=0.22\textwidth]{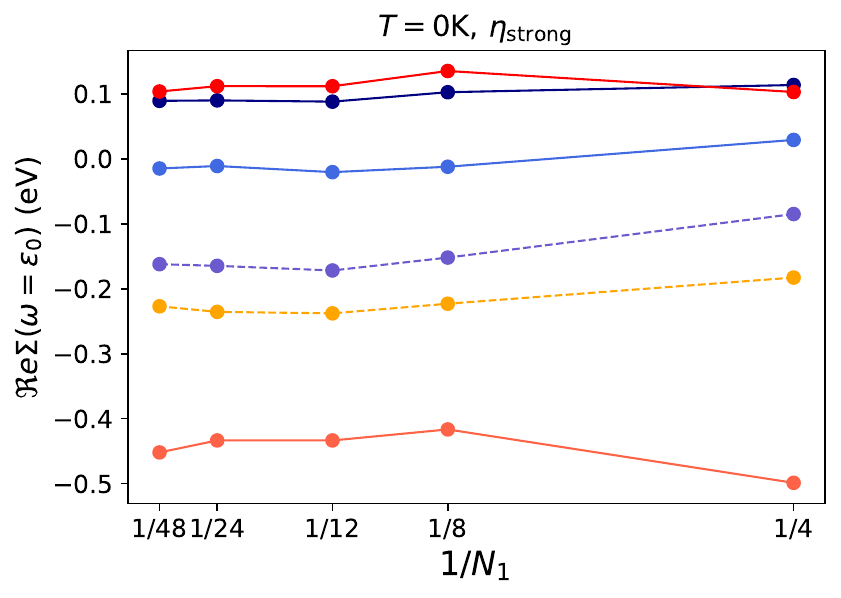}
\includegraphics[width=0.22\textwidth]{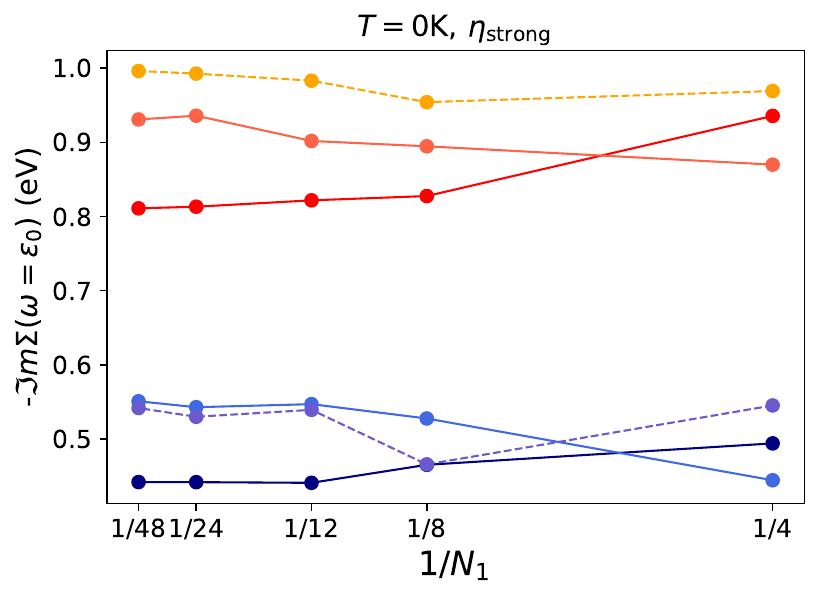}
\vspace{-2mm}
\caption{Convergence of the self-energy with the supercell size $N_1$ for several $\mathbf{k}$ points, at $\eta_\mathrm{exp}$ and $\eta_\mathrm{strong}$ ($\d=0.1$ eV is fixed). At $\eta_\mathrm{exp}$, the real part is not well converged for all $\mathbf{k}$ points, while for the imaginary part arguably at least $N_1=24$ to get a good result (tuning $\d$ for each SC would give more similar results for different SCs). Instead at $\eta_\mathrm{strong}$, convergence is good at already $N_1=8$.}
\label{fig:conv_N}
\end{figure}

In Fig.~\ref{fig:conv_N}, we show how the real and imaginary part of self-energy at the bare value converges as a function of the SC size $N_1$, at the experimental and strong couplings. For $\eta_\mathrm{strong}$, convergence is achieved for smaller supercells. So the stronger the coupling, the more important a NP method is, and the easier it will likely be to achieve convergence. This is important, since SC methods are usually too computationally demanding to be applied to very large SCs.

\subsection{Asymmetry}
\label{sec:asymmetry}

\begin{figure}
\includegraphics[width=0.4\textwidth]{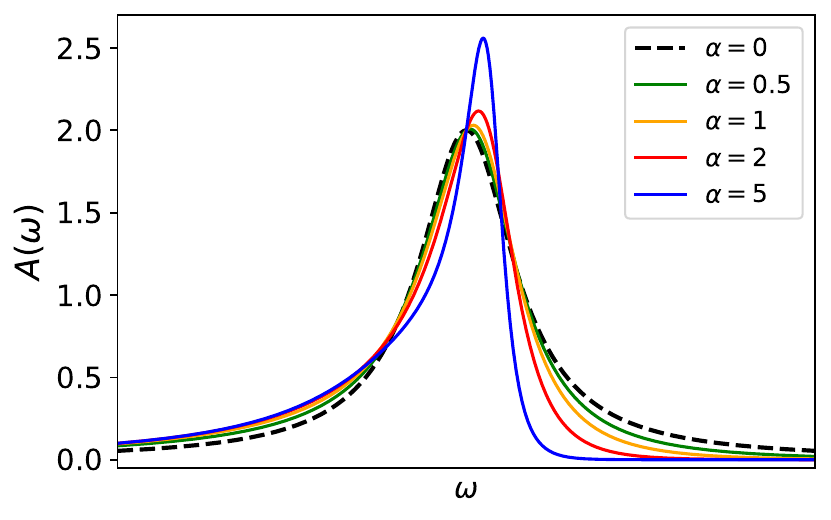}
\vspace{-2mm}
\caption{Spectral function for different values of the asymmetry parameter $\a$. $\a=0$ corresponds to a symmetric Lorentzian. Up to $\a=$ 1 the asymmetric can be considerate moderate, while $\a=5$ gives a very asymmetric shape.}
\label{fig:asymmetry}
\end{figure}

\begin{figure}
\subfigure[]{\includegraphics[width=0.2\textwidth]{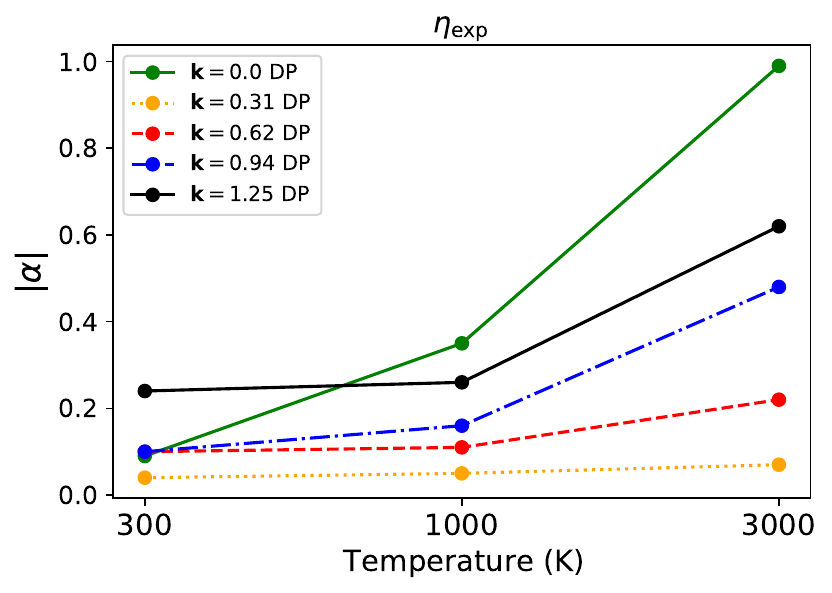}}
\subfigure[]{\includegraphics[width=0.2\textwidth]{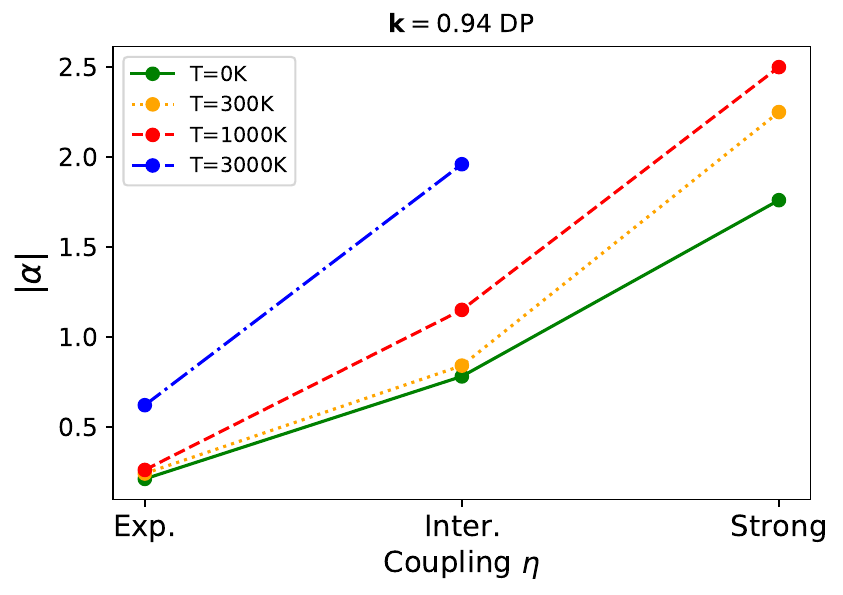}}
\vspace{-4mm}
\caption{(a) Asymmetry parameter $|\alpha|$ vs Temperature for $\eta_\mathrm{exp}$. Asymmetry increases with temperature for all points considered and results are similar at $\eta_\mathrm{inter}$ and $\eta_\mathrm{strong}$. (b) $|\alpha|$ vs Coupling. Asymmetry also increases when increasing the coupling. Thus, the asymmetry of an experimental spectral function could be an indicator of strong coupling, and that a non-perturbative method is required to describe the measurements.}
\label{fig:alpha_T_and_eta}
\end{figure}

With the rescaling $a, \o, \o_0, A, \Gamma_0 \rightarrow \lambda a, \lambda \o, \lambda \o_0, \lambda A, \lambda \Gamma_0$ in Eq.~\eqref{eq:A_as}, the spectral function does not change. This implies that if $a$ and $\Gamma_0$ are rescaled simultaneously in this way, the amount of asymmetry does not change. Fig.~\ref{fig:asymmetry} illustrates how asymmetric the peak is for several values of $\a=a\Gamma_0$ and Fig.~\ref{fig:alpha_T_and_eta} how $|\alpha|$ increases with temperature and coupling.

\subsection{Comparison between including and not including DW in the Hamiltonian}

\begin{figure*}
\centering
\includegraphics[width=0.4\textwidth]{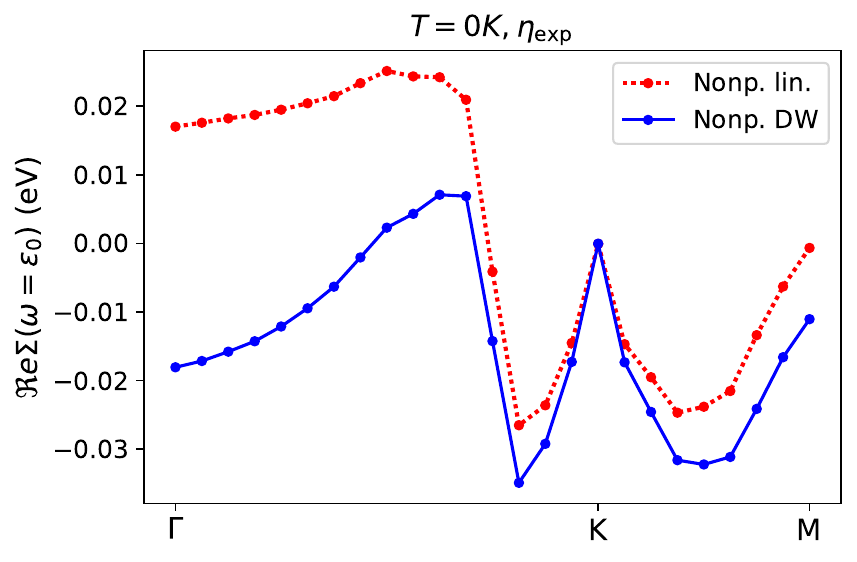}
\includegraphics[width=0.4\textwidth]{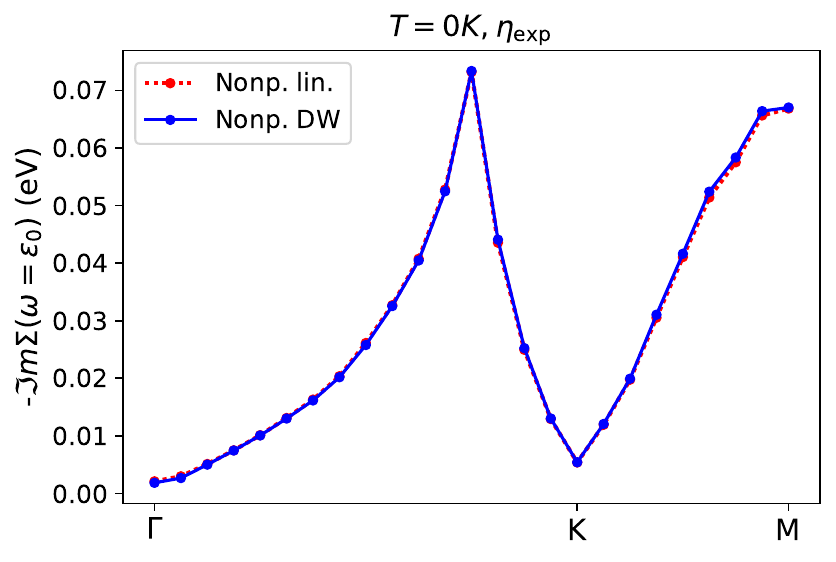}
\includegraphics[width=0.4\textwidth]{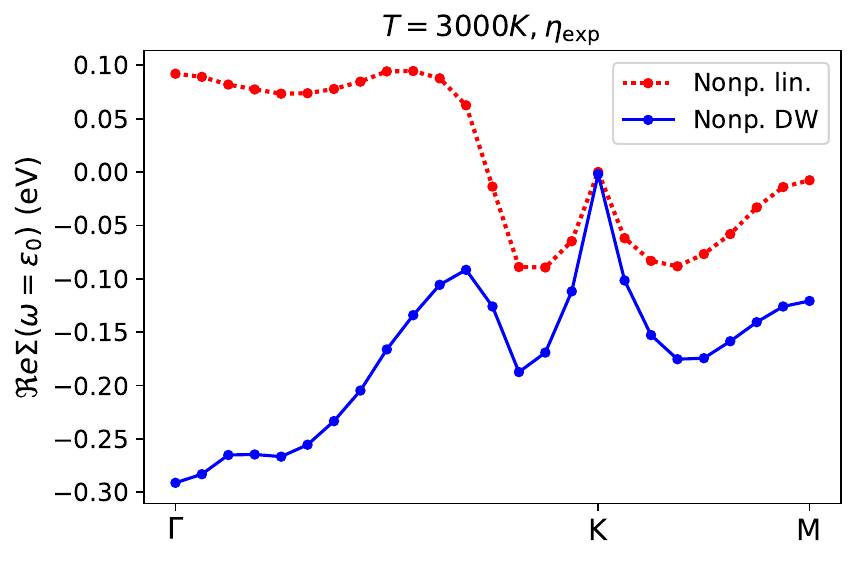}
\includegraphics[width=0.4\textwidth]{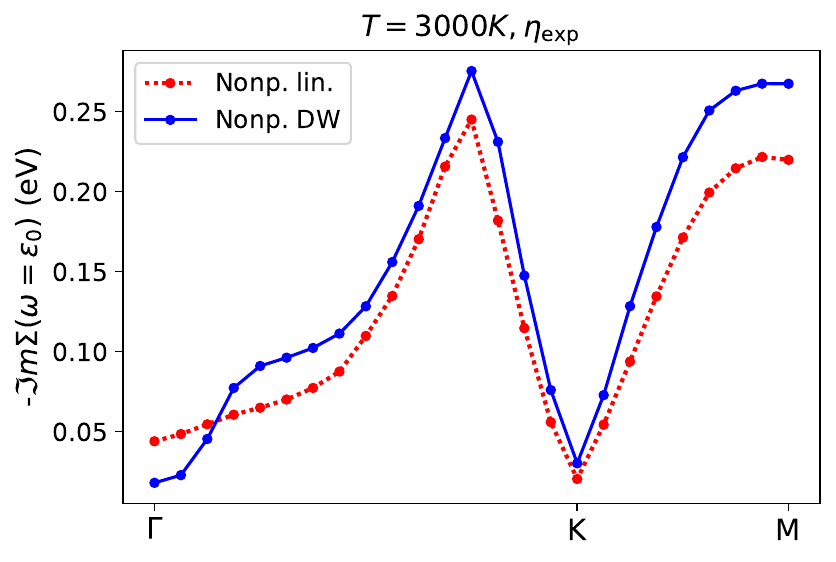}\\
(a)\\
\includegraphics[width=0.4\textwidth]{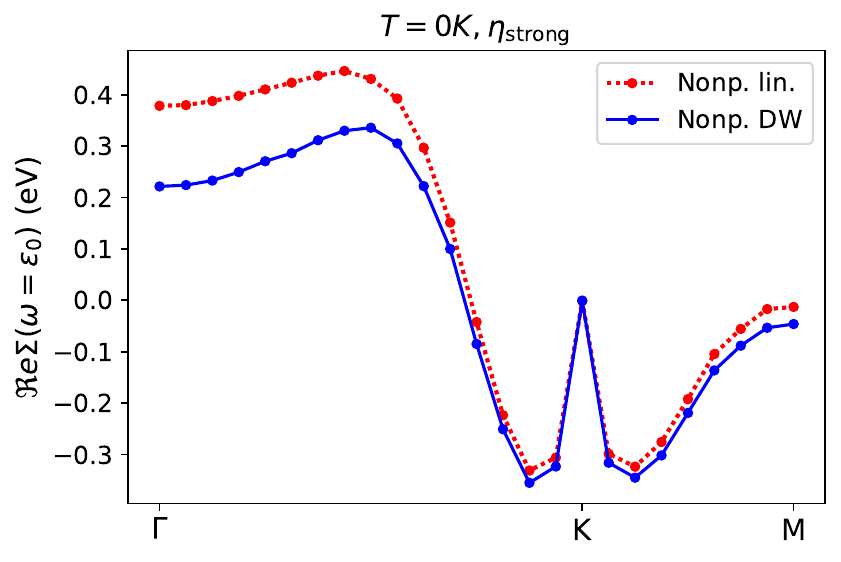}
\includegraphics[width=0.4\textwidth]{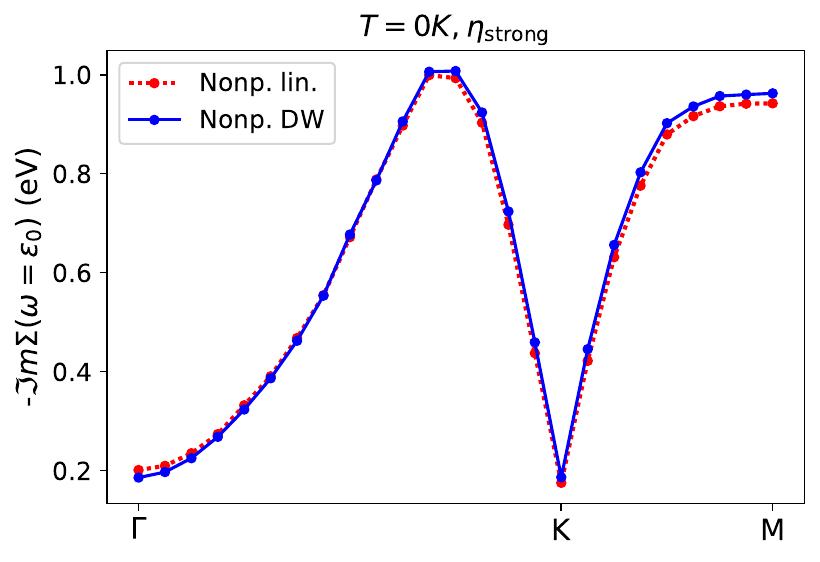}
\includegraphics[width=0.4\textwidth]{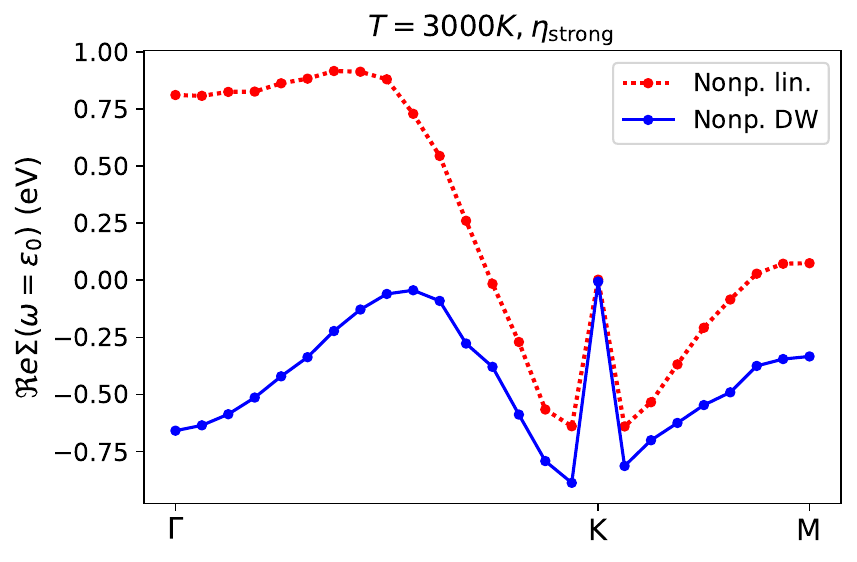}
\includegraphics[width=0.4\textwidth]{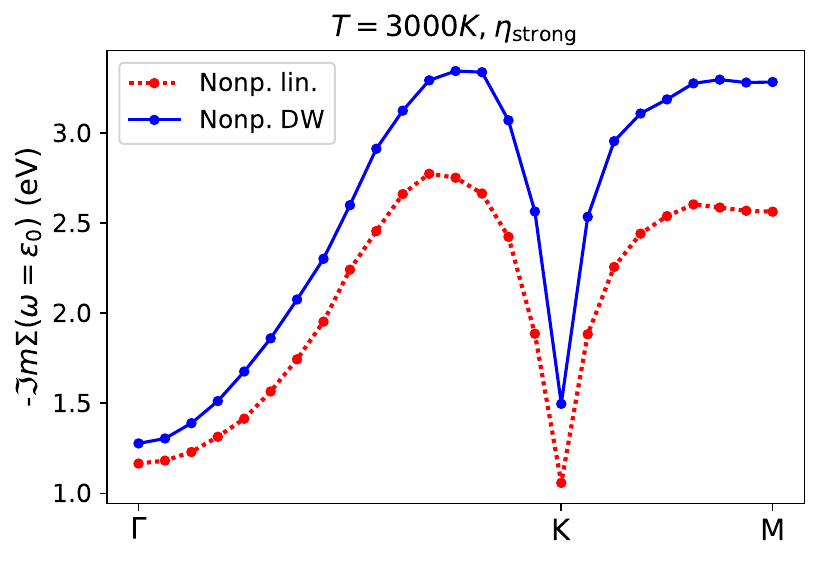}\\
(b)
\caption{Comparison between NP calculations with and without including quadratic (DW) and higher order terms in the Hamiltonian at (a) $\eta_\mathrm{exp}$ and (b) $\eta_\mathrm{strong}$, for both $T=0$ K and $T=3000$ K (that is, using Eqs.~\eqref{eq:hopping} (blue) and ~\eqref{eq:t_linear} (red)). For the real part, contributions from the higher order terms are negative for all $\mathbf{k}$. They do not affect the imaginary part at $T=0$ K (at $\eta_\mathrm{exp}$ and just barely at $\eta_\mathrm{strong}$), but they do at $T=3000$ K (through terms that involve $g$ and $g^\textrm{DW}$). Also, as mentioned earlier, we notice how the DW relative contribution to the Fermi velocity is larger at higher temperatures (for both $\eta_\mathrm{exp}$ and $\eta_\mathrm{strong}$).}
\label{fig:T0_exp_N48_DW}
\end{figure*}

In the main section, we mentioned the contributions from the Fan and DW to the change in the Fermi velocity. Here, we compare calculations linearizing the hopping parameter as in Eq.~\eqref{eq:t_linear}, and using the full distance Eq.~\eqref{eq:hopping} (that is, including DW and higher order terms). See Fig.~\ref{fig:T0_exp_N48_DW}. For the imaginary part, the DW term can be neglected at $T = 0$ K, but not at higher temperatures, which means it contributes through higher order terms ($\Im m \Sigma^\mathrm{DW}=0$, so the DW term does not contribute to lowest order). At stronger coupling, the DW piece reduces $\Re e\Sigma$ at both temperatures, just as for $\eta_\mathrm{exp}$. It is interesting to notice how the DW piece is still barely relevant at $T=0$ K for the imaginary part. At $T=3000$ K, DW contributions play a larger relative role compared to $\eta_\textrm{exp}$ and increase the width of all $\mathbf{k}$ points.

\clearpage
\newpage

\bibliography{bibliography}

\end{document}